\documentclass[10pt]{article}
\usepackage{amssymb}
\usepackage{amscd}
\usepackage{latexsym}
\usepackage{boldtensors}
\usepackage{amsmath}
\usepackage{graphicx}
\renewcommand{\theequation}{\arabic{section}.\arabic{equation}}
\newcommand{\newsec}{\setcounter{equation}{0}\section}%
\newcommand{\Zz}{{\mathbb Z}}

\newcommand{\Rr}{{\mathbb R}}

\def\be{\begin{equation}}
\def\ee{\end{equation}}
\def\bea{\begin{eqnarray}}
\def\eea{\end{eqnarray}}
\def\Tr{{\rm \,Tr\,}}
\def\tr{{\rm \,tr\,}}
\def\d{{\,\rm d}}

\def\0{{\bf 0}}

\def\k{{\bf k}}

\def\p{{\bf p}}
\def\s{{\bf s}}

\def\v{{\bf v}}
\def\x{{\bf x}}
\def\y{{\bf y}}
\def\z{{\bf z}}

\def\P{{\bf P}}

\def\S{{\bf S}}

\def\h2m{\frac{\hbar^2}{2m}}
\def\p0{{P_{\beta H^0_N}}}

\def\boldom{{~\omega}}
\newtheorem{theorem}{Theorem}[section]

\newtheorem{lemma}{Lemma}[section]
\newtheorem{definition}{Definition}[section]
\newtheorem{proposition}{Proposition}[section]

\begin{document}
\title{
\large\bf Proof of phase transition in homogeneous systems of interacting bosons\footnote{Work supported by OTKA Grant No. K109577.}}
\author{Andr\'as S\"ut\H o\\Wigner Research Centre for Physics\\Hungarian Academy of Sciences\\P. O. B. 49, H-1525 Budapest, Hungary\\
E-mail: suto@szfki.hu\\}
\date{December 5, 2017}
\maketitle
\thispagestyle{empty}
\begin{abstract}
\noindent

Using the rigorous path integral formalism of Feynman and Kac we prove London's eighty years old conjecture that during the superfluid transition in liquid helium Bose-Einstein condensation (BEC) takes place. The result is obtained by proving first that at low enough temperatures macroscopic permutation cycles appear in the system, and then showing that this implies BEC. We find also that in the limit of zero temperature the infinite cycles cover the whole system, while BEC remains partial. For the Bose-condensed fluid at rest we define a macroscopic wave function. Via the equivalence of 1/2 spins and hard-core bosons the method extends to lattice models. We show that at low enough temperatures the spin-1/2 axially anisotropic Heisenberg models, including the isotropic ferro- and antiferromagnet and the XY model, undergo magnetic ordering.

\vspace{2mm}
\noindent

\vspace{2mm}
\noindent\textbf{Contents}

\renewcommand\theenumii{\arabic{enumi}.\arabic{enumii}}
\renewcommand\labelenumii{\theenumii}

\begin{enumerate}
\item
Introduction
\item
Path integral formulas
\item
The ideal Bose gas revisited
\item
Cycle percolation
\begin{enumerate}
\item
Representative trajectories
\item
Fluids
\begin{description}
\item[{\rm 4.2.1}]
Liquid helium
\item[{\rm 4.2.2}]
Bose gas with nonnegative interaction
\end{description}
\item
Solids
\end{enumerate}
\item
Bose-Einstein condensation
\item
Reduced density matrix and macroscopic wave function
\item
Phase transition in isotropic and axially anisotropic spin-1/2 Heisenberg models
\item
Summary
\item
Appendix. Some upper bounds
\item
References
\end{enumerate}

\end{abstract}

\newsec{Introduction}

In 1924 Bose gave a deceptively simple statistical physical derivation of Planck's radiation formula [Bos]. To obtain the good result all he had to do was to distribute the light quanta among the cells of volume $h^3$ of the phase space somewhat differently than usual. In an endnote the German translator Einstein praised the work and promised to apply its method to an ideal atomic gas, that he indeed did in three papers [E1-3]. The surprising result, today known as Bose-Einstein condensation (BEC), was not received with much enthusiasm. Distinguished colleagues as Halpern, Schr\"odinger or Smekal had difficulty to understand the new "cell counting" that Bose, and Einstein in [E1], used instead of Boltzmann's. Einstein answered the objections in papers [E2,3] and also in letters, see e.g. [Schr] and the very clear response [E4]. Somewhat later another blow came from Uhlenbeck. To quote London [Lon2],

\noindent
\emph{"This very interesting discovery, however, has not appeared in the textbooks, probably because Uhlenbeck in his thesis {\rm [Uh]} questioned the correctness of Einstein's argument. Since, from the very first, the mechanism appeared to be devoid of any practical significance, all real gases being condensed at the temperature in question, the matter has never been examined in detail; and it has been generally supposed that there is no such condensation phenomenon."}

The regard onto Einstein's work changed in 1938 with the discovery of superfluidity [All, Kap]. Fritz London promptly reacted [Lon1], and in a follow-up paper [Lon2] he detailed his view, that superfluidity must have to do with BEC. To support his idea, he computed the critical temperature of the ideal gas with the mass of the He4 atom and the density of liquid helium, and found it not very far off, 1K above the $\lambda$ point separating the He I and He II phases. Prior to that he had to reexamine the controversy between Einstein and Uhlenbeck, and take Einstein's side. In his derivation Einstein arrived at an equation [in today's notation our Eq.~(\ref{N-mu})] connecting the number of particles $N$ to the chemical potential $\mu$, that one must solve for the latter. $\mu$ appears in an infinite sum over the allowed discrete values of the single-particle momentum. Einstein approximated the sum with an integral and observed the convergence of the integral in three dimensions at $\mu=0$, the largest possible value of the chemical potential: as if $N$ could not go beyond a maximum. At the beginning of his second paper he resolved this paradox by assigning the surplus particles to the zero momentum mode. Uhlenbeck, certainly unaware of Ehrenfest's earlier and identical criticism, argued that without the approximation by an integral the paradox disappears, the original equation can be solved with a $\mu<0$ for arbitrarily large $N$. At that time two crucial mathematical elements were missing from the weaponry of theoretical physicists: a clear notion of the thermodynamic limit and its importance to see a sharp phase transition in the framework of statistical physics, and the appearance of the Dirac delta in probability theory, the fact that in the limit of a sequence of discrete probability distributions an atomic measure can emerge on a continuous background. Einstein's intuition worked correctly, he tacitly performed the thermodynamic limit. The separate treatment of the zero momentum state bothered physicists for a long time, including Feynman, who proposed an alternative derivation based on the statistics of permutation cycles [Fe2]. Simultaneously with London's publications, in a paper written with Kahn, Uhlenbeck also admitted that Einstein was right [Kah]. Tisza published his two-fluid theory about the same time [T1,2], and attributed the specific transport properties of helium II to BEC. However, the argument against describing a strongly interacting dense system of atoms with an ideal gas remained, and set the task: prove BEC in the presence of interaction.

The first consistent theory of superfluidity was given by Landau in 1941 [Lan]. This extremely influential work denied all connection with BEC (clearly, a position taken against London and even more Tisza, who was earlier in his group in Harkov; see also Kadanoff [Kad]). The research on BEC for interacting bosons started after World War II and produced a huge number of papers that we can review only in great lines. Monographs about it and its connection with superfluidity extend over decades, some of them are [Lon4, No, Gr, Sew, Pet, Pi1, Li5, Leg, Ued, Ver, Kag]; those written after 2000 usually cover also the theory of trapped dilute ultra-cold gases of alkaline atoms, that we will not discuss. Maybe the first and certainly one of the most important contributions was that of Bogoliubov, who described superfluidity on the basis of BEC of weakly interacting bosons [Bog1]; for a recent review see [Z]. His theory had an immense impact on the forthcoming research in quantum statistical physics. Bogoliubov initiated the algebraic approach, the use of second quantization. Writing the Hamiltonian in terms of creation and annihilation operators proved to be very fruitful, because it opened the way for diverse approximations. The interaction, which now must be integrable, appears in a quartic expression. Dropping everything not reducible to a quadratic form makes it possible to diagonalize the Hamiltonian (for a discussion of some problems arising with the truncation of the Hamiltonian see the Introduction of [Su10]). These models, including those expressible with number operators and known under the names of mean-field, perturbed mean-field, full diagonal or imperfect Bose gas model, lead naturally to BEC [Hu4, Dav, Fa1, Buf, Ber1, Lew1, Ber2, Do1, Do2]. One of the deep approximations Bogoliubov made was the $c$-number substitution of the operators creating and annihilating a particle of zero momentum; its justification was the subject of later papers [G5, Su8, Su9, Li6]. Another contribution of Bogoliubov, his inequality and $1/q^2$-theorem became the main tool to prove the absence of BEC -- and the breakdown of a continuous symmetry in general -- at positive temperatures in one- and two-dimensional quantum-mechanical models [Bog2].

Analytical methods permit to get closer to the problem of liquid helium. A popular one of the fifties was the use of pseudo-potentials [Hu3, Hu4]. However, the {\em par excellence} analytical method is functional integration. Feynman devised it to solve the time-dependent Schr\"odinger equation and to compute $\left\langle\x\left|e^{-it H/\hbar}\right|\x\right\rangle$ [Fe1]. The mathematical justification for imaginary time $it=\hbar\beta$, where $\beta$ is real positive, as a functional integral with the Wiener measure came from Kac [Kac1,2]. The acknowledgement in [Fe2] reveals that it must have been Kac who convinced Feynman to apply what we call today the Feynman-Kac formula to the $\lambda$-transition of liquid helium. The result was three seminal papers [Fe2-4], the first of which was devoted to a first-principle proof of the $\lambda$-transition. Although Feynman failed in this part of his program -- he went too far in the reduction of the problem to that of the ideal gas --, he seized a point which gained importance with time: the appearance of long permutation cycles during the transition.

Another cardinal contribution from the fifties was due to Oliver Penrose and Onsager [Pen]. These authors found the connection between the expected number of particles in the condensate and the largest eigenvalue of the one-particle reduced density matrix $\sigma_1$. For an alternative characterization of BEC they used the non-decay of the off-diagonal element of the integral kernel of $\sigma_1$, which later was baptized off-diagonal long-range order [Ya]. They gave the first and surprisingly precise estimate of the condensate fraction in the ground state of liquid helium, $\sim 8\%$, only slightly modified later by a sophisticated numerical calculation [Ce] and deduction from experiment [Sn]. They extended their study to positive temperatures and found a connection between BEC and the fraction of particles in "large" permutation cycles.

In 1960 the start of the Journal of Mathematical Physics marked the adulthood of a new discipline. While the theory of BEC seemed more or less settled for most physicists, their more math-minded colleagues saw there a field to explore. The mathematical results of this decade are mostly negative, they prove the absence of BEC in different situations: at low fugacity, in one and two dimensions at positive temperatures, and in one dimension in the ground state. In three thorough and difficult papers Ginibre adapted the method of combining Banach space technics with the Kirkwood-Salzburg equation [Bog3, R2, R4, R5, Bog4] to quantum statistics, and proved the existence, analyticity and exponential clustering of the reduced density matrices at low fugacity in the thermodynamic limit [G2-4]. This remains until today the most elaborate application of the path integral method in statistical physics. For the quantal version of Tonks' hard rod model [Ton] Girardeau established the ground state, easily deducible from that of the free Fermi gas [Gi]. However, even the explicit knowledge of the ground state did not permit to decide quickly about BEC; finally, it was shown not to exist [Schu, Len]. The analogous model with soft-delta interaction is more difficult, it was solved with Bethe Ansatz by Lieb and Liniger for the ground state [Li1] and by Lieb for the excited states [Li2]. There is probably no BEC in this model either, approximate methods predict an algebraic decay of the off-diagonal correlation [Hal, Cr, Ko]. The general one-dimensional case was studied much later, with the conclusion that neither diagonal nor off-diagonal long-range order can exist in the ground state provided that the compressibility is finite [Pi2]. The question of the breakdown of a continuous symmetry at positive temperatures in one and two dimensions was discussed in generality by Wagner [W]. BEC can be considered as an instance of such a breakdown: that of gauge invariance [Fa2, Su8, Su9, Li6]. Its absence was shown by Hohenberg [Ho], the completion of the proof with the extension of Bogoliubov's inequality to unbounded operators was done later [Bou].

In the seventies two schools, one in Leuven directed by Andr\'e Verbeure and another in Dublin under John Lewis' leadership started a research program in quantum statistical physics, with special emphasis on different aspects of BEC. The main tool of the Leuven School was operator algebraic. The Dublin School dominantly worked on BEC in the free Bose gas with different domain shapes, boundary conditions, imperfection, for bosons with spins, etc. The intertwined activity of these two schools is nicely reviewed in Verbeure's book [Ver]. The most important result of the seventies was the long-waited first proof of BEC in a system of interacting bosons in three dimensions. This was done on the cubic lattice for hard-core bosons at half-filling by Dyson, Lieb and Simon [Dy]. The extension to the ground state on the square lattice took another decade [Ken, Kub].

In the mathematical literature of the interacting Bose gas most often either the pair potential or its Fourier transform is chosen to be nonnegative. This guaranties stability; moreover, the estimates are easier, bounds are sharper with a sign-keeping potential. Another reason is specific to dilute systems: the effect of a nonnegative spherical pair potential in the dilute limit reduces to $s$-wave scattering, so the interaction can be characterized by a single parameter, the $s$-wave scattering length $a$. This made it possible to obtain rigorously the ground state energy [Li3,4] and the free energy [Se1, Yi] in the dilute gas limit. Yet another question in which the positivity of the interaction played a role is the sign of the shift $\Delta T_c=T_c-T^0_c$ of the critical temperature in the dilute limit. Here $T_c$ and $T^0_c$ are the critical temperatures of the interacting and the ideal gas, respectively. There is an old intuitive argument by Huang, saying that ${\bf r}$-space repulsion implies $\k$-space attraction [Hu1,2]; so BEC is easier for repulsive bosons, $T_c>T^0_c$. In a long debate a consensus has formed that the shift was positive, and $\Delta T_c\approx 1.3\sqrt{a\rho^{1/3}}$ in 3D, where $\rho$ is the number density. Although the question is pertinent, it cannot be decided without proving $T_c>0$, which was done only for the free Bose gas in a nonnegative external field, where $\Delta T_c>0$ follows from the min-max principle [Kac3]. A rigorous upper bound on $\Delta T_c$ and a review of the related literature can be found in [Se2]. The only case offering the possibility of a comparison with experiment is the $\lambda$-transition of liquid helium. Here the shift is negative [Lon2], but the system is dense, and the pair potential acting in it has an attractive tail. As a matter of fact, it would be possible to prove $\Delta T_c<0$ without proving $T_c>0$. One may then think that even BEC should not be more difficult to prove for, say, the Lennard-Jones potential than for a purely repulsive one.

Permutation cycles necessarily appear in quantum statistical physics whenever bosonic or fermionic systems are treated within the first quantized formalism. Yet, their analysis has not been uniform in time. There has been a revival of interest starting in the nineties. Motivated by a paper of Aizenman and Lieb [Ai1] on the Hubbard model, the present author gave a precise probabilistic formulation for their application to Bose systems [Su1]. Simultaneously, they reappeared in the description of certain quantum spin models [Tot, Ai2]. Dealing with permutation cycles was the central problem in Ceperley's very careful numerical work on liquid helium [Ce]. Further studies of Bose systems based on the analysis of the distribution of permutation cycles followed [Bun, Scha, Su2, Do3, Ben, Uel1, Uel2], with specific results on the ideal gas and mean-field type models. Later on, the investigation has been extended to the statistics of cycles in interesting new models of random permutations [Bet1-3].

The results of the present work are obtained through a study of permutation cycles for interacting bosons as they come about in the Feynman-Kac representation of the partition function. The key ideas are as follows. First, we write down a recurrence relation expressing the $N$-particle partition function $Q_N$ as a sum over $n$ of $Q_{N-n}$, where $n$ is the length of the cycle containing a marked particle, say, particle 1. The other entries of the summand are the single-particle partition function at inverse temperature $n\beta$ which is easily estimable, and an average over trajectories that can be considered as the configurational partition function for $n$ interacting particles in the annealed random field of $N-n$ other particles. The second idea is to replace the average by a single representative term. The choice of a representative set of trajectories demands some reflection, it must be typical in some sense for a fluid phase. If one divides the recurrence equation by $Q_N$, one obtains a resolution of the unity that can be interpreted as the sum of the probabilities that particle 1 is in a cycle of length $n$. Taking the thermodynamic limit under the summation sign provides the probability that in the infinite system 1 is in a finite cycle. The third idea is to replace the limit of $Q_{N-n}/Q_N$ by $z^n$ where $z$ is the fugacity computed in the canonical ensemble and depending on the density and the temperature. This dependence is unknown; we only know that in the case of positive interactions, for $z<1$ (actually for $z\leq 1$), in the infinite system each particle is in a finite cycle with probability 1, and there is no BEC [Uel2]. The formula thus obtained opens the way to prove macroscopic cycle percolation -- the appearance of infinite cycles in the thermodynamic limit, which contain a positive fraction of the total number of particles -- in systems of bosons that remain in a fluid phase down to zero temperature. Examples are the helium liquid and repulsively interacting dilute Bose gases. The proof cannot be done without making an ergodic hypothesis concerning the imaginary-time dynamics of particles in fluids. After this, one can turn to the problem of BEC. To solve it, one must find out the physical meaning of the permutation cycles that appear until now as mere mathematical objects. The clue is given by the ideal gas where particles in the same cycle have the same momentum. In the interacting case it is possible to bring the partition function to a form in which the particles composing the cycle of 1 are in the same one-particle state $\varphi$ with a certain probability $p_\varphi$ that spreads over an infinite set of states. Then, we can prove that BEC implies cycle percolation and that, under a conjecture about the spectrum of a Schr\"odinger operator on the torus, macroscopic cycle percolation implies BEC. The conclusion is that macroscopic cycle percolation and BEC are simultaneous phenomena. The condensate fraction is obtained as $\sum|\langle\varphi,\varphi_\0\rangle|^2 p_\varphi$, where $\varphi_\0$ is the zero momentum plane wave and $\sum p_\varphi$ is the probability that particle 1 is in an infinite cycle. Part of the result is that $\sum p_\varphi$ goes to 1 as the temperature goes to zero while $\sum|\langle\varphi,\varphi_\0\rangle|^2 p_\varphi$ remains smaller than 1. This suggests that superfluidity, which becomes complete at zero temperature, is more fundamentally connected to cycle percolation and the fact that a macroscopic number of particles are in the same state {\em different} from $\varphi_\0$, than to BEC. The idea of the proof outlined above can be applied to prove a phase transition in certain spin-1/2 models on hypercubic lattices in three dimensions and above. Half spins on a lattice are equivalent to hard-core bosons. If the Hamiltonian, written in terms of a lattice gas, is invariant under particle-hole transformation, then at all temperatures the canonical free energy takes its minimum as a function of the density at half filling. Thus, the fugacity is identically 1 at half filling for all temperatures. The condition of invariance under particle-hole transformation is fulfilled by the whole family of axially anisotropic Heisenberg models, including the isotropic ferromagnetic and antiferromagnetic Hamiltonians and the XY model. Half filling corresponds to restriction to the subspace where the third component of the total spin is zero. We prove the phase transition by showing that within this subspace at low enough temperatures there is either axial ordering with phase separation or planar ordering -- the latter corresponds to BEC. For the XY and antiferromagnetic models our finding is not new, the proof of their phase transition was among the results of Dyson, Lieb and Simon [Dy]. The proof of ordering of the ferromagnets is new: in the classical case Fr\"ohlich, Simon and Spencer could do it by reflection positivity [Fr], but this method is not applicable to the quantum ferromagnet. The isotropic model and all those with dominant planar components are lattice analogues of liquids of light bosons of various mass, ready to undergo BEC. Those with a dominant vertical component are analogues of liquids of heavier bosons that will crystallize under cooling.

The paper is organized as follows. Section 2 presents the Feynman-Kac formula for the canonical partition function and fixes the notations. In Section 3 we return to the ideal Bose gas and rederive some of our old results [Su1, Su2] about the connection between cycle percolation an BEC. The proof we give here for the macroscopic cycle percolation serves as a reference in later sections. The longest part of the paper is Section 4. It starts with a precise definition of cycle percolation and macroscopic cycle percolation. Section 4.1 introduces the recurrence relation for the partition function and describes the family of sets of trajectories from which the representative set is taken. Section 4.2 contains the proof of macroscopic cycle percolation in Bose liquids. In Section 4.3 we comment on Bose solids. The proof of a crystalline phase transition is probably the deepest open problem of equilibrium statistical physics. Crystallization must be a classical event, because it occurs at relatively high temperatures. Yet, one may ask how should one conceive a solid at very low temperatures when quantum effects are surely present. We suggest that it should be considered as an instance of many-body localization. Section 5 connects cycle percolation and BEC in interacting systems, thereby extending the result valid for the ideal gas. In Section 6 we examine the relation between the condensate density and the off-diagonal long-range order parameter, and define a macroscopic wave function. The main result of the section is Eq.~(\ref{cond-fraction-macroscopic}) that expresses the condensate fraction as the square of the $L^1$-norm of a macroscopic wave function multiplied with the probability of cycle percolation. Section 7 contains the proof of phase transition in spin-1/2 lattice models. The paper ends with a Summary and an Appendix.

\newsec{Path integral formulas}

We consider $N$ bosons on a $d\geq 3$-dimensional torus of side length $L$, i.e., in the cube $ \Lambda=(-L/2,L/2]^d$ taken with periodic boundary conditions. They interact via a pair potential $u$ that we suppose to depend only on the inter-particle distance and be tempered,
\be\label{tempered}
|u(\x)|\leq C/|\x|^{d+\eta},\quad |\x|\geq R
\ee
with some $C,\eta,R>0$. Further conditions will be given later. Working on the torus makes it necessary to periodize $u$,
\be\label{uL}
u_L(\x)=\sum_{\z\in\Zz^d}u(\x+L\z).
\ee
Owing to temperedness, the sum is convergent. The Hamiltonian is
\be
H_{N,L}=-\frac{\hbar^2}{2m}\sum_{i=1}^N\frac{\partial^2}{\partial\x_i^2}+U_{N,L}=H^0_{N,L}+U_{N,L}, \quad U_{N,L}(\x_1,\ldots,\x_N)=\sum_{i<j}u_L(\x_j-\x_i).
\ee
The canonical partition function in path integral representation reads
\be\label{Q_N}
Q_{N,L}=\Tr P_+e^{-\beta H_{N,L}}=\frac{1}{N!}\sum_{\pi\in S_N} \prod_{i=1}^N\int_\Lambda\d\x_i\int W^\beta_{\x_i\x_{\pi(i)}}(\d\boldom_i)\ e^{-\beta {\cal U}\left(\boldom^{N}\right)},
\ee
where $\boldom^{N}=(\boldom_1,\ldots,\boldom_N)$,
\be\label{calU}
{\cal U}\left(\boldom^{N}\right)=\beta^{-1}\int_0^\beta U_{N,L}\left(\boldom^{N}(t)\right)\d t= \beta^{-1}\int_0^\beta U_{N,L}(\boldom_1(t),\ldots,\boldom_{N}(t))\d t
\ee
and $P_+$ denotes the orthogonal projection to the symmetric subspace of the $N$-particle Hilbert space. To simplify the notation, we shall drop the subscript $L$ from $H_{N,L}$, $U_{N,L}$, and $Q_{N,L}$. In Eq.~(\ref{Q_N}), $S_N$ is the group of permutations of $N$ elements and $W^\beta_{\x\y}$ is the Wiener measure for Brownian bridges on the torus: Let $f$ be a $\Lambda$-periodic functional over the space of trajectories in $\Rr^d$ parametrized by $t\in [0,\beta]$, where periodicity is understood in the sense that $f$ depends only on $\{\boldom(t)(\rm{mod}\,\Lambda)\}_{t\in [0,\beta]}$. Then by definition
\be\label{W}
\int W^\beta_{\x\y}(\d\boldom)f(\boldom)=\sum_{\z\in\Zz^d}\int P^\beta_{\x,\y+L\z}(\d\boldom)f(\boldom).
\ee
Here $P^\beta_{\x\y}$ is the Brownian bridge measure in $\Rr^d$; it is defined by extension from functionals that depend only on a finite number of points of the trajectories [G1]. In the simplest case $f$ depends only on a single point $\boldom(t_0)$; then
\be\label{one-point}
\int P^\beta_{\x\y}(\d\boldom)f(\boldom(t_0))=\int\d\x'\psi_{t_0}(\x'-\x)f(\x')\psi_{\beta-t_0}(\y-\x'),
\ee
where
\be
\psi_t(\x)=\lambda_t^{-d}e^{-\pi \x^2/\lambda_t^2}.
\ee
In this equation $\lambda_t$ is the thermal de Broglie wave length at inverse temperature $t$,
\be\label{lambda}
\lambda_t=\sqrt{2\pi\hbar^2t/m}.
\ee
Setting $f=1$ and using the contraction property
\be
\int\d\x'\psi_{t_0}(\x'-\x)\psi_{\beta-t_0}(\y-\x')=\psi_\beta(\y-\x),
\ee
one obtains the (positive) measures of the set of trajectories $\{\boldom|\boldom(0)=\x,\boldom(\beta)=\y\}$:
\be\label{norms}
\int P^\beta_{\x\y}(\d\boldom)=\lambda_\beta^{-d}e^{-\pi(\x-\y)^2/\lambda_\beta^2},\quad \int W^\beta_{\x\y}(\d\boldom)=\lambda_\beta^{-d}\sum_{\z\in\Zz^d}e^{-\pi(\x-\y+L\z)^2/\lambda_\beta^2}.
\ee
When a trajectory figuring in the right member of Eq.~(\ref{W}) is represented on the torus, $\z$ corresponds to its winding vector. Note that both measures are determined by the kinetic energy operator $H^0_1=-\frac{\hbar^2}{2m}\Delta$ through the integral kernel of $e^{-\beta H^0_{1,L}}$:
\be\label{P-W-continuum}
\int W^\beta_{\x\y}(\d\boldom)=\left\langle\y\left|e^{-\beta H^0_{1,L}}\right|\x\right\rangle,\quad\int P^\beta_{\x\y}(\d\boldom) =\lim_{L\to\infty}\left\langle\y\left|e^{-\beta H^0_{1,L}}\right|\x\right\rangle.
\ee
From here Eqs.~(\ref{norms}) can also be obtained by inserting the complete system of eigenvectors of $H^0_{1,L}$ on the torus and using the Poisson summation formula.

Averaging over the symmetric group in Eq.~(\ref{Q_N}) is facilitated by the fact that the summand is constant within each conjugation class, so $Q_N$ can be written as a weighted sum over the conjugation classes. There are different ways to do this, and each corresponds to a specific resolution of the unity. Ginibre's choice [G1-4] was
\be
1=\sum_{n=1}^N\frac{1}{n!}\sum_{p_1=1}^N\cdots\sum_{p_n=1}^N\delta_{\sum p_i,N}\prod_{i=1}^n\frac{1}{p_i},
\ee
ours is
\be\label{resol}
1=\frac{1}{N}\left(1+\sum_{n_1=1}^{N-1}\frac{1}{N-n_1}\left(1+\sum_{n_2=1}^{N-n_1-1}\frac{1}{N-n_1-n_2}
\left(1+\cdots\right)\right)\right),
\ee
obtained by "skew" simplification from
\be\label{resol2}
1=\frac{1}{N!}\sum_{n_1=1}^N\frac{(N-1)!}{(N-n_1)!}\sum_{n_2=1}^{N-n_1}\frac{(N-n_1-1)!}{(N-n_1-n_2)!}\cdots.
\ee
To arrive at Eq.~(\ref{resol2}) we write $\pi=\pi_1\pi_2\pi_3\cdots$ where $\pi_1$ is the cycle that contains 1, $\pi_2$ is the cycle that contains the smallest number not in $\pi_1$, $\pi_3$ is the cycle that contains the smallest number not in $\pi_1$ and $\pi_2$, and so on. The number of permutations with 1 in a cycle of length $n_1$ is
$${N-1\choose n_1-1} (n_1-1)!=\frac{(N-1)!}{(N-n_1)!},$$
and the other factors are obtained similarly. If we insert $\exp\left\{-\int_0^\beta U_N(\boldom_1(t),\ldots,\boldom_N(t))\d t\right\}$ into Eq.~(\ref{resol}) under the summation signs, factorize the exponential, and perform the integrations, we find (see also [Uel1,2])
\bea\label{recur-gen}
Q_N &=&\frac{1}{N}\sum_{n=1}^N \int_\Lambda\d\x\int W^{n\beta}_{\x\x}(\d\boldom_0)e^{-\beta{\cal U}_{n}(\boldom_0)} Q_{N-n}(\boldom_0) \nonumber\\
&=&\sum_{n=1}^N \frac{L^d}{N}\int W^{n\beta}_{\0\0}(\d\boldom_0)e^{-\beta{\cal U}_{n}(\boldom_0)} Q_{N-n}(\boldom_0).
\eea
For the second equality we used translation invariance. In Eq.~(\ref{recur-gen})
\be\label{calU}
{\cal U}_{n}(\boldom_0)=\sum_{k=0}^{n-1}{\cal U}_{n,k}(\boldom_0),\qquad
{\cal U}_{n,k}(\boldom_0)=\frac{1}{\beta}\int_0^\beta\frac{1}{2}\sum_{l\neq k} u_L\left(\boldom_0(l\beta+t)-\boldom_0(k\beta+t)\right)\d t,
\ee
\be\label{QN-n}
Q_{N-n}(\boldom_0)=\frac{1}{(N-n)!}\sum_{\pi\in S_{N-n}}\prod_{i=1}^{N-n}\int_\Lambda\d\x_i\int W^\beta_{\x_i\x_{\pi(i)}}(\d\boldom_i)\ e^{-\beta{\cal U}\left(\boldom^{N-n}\right)}e^{-\beta V_n\left(\boldom_0,\boldom^{N-n}\right)},
\ee
and
\be
V_n\left(\boldom_0,\boldom^{N-n}\right)=\sum_{k=0}^{n-1} V_{n,k}\left(\boldom_0,\boldom^{N-n}\right),\qquad V_{n,k}\left(\boldom_0,\boldom^{N-n}\right)= \frac{1}{\beta}\int_0^\beta \sum_{i=1}^{N-n}u_L\left(\boldom_0(k\beta+t)-\boldom_i(t)\right)\d t.
\ee
Without the last exponential factor, $Q_{N-n}(\boldom_0)$ is just $Q_{N-n}$.

Bose-Einstein condensation can be seen on the non-vanishing of the condensate fraction $\langle N_\0\rangle/N$ in the thermodynamic limit. Here $\langle N_\k\rangle$ is the canonical expectation value of the occupation number operator $N_\k$ for the single-particle state $\varphi_\k(\x)=L^{-d/2}e^{i\k\cdot\x}$. The eigenvalues of the  one-particle reduced density matrix             \be
\sigma_1=\frac{N}{Q_N}\Tr_{2\ldots N}P_+e^{-\beta H_{N}}
\ee
are $\langle N_\k\rangle$, $\k\in\frac{2\pi}{L}\Zz^d$ (see e.g. [Su6]); among them $\langle N_\0\rangle$ is the largest, and $\langle N_\0\rangle/N$ is just the order parameter $A_1$ introduced by Penrose and Onsager [Pen]. Indeed, from the spectral resolution of $\sigma_1$,
\be\label{sigma1}
\sigma_1=\sum_{\k\in\frac{2\pi}{L}\Zz^d}\langle N_\k\rangle |\k\rangle\langle\k|,
\ee
one obtains
\be
\frac{\langle N_\0\rangle}{N}=\frac{1}{NL^d}\int_\Lambda\d\x\int_\Lambda\d\y\langle\x|\sigma_1|\y\rangle=A_1.
\ee
Above, $|\k\rangle\langle\k|$ is the orthogonal projection to the subspace of $\varphi_\k$, and $\langle\x|\sigma_1|\y\rangle$ is the integral kernel of $\sigma_1$. The path integral representation of $\langle N_\0\rangle/N$ is
\be\label{N0-path}
\frac{\langle N_\0\rangle}{N}=\frac{1}{NQ_N}\sum_{n=1}^N\int_\Lambda\d\x\int W^{n\beta}_{\0\x}(\d\boldom_0)e^{-\beta{\cal U}_n(\boldom_\0)}Q_{N-n}(\boldom_\0).
\ee
\newsec{The ideal Bose gas revisited}
If there is no interaction, Eq.~(\ref{recur-gen}) becomes
\be\label{recur-free}
Q^0_N=\frac{1}{N}\sum_{n=1}^N q_n Q^0_{N-n}.
\ee
In this equation $q_n$ is the one-particle partition function at inverse temperature $n\beta$,
\be\label{qn}
q_n=L^d\int W^{n\beta}_{\0\0}(\d\boldom)=\sum_{\z\in\Zz^d}e^{-\pi n\lambda_\beta^2 \z^2/L^2} =\frac{1}{n^{d/2}}\left(\frac{L}{\lambda_\beta}\right)^d\sum_{\z\in\Zz^d}e^{-\pi (L/\lambda_\beta)^2\z^2/n},
\ee
which is a monotone decreasing function of $n$. Together with the initial condition $Q^0_0=1$, Eq.~(\ref{recur-free}) defines recursively $Q^0_N$. From this equation one can easily reproduce most of the results about cycle percolation (the appearance of infinite permutation cycles) and its connection with Bose-Einstein condensation, obtained earlier in [Su1,2].

With the help of the single-particle energies $\epsilon_\k=\hbar^2\k^2/2m$, $\k\in(2\pi/L)\Zz^d$, the canonical partition function can still be written as
\be
Q^0_N=\sum_{\sum_{\k\neq\0} n_\k\leq N}e^{-\beta\sum n_\k\epsilon_\k}=\sum_{M=0}^N \widehat{Q}^0_M,
\ee
where
\be
\widehat{Q}^0_M=\sum_{\sum_{\k\neq\0} n_\k=M}e^{-\beta\sum n_\k\epsilon_\k},
\ee
and the summations run over sets $\{n_\k\}_{\k\in(2\pi/L)\Zz^d\setminus\{\0\}}$ of nonnegative integers. This shows that $Q^0_N>Q^0_{N-1}$, a crucial property for the proof of cycle percolation [Su1], which can also be obtained from (\ref{recur-free}).

\begin{lemma}
Let $A_{-1}=0$, $A_0=1$, $a_1,a_2,\ldots$ arbitrary numbers, and define recursively $A_N$ by
\be\label{A}
A_N=\frac{1}{N}\sum_{n=1}^N a_nA_{N-n}.
\ee
Then
\be\label{deltaA}
A_N-A_{N-1}=\frac{1}{N}\sum_{n=1}^N(a_n-1)(A_{N-n}-A_{N-n-1}).
\ee
\end{lemma}

\vspace{5pt}
\noindent
{\bf Proof.} This follows by a simple computation. $\quad\Box$

\vspace{5pt}
Now if $a_n>1$, then $A_N>A_{N-1}$ can be proven by induction; and, because $q_n>1$, this applies to $Q_N^0$. $Q^0_N-Q^0_{N-1}=\widehat{Q}^0_N$, and keeping only the $n=1$ term and iterating we obtain
\be
\widehat{Q}^0_N>\frac{(q_1-1)^N}{N!}\quad(N\geq 1).
\ee
With $\widehat{Q}^0_0=1$ one then concludes that for fixed $L$,
\be
\lim_{N\to\infty}Q^0_N>e^{q_1-1}.
\ee
The limit is finite [Su3], we have
\be
\lim_{N\to\infty}Q^0_N=\prod_{\z\in\Zz^d\setminus\{\0\}}\left[1-e^{-\pi(\lambda_\beta/L)^2\z^2}\right]^{-1} =\exp\left\{\sum_{n=1}^\infty\frac{q_n-1}{n}\right\}\asymp e^{\zeta(1+d/2)(L/\lambda_\beta)^d}\quad (L/\lambda_\beta\gg 1).
\ee
Here $\zeta(x)=\sum_{n=1}^\infty n^{-x}$, the Riemann zeta function. The multiplier of $L^d$ in the exponent is $-\beta$ times the free energy density $f^0(\rho,\beta)$ for the density $\rho$ above its critical value,
\be\label{f0-limit}
f^0(\rho,\beta)= -\left(\frac{m}{2\pi\hbar^2}\right)^{d/2}\frac{\zeta(1+d/2)}{\beta^{1+d/2}},
\quad \rho\geq\rho_c^0(\beta)=\zeta(d/2)/\lambda_\beta^d.
\ee
Because Eq.~(\ref{deltaA}) has the same form as Eq.~(\ref{A}), some information about $Q^0_{N+1}-2Q^0_N+Q^0_{N-1}$ can also be obtained, e.g., to locate the point where $Q^0_N$ turns from convex to concave, and to show that $L^d\rho^0_c(\beta)$ is on the concave part. (Recall that $\ln Q^0_N$ is concave [Lew1], [Su4].)

Let $\xi_1$ denote the length of the cycle containing 1. Its probability distribution is
\be
P^0_{N,L}(\xi_1=n)=\frac{q_n Q^0_{N-n}}{NQ^0_N}.
\ee

\begin{theorem}\label{thm-ideal-1}
In the ideal Bose gas there is BEC if and only if there exists an $\varepsilon>0$ such that
\be\label{cond-P0-limit}
\lim_{N,L\to\infty} P^0_{N,L}\left(\xi_1\geq \varepsilon L^2\right)>0.
\ee
\end{theorem}

\vspace{5pt}
\noindent
{\bf Proof.} (i) Suppose first that (\ref{cond-P0-limit}) holds true. From the second of Eqs.~(\ref{norms}) one finds
\be
\int_\Lambda \d\x\int W^{n\beta}_{\0\x}(\d\boldom)=1.
\ee
Therefore
\bea\label{cond-fraction-ideal}
\frac{\langle N_\0\rangle}{N}
&=&\frac{1}{NQ^0_N}\sum_{n=1}^NQ^0_{N-n}=\sum_{n=1}^N\frac{1}{q_n}P^0_{N,L}(\xi_1=n)\nonumber\\                                &\geq&\sum_{n\geq\varepsilon L^2}\frac{1}{q_n}P^0_{N,L}(\xi_1=n)\geq \frac{P^0_{N,L}\left(\xi_1\geq \varepsilon L^2\right)}{\sum_{\z\in\Zz^d}\exp\{-\pi\varepsilon\lambda_\beta^2\z^2\}},
\eea
where we used the monotonic decrease of $q_n$. Taking the limit we find $\lim_{N,L\to\infty}\langle N_\0\rangle/N>0$.

(ii) Suppose now that for any $\varepsilon>0$,
$
\lim_{N,L\to\infty}P^0_{N,L}\left(\xi_1\geq\varepsilon L^2\right)=0.
$
We have
\bea
\frac{\langle N_\0\rangle}{N}&=&\sum_{n<\varepsilon L^2}\frac{1}{q_n}P^0_{N,L}(\xi_1=n)+\sum_{n\geq\varepsilon L^2}\frac{1}{q_n}P^0_{N,L}(\xi_1=n)\nonumber\\
&\leq& \frac{P^0_{N,L}\left(\xi_1<\varepsilon L^2\right)}{\sum_{\z\in\Zz^d}\exp\{-\pi\varepsilon\lambda_\beta^2\z^2\}}
+P^0_{N,L}\left(\xi_1\geq\varepsilon L^2\right).
\eea
Taking the limit,
\be
\lim_{N,L\to\infty}\frac{\langle N_\0\rangle}{N}\leq \frac{1}{\sum_{\z\in\Zz^d}\exp\{-\pi\varepsilon\lambda_\beta^2\z^2\}}
\ee
for $\varepsilon$ arbitrarily small, therefore $\lim_{N,L\to\infty}\langle N_\0\rangle/N=0$. $\quad\Box$

\vspace{10pt}
This theorem does not tell us how $N$ and $L$ should go to infinity so that cycle percolation takes place. The following proposition will do it. Recall that $\rho_c^0(\beta)$ is the maximum density of particles of nonzero momentum. Thus, if $\rho>\rho_c^0(\beta)$ and $\rho_0$ denotes the density of particles with $\k=\0$, then $\rho_c^0(\beta)=\rho-\rho_0$ or
\be\label{zeta}
\zeta(d/2)=(\rho-\rho_0)\lambda_\beta^d.
\ee

\begin{proposition}\label{thm-ideal-2}
Let $\rho\lambda_\beta^d>\zeta(d/2)$, and choose any positive $\varepsilon<1-\frac{\zeta(d/2)}{\rho\lambda_\beta^d}=\frac{\rho_0}{\rho}$. Then
\be\label{P-eps}
\lim_{N,L\to\infty,N/L^d=\rho}P^0_{N,L}\left(\xi_1\geq\varepsilon N\right)\geq \frac{\rho_0}{\rho}-\varepsilon.
\ee
\end{proposition}

\vspace{5pt}
\noindent
{\bf Remark.} Earlier we proved with a more detailed argument that in fact
\be\label{earlier(41)}
\lim_{N,L\to\infty,N/L^d=\rho}P^0_{N,L}(\xi_1>\varepsilon N)=\frac{\rho_0}{\rho}-\varepsilon;
\ee
see Eq.~(41) of [Su2].

\vspace{5pt}
\noindent
{\bf Proof.} Let $\rho=N/L^d$.
\be
P^0_{N,L}(\xi_1<\varepsilon N)\leq \sum_{n<\varepsilon N}\frac{q_n}{N}=\varepsilon+\frac{1}{\rho}\sum_{n<\varepsilon N}\frac{1}{L^d}\sum_{\z\in\Zz^d\setminus\{\0\}}e^{-\pi n\lambda_\beta^2 \z^2/L^2},
\ee
therefore
\be
P^0_{N,L}(\xi_1\geq\varepsilon N)\geq 1-\varepsilon-\frac{1}{\rho}\sum_{n<\varepsilon N}\frac{1}{L^d}\sum_{\z\in\Zz^d\setminus\{\0\}}e^{-\pi n\lambda_\beta^2 \z^2/L^2},
\ee
whose limit is
\be
\lim_{N,L\to\infty,N/L^d=\rho}P^0_{N,L}(\xi_1\geq\varepsilon N)\geq 1-\varepsilon-\frac{1}{\rho} \sum_{n=1}^\infty\int_{\Rr^d} e^{-\pi n\lambda_\beta^2\x^2}\d\x=1-\varepsilon-\frac{\zeta(d/2)}{\rho\lambda_\beta^d}. \quad\Box
\ee

\vspace{10pt}
The above proposition strengthens Theorem~\ref{thm-ideal-1} in one direction: BEC implies (\ref{P-eps}) which is much stronger than (\ref{cond-P0-limit}). For later use we now prove without reference to BEC that the length of the cycles that become infinite in the infinite system diverges as fast as $N$.

The limit of $P^0_{N,L}$ as $N, L\to\infty$, $N/L^d=\rho$, exists. It is
\be
P^0_{\rho,\beta}(\xi_1=n)=\frac{e^{n\beta\mu^0(\rho,\beta)}}{\rho\lambda_\beta^d n^{d/2}}
\ee
where $\mu^0(\rho,\beta)$ is the solution for $\mu$ of Einstein's equation
\be\label{N-mu}
N=\sum_{\k\in\frac{2\pi}{L}\Zz^d}\frac{1}{e^{\beta(\epsilon_\k-\mu)}-1}
\ee
in the thermodynamic limit, that is, of
\be
\rho=\frac{1}{(2\pi)^d}\int_{\Rr^d}\frac{1}{e^{\beta(\epsilon(\k)-\mu)}-1}\d\k
\ee
if $\rho\leq\rho_c^0(\beta)$. If $\rho\geq\rho_c^0(\beta)$ then $\mu^0(\rho,\beta)\equiv 0$. (Note the identity
$
\int_{\Rr^d}[e^{\pi\k^2}-1]^{-1}\d\k=\zeta(d/2)
$
for $\d\geq 3$.) The probability $\sum_{n=1}^\infty P^0_{\rho,\beta}(\xi_1=n)$ that 1 is in a finite cycle in the infinite system can be smaller than 1, this is precisely the case of cycle percolation that we discuss systematically in the next section. To appreciate the assertion of the following proposition imagine that $\sum_{n=1}^\infty P^0_{\rho,\beta}(\xi_1=n)<1$.

\begin{proposition}\label{suppl-ideal}
If $K_N\to\infty$ and $K_N/N\to 0$, then
\be
\lim_{N,L\to\infty,N/L^d=\rho}P^0_{N,L}(\xi_1\leq K_N)=\sum_{n=1}^\infty P^0_{\rho,\beta}(\xi_1=n).
\ee
\end{proposition}

\vspace{10pt}
\noindent
{\bf Proof.}
We use the rightmost form of $q_n$ in (\ref{qn}).
\be\label{P0NL(KN)}
P^0_{N,L}(\xi_1\leq K_N)=\frac{1}{\rho\lambda_\beta^d}\sum_{n=1}^{K_N}\frac{1}{n^{d/2}}\frac{Q^0_{N-n}}{Q^0_N} +\frac{1}{\rho\lambda_\beta^d}\sum_{n=1}^{K_N}\frac{1}{n^{d/2}}\frac{Q^0_{N-n}}{Q^0_N} \sum_{\z\neq\0}e^{-\pi L^2\z^2/n\lambda_\beta^2}.
\ee
We show that the second sum goes to zero for $K_N=o(N)$ as $L\propto N^{1/d}\to\infty$.
\bea
\sum_{n=1}^{K_N}\frac{1}{n^{d/2}}\frac{Q^0_{N-n}}{Q^0_N} \sum_{\z\neq\0}e^{-\pi L^2\z^2/n\lambda_\beta^2} &\leq& d\,\frac{\lambda_\beta}{L}\sum_{n=1}^{K_N}\frac{1}{n^{(d-1)/2}} \left[1+\frac{\sqrt{n}\lambda_\beta}{L}\right]^{d-1}\nonumber\\
&=& d\sum_{k=0}^{d-1}{d-1\choose k}\left(\frac{\lambda_\beta}{L}\right)^{d-k}\sum_{n=1}^{K_N}\frac{1}{n^{k/2}}.
\eea
The sum over $n$ is $K_N$ if $k=0$, $O(\sqrt{K_N})$ if $k=1$, $O(\ln K_N)$ if $k=2$, and $O(1)$ if $k\geq 3$. $K_N=o(N)$ is necessary only for the vanishing of the $k=0$ term. So for any fixed $M$,
\bea
P^0_{\rho,\beta}(\xi_1\leq M)
&\leq& \lim_{N,L\to\infty,N/L^d=\rho}P^0_{N,L}(\xi_1 \leq K_N)=\frac{1}{\rho\lambda_\beta^d}\lim_{N,L\to\infty,N/L^d=\rho} \sum_{n=1}^{K_N}\frac{1}{n^{d/2}}\frac{Q^0_{N-n}}{Q^0_N}\nonumber\\
&\leq& \min\left\{1,\frac{\zeta(d/2)}{\rho\lambda_\beta^d}\right\}=\sum_{n=1}^\infty P^0_{\rho,\beta}(\xi_1=n).
\eea
Taking the limit $M\to\infty$ we obtain the result.$\qquad\Box$

\vspace{10pt}
\noindent
This proposition tells us that by letting $K_N$ increase slower than $N$ we do not pick up any probability coming from infinite cycles. We summarize:

\begin{theorem}\label{thm-ideal-3}
\be\label{ideal-concise}
1-\sum_{n=1}^\infty P^0_{\rho,\beta}(\xi_1=n)
=1-\frac{1}{\rho\lambda_\beta^d}\sum_{n=1}^\infty\frac{e^{n\beta\mu^0(\rho,\beta)}}{n^{d/2}} =\lim_{\varepsilon\to 0}\ \lim_{N,L\to\infty,N/L^d=\rho}P^0_{N,L}(\xi_1>\varepsilon N)=\frac{\rho_0}{\rho}.
\ee
\end{theorem}
This is the most concise formulation of the connection between cycle percolation and BEC in the ideal Bose gas: the probability that a marked particle is in an infinite cycle is equal to the probability that it is in a macroscopic cycle, and this equals the condensate fraction.

\newsec{Cycle percolation}

The term "cycle percolation" was applied in [Su1,2] to the appearance of infinite permutation cycles in the thermodynamic limit of bosonic systems, shown to exist in the ideal Bose gas. The cycle containing 1 is the analogue of the connected occupied cluster containing the origin in the percolation problem. The conditional probability that the origin belongs to a finite cluster, provided that it belongs to any, is 1 below the percolation threshold and is smaller than 1 above it [Kun]. The missing probability is that of the event that the cluster containing the origin is infinite. Cycle percolation also shows up as a deficiency of probability. If $\xi_1$ denotes the length of the cycle containing 1 (that is, the number of particles in it), $P_{N,L}$ its probability distribution
\be
P_{N,L}(\xi_1=n)= \frac{L^d}{NQ_N}\int W^{n\beta}_{\0\0}(\d\boldom_0)e^{-\beta{\cal U}_{n}(\boldom_0)} Q_{N-n}(\boldom_0),
\ee
and
\be
P_{\rho,\beta}(\xi_1=n)=\lim_{N,L\to\infty,N/L^d=\rho} P_{N,L}(\xi_1=n)
\ee
whose existence we anticipate, then one has
\be
1=\lim_{N,L\to\infty,N/L^d=\rho}\sum_{n=1}^N P_{N,L}(\xi_1=n)=\sum_{n=1}^M P_{\rho,\beta}(\xi_1=n)+\lim_{N,L\to\infty,N/L^d=\rho}\sum_{n=M+1}^N P_{N,L}(\xi_1=n),
\ee
valid for any $M$. Thus,
\be\label{Sum-P}
\sum_{n=1}^\infty P_{\rho,\beta}(\xi_1=n)\leq 1,
\ee
with a strict inequality in the regime of cycle percolation. Further analogy between non-interacting percolation and the ideal Bose gas is the algebraic vs exponential decay of the respective probabilities in and out of the percolation regime. The most peculiar property of cycle percolation in the ideal Bose gas is the countably infinite number of infinite cycles, each containing a positive fraction of particles [Su2]. Similar result was obtained in percolation on Bethe lattices [Fi].

\begin{definition}
(i) We speak about cycle percolation if
\be\label{CP-def}
P_{\rho,\beta}(\xi_1<\infty)\equiv\sum_{n=1}^\infty P_{\rho,\beta}(\xi_1=n) < 1.
\ee
We call
\be
P_{\rho,\beta}(\xi_1=\infty)\equiv 1-P_{\rho,\beta}(\xi_1<\infty)=\lim_{M\to\infty}\ \lim_{N,L\to\infty,N/L^d=\rho} P_{N,L}(\xi_1>M)
\ee
the probability of cycle percolation.

\noindent
(ii) Cycle percolation is macroscopic if
\be\label{macr-CP-def}
P_{\rho,\beta}(\xi_1=\infty)=\lim_{\varepsilon\to 0}\ \lim_{N,L\to\infty,N/L^d\to\rho}P_{N,L}(\xi_1>\varepsilon N),
\ee
i.e., the probability that the length of the cycle of 1 tends to infinity slower than $N$ goes to zero as $N$ increases.
\end{definition}

Thus, in the ideal gas cycle percolation is macroscopic. We show that in general Eq.~(\ref{CP-def}) is equivalent to having a non-vanishing asymptotic probability for the event that the length of the cycle of 1 goes to infinity with $N$.

\begin{lemma}\label{aNM}
Let $\{a_{N,M}\}_{1\leq M\leq N}$ be an infinite sequence with the properties
\be\label{cond1}
0< a_{N,1}< a_{N,2}<\cdots< a_{N,N}=1,\quad a_M=\lim_{N\to\infty}a_{N,M}\quad\mbox{exists},\quad a_M<a_{M+1}.
\ee
Let
$
a=\lim_{M\to\infty}a_M.
$

\noindent
(i) For any sequence of integers $K_N\leq N$ going to infinity with $N$,
\be\label{KN<}
\liminf_{N\to\infty}a_{N,K_N}\geq a.
\ee
(ii) If $a<1$ then for every $b\in [a,1)$ there exists a sequence $M^b_N$ going to infinity such that
\be
\limsup_{N\to\infty}a_{N,M^b_N-1}\leq b\leq\liminf_{N\to\infty}a_{N,M^b_N}.
\ee

\end{lemma}

\vspace{10pt}
\noindent
{\bf Proof.} (i) For any fixed $M$, $a_{N,K_N}>a_{N,M}$ if $N$ is large enough. Then
\be
\liminf_{N\to\infty}a_{N,K_N}\geq\lim_{N\to\infty}a_{N,M}=a_M,
\ee
and sending $M$ to infinity yields (\ref{KN<}).

(ii) Define
\be
M^b_N=\min\{M\leq N: a_{N,M}\geq b\}.
\ee
The set whose minimum must be taken is nonempty, since $a_{N,N}=1> b$. $M^b_N$ may not be monotone increasing, but it tends to infinity: supposing that it is a constant $M_0$ on a subsequence $N_i$ would lead to a contradiction,
$$a_{M_0}<a\leq b\leq\liminf_{i\to\infty}a_{N_i,M^b_{N_i}}=\liminf_{i\to\infty}a_{N_i,M_0}
=a_{M_0}.$$
The claim is verified due to $a_{N,M^b_{N}-1}<b\leq a_{N,M^b_{N}}$. $\qquad\Box$

\vspace{10pt}
\noindent
{\bf Examples.} (i) Let
\be
a_{N,M}=a_M+\frac{M}{N}(1-a_N),\quad 1\leq M\leq N,
\ee
where $a_M$ is strictly increasing and tends to $a\leq 1$ as $M$ goes to infinity. If $K_N\to\infty$ but $K_N/N\to 0$, then $a_{N,K_N}\to a$. If $a<1$, then for any positive $\varepsilon<1-a$,
\be
\lim_{N\to\infty}a_{N,\lfloor\varepsilon N/(1-a)\rfloor}=a+\varepsilon.
\ee
(ii) Let $\{a_M\}$ be as before, $0<\delta<1$, and
\be
a_{N,M}=\left\{\begin{array}{ll}
a_M+\frac{M}{N^\delta}(1-a_N),& M< N^\delta\\
a_M+1-a_N, & M\geq N^\delta.
\end{array}\right.
\ee
If $K_N\to\infty$ but $K_N/N^\delta\to 0$, then $a_{N,K_N}\to a$. If $a<1$, then for any positive $\varepsilon<1-a$,
\be
\lim_{N\to\infty}a_{N,\lfloor\varepsilon N^\delta/(1-a)\rfloor}=a+\varepsilon.
\ee
On the other hand, if $K_N\geq N^\delta$, then
\be
\lim_{N\to\infty}a_{N,K_N}=1.
\ee

\begin{proposition}\label{PKN}
$P_{\rho,\beta}(\xi_1<\infty) < 1$ if and only if $\lim_{N,L\to\infty,N/L^d=\rho}P_{N,L}(\xi_1\leq K_N)<1$ for some sequence $K_N\to\infty$. In detail, we have the following.

\noindent
(i) For any sequence of integers $K_N\leq N$ going to infinity with $N$,
\be\label{leqKNgeq}
\liminf_{N,L\to\infty,N/L^d=\rho}P_{N,L}(\xi_1\leq K_N)\geq P_{\rho,\beta}(\xi_1<\infty).
\ee
(ii) If  $P_{\rho,\beta}(\xi_1<\infty)<1$ then for each $b\in [P_{\rho,\beta}(\xi_1<\infty),1)$ there exists a sequence $M^b_N$ tending to infinity with $N$ such that
\be\label{about-b}
\limsup_{N,L\to\infty,N/L^d=\rho}P_{N,L}(\xi_1\leq M^b_N-1)\leq b\leq \liminf_{N,L\to\infty,N/L^d=\rho}P_{N,L}(\xi_1\leq M^b_N).
\ee
\end{proposition}

\vspace{10pt}
\noindent
{\bf Proof.}
$P_{N,L}(\xi_1=n)$ and its limit $P_{\rho,\beta}(\xi_1=n)$ are nonzero for any $n$ fixed. This implies that $P_{N,L}(\xi_1\leq M)$ and $P_{\rho,\beta}(\xi_1\leq M)$ are strictly increasing with $M$. Then, the previous lemma applies with
\be
a_{N,M}=P_{N,L}(\xi_1\leq M),\quad a_M=\lim_{N\to\infty}a_{N,M}=P_{\rho,\beta}(\xi_1\leq M),\quad a=\lim_{M\to\infty}a_M=P_{\rho,\beta}(\xi_1<\infty). \quad\Box
\ee

\vspace{10pt}
\noindent
If $\lim_{N,L\to\infty,N/L^d=\rho}P_{N,L}(\xi_1= K_N)=0$ for any $K_N\to\infty$ then $M^b_N$ can be given only up to an error of $o(M^b_N)$. We illustrate this on the example of the ideal Bose gas.

\begin{proposition}
For the ideal Bose gas let
\be
P^0_{\rho,\beta}(\xi_1<\infty)<1,\quad b=P^0_{\rho,\beta}(\xi_1<\infty)+\varepsilon\quad \left[0<\varepsilon<1-P^0_{\rho,\beta}(\xi_1<\infty)\right].
\ee
Then $M^b_N=\varepsilon N+o(N)$.
\end {proposition}

\vspace{10pt}
\noindent
{\bf Proof.} If $K_N\to\infty$ then
\be
\lim_{N,L\to\infty,N/L^d=\rho}P^0_{N,L}(\xi_1= K_N)=0.
\ee
Indeed, from the middle term of Eq.~(\ref{qn}),
\be
P^0_{N,L}(\xi_1=n)\leq \frac{q_{n}}{N}=\frac{1}{N}\left[1+2\sum_{z=1}^\infty e^{-\pi n\lambda_\beta^2 z^2/L^2}\right]^d\leq \frac{1}{N}\left[1+\frac{L}{\lambda_\beta\sqrt{n}}\right]^d.
\ee
If $n=K_N$, and $N,L,K_N\to\infty$, $N/L^d=\rho$, the upper bound, and therefore the probability that $\xi_1=K_N$, goes to zero. Now
\be\label{low-b-up}
P^0_{N,L}(\xi_1\leq M^b_N-1)<b\leq P^0_{N,L}(\xi_1\leq M^b_N)
\ee
where $M^b_N$ goes to infinity. Therefore
\be
P^0_{N,L}(\xi_1\leq M^b_N)-P^0_{N,L}(\xi_1\leq M^b_N-1)\leq P^0_{N,L}(\xi_1=M^b_N)\to 0.
\ee
So both the upper and lower bounds in (\ref{low-b-up}) tend to $b$,
\be\label{=b=}
\lim_{N,L\to\infty,N/L^d=\rho}P^0_{N,L}(\xi_1\leq M^b_N-1)=\lim_{N,L\to\infty,N/L^d=\rho}P^0_{N,L}(\xi_1\leq M^b_N)=b.
\ee
From Proposition~\ref{suppl-ideal} it follows that the interval $\left(P^0_{\rho,\beta}(\xi_1<\infty),1\right]$ can be covered only with the limit points of $P^0_{N,L}(\xi_1\leq K_N)$ for different $K_N\propto N$. Combining Eqs.~(\ref{earlier(41)}), (\ref{ideal-concise}) and (\ref{=b=}),
\be
\lim_{N,L\to\infty,N/L^d=\rho}P^0_{N,L}(\xi_1\leq\varepsilon N)=P^0_{\rho,\beta}(\xi_1<\infty)+\varepsilon=b=\lim_{N,L\to\infty,N/L^d=\rho}P^0_{N,L}(\xi_1\leq M^b_N),
\ee
so $|M^b_N-\varepsilon N|=o(N)$. $\quad\Box$


\subsection{Representative trajectories}\label{Gordian}

The subsequent discussion applies to interactions with the following properties.
\begin{enumerate}
\item
$u$ is
superstable: If $u$ is stable then there is some $A>0$ such that $\sum_{1\leq i<j\leq N}u(\x_j-\x_i)\geq -AN$ for any $N$ and any $\x_1,\ldots,\x_N$. Superstability means that the potential energy per unit volume increases with $\rho$ eventually at least as $\rho^2$. In a formula, if $\Lambda$ is large enough, then for any $N$ and any $\x_1,\ldots,\x_N\in\Lambda$,
\be\label{super}
\sum_{1\leq i<j\leq N}u(\x_j-\x_i)\geq -AN+BN^2/L^d,
\ee
where $A,B>0$ [R1-3]. The negative linear term cannot be dropped even if $u\geq 0$, otherwise the inequality would fail for small $N$ or at low density (think of a finite-range $u$). When $u$ is positive definite, the largest possible $B$ is $\frac{1}{2}\int u(\x)\d\x$, in which case the smallest $A$ is $\frac{1}{2}u(\0)$ [Su5].

\item
$u$ can diverge only continuously (possibly at the origin or at the boundary of a hard core).

\item
$u$ is tempered, c.f. Eq.~(\ref{tempered}); in particular, it is integrable outside the origin or the hard core. If $u$ is attractive at large distances, it satisfies (\ref{tempered}) with $\eta\geq d$.
\end{enumerate}

We start by rewriting the partition function in the form of a recurrence relation,
\be\label{QNint}
Q_N=\frac{1}{N}\sum_{n=1}^Nq_nQ_{N-n}G_N(n),
\ee
where
\bea\label{QN-2}
G_N(n) &=& \int\mu^{n\beta}_1(\d\boldom_0)\int\mu^\beta_{N-n}\left(\d\boldom^{N-n}\right) e^{-\beta H(\boldom_0,\boldom^{N-n})}\nonumber\\
&=& \left\langle e^{-\beta H(\boldom_0,\boldom^{N-n})}\right\rangle_{\mu^{n\beta}_1\times\mu^\beta_{N-n}} \equiv e^{-\beta\Psi_{n,N-n}},
\eea
\be\label{self+int}
H(\boldom_0,\boldom^{N-n})={\cal U}_{n}(\boldom_0)+ V_{n}\left(\boldom_0,\boldom^{N-n}\right),
\ee
and $\mu^{n\beta}_1$ and $\mu^\beta_{N-n}$ are normalized measures. If $\Omega^\beta$ denotes the family of $\left(<\frac{1}{2}\right)$-H\"older continuous one-particle trajectories parametrized by $t\in [0,\beta]$, and
\be
\Omega^\beta_{\x\y}=\left\{\boldom\in\Omega^\beta| \boldom(0)=\x, \boldom(\beta)=\y\right\}, \quad \Omega^\beta_X =\bigcup_{\pi\in S(X)}\prod_{\x\in X}\Omega^\beta_{\x\pi(\x)}\quad(X\subset\Lambda), \quad \Omega^\beta_M=\bigcup_{X\subset\Lambda,|X|=M}\Omega^\beta_X,
\ee
where $S(X)$ is the set of permutations of $X$, then $\mu^{n\beta}_1$ is a probability measure on $\Omega^{n\beta}_{\0\0}$ given by
\be
\mu^{n\beta}_1(\d\boldom_0)= W^{n\beta}_{\0\0}(\d\boldom_0)\left[\int W^{n\beta}_{\0\0}(\d\boldom)\right]^{-1},
\ee
and $\mu^\beta_M$ is the Gibbs measure on $\Omega^\beta_M$,
\be\label{Gibbs-omM}
\mu^\beta_M\left(\d\boldom^M\right) =\frac{e^{-\beta {\cal U}(\boldom^M)}\prod_{i=1}^M\d\x_i\ W^\beta_{\x_i\x_{\pi(i)}}(\d\boldom_i)}{Q_M M!}.
\ee
Integration with $\mu^\beta_M\left(\d\boldom^M\right)$ includes summation over $S_M$.

$G_N(n)$ can be interpreted as the partition function of a classical system of $n+(N-n)$ particles, characterized by the \emph{a priori} measure $\mu^{n\beta}_1\times\mu^\beta_{N-n}$ and the energy functional $H(\boldom_0,\boldom^{N-n})$. The system can also be viewed as $n$ trapped particles in the annealed random field of $N-n$ other particles. The trap and the random field are represented by $\mu^{n\beta}_1$ and $\mu^\beta_{N-n}$, respectively, and $\Psi_{n,N-n}$ is the annealed free energy.

\vspace{10pt}
\noindent
We proceed with an additional restriction.

\vspace{10pt}
\noindent
{\bf Homogeneous phase condition.} {\em In the domain of $\rho$ and $\beta$ under investigation, in infinite volume each pure state is translation invariant and of the same density, $\rho$.}

\vspace{10pt}
\noindent
A fluid -- gas or liquid -- in statistical physics is a pure translation invariant state; so the condition is about a fluid of a given density.
The infinite-volume state obtained with periodic boundary condition is translation invariant. However, if it is a mixture of pure states of different densities, very different sets of trajectories can considerably contribute to the average composing $G_N(n)$. On the way we wish to go further this situation must be avoided. A Bose-condensed fluid can be regarded as a uniform mixture of pure states of different gauges. Because all these phases are translation invariant and of the same density, the condition is fulfilled. This perception of the system is more pertinent for the Bose-condensed hard-core lattice gas. In spin language BEC means planar spin ordering, and the uniform average over different spin orientations corresponds to averaging over gauges.


Due to the supposed continuity property of $u$, we can apply the first mean-value theorem and write
\be\label{GN}
G_N(n)=e^{-\beta H\left(\widetilde{\boldom}_0,\widetilde{\boldom}^{N-n}\right)}
\ee
with some specific but not unique set of trajectories $\widetilde{\boldom}_0,\widetilde{\boldom}_1,\ldots,\widetilde{\boldom}_{N-n}$, that we call representative. Hence, the annealed free energy is
\be
\Psi_{n,N-n}=H\left(\widetilde{\boldom}_0,\widetilde{\boldom}^{N-n}\right)
={\cal U}_{n}(\widetilde{\boldom}_0)+ V_{n}\left(\widetilde{\boldom}_0,\widetilde{\boldom}^{N-n}\right).
\ee
In the choice of $\left(\widetilde{\boldom}_0,\widetilde{\boldom}^{N-n}\right)$ one must take account of some consistency criteria, compatible with the homogeneous phase condition.

\vspace{10pt}
\noindent
{\bf Characterization of the representative trajectories.}
(i) A saddle-point or measure-concentration argument should be valid. If we divide the range of $H(\boldom_0,\boldom^{N-n})$ into intervals $I_l$ of length $|I_l|\ll 1/\beta$, we can write
\bea
G_N(n)=\sum_l \left\langle e^{-\beta H(\boldom_0,\boldom^{N-n})}\left|H(\boldom_0,\boldom^{N-n})\in I_l\right.\right\rangle_{\mu^{n\beta}_1\times\mu^\beta_{N-n}} \nonumber\\
\times \left(\mu^{n\beta}_1\times\mu^\beta_{N-n}\right)\left(H(\boldom_0,\boldom^{N-n})\in I_l\right).
\eea
There is a corresponding partition
\be
{\cal K}_l=\left\{\left(\boldom_0,\boldom^{N-n}\right)\in \Omega^{n\beta}_{\0\0}\times\Omega^\beta_{N-n}\left| H(\boldom_0,\boldom^{N-n})\in I_l\right.\right\}
\ee
of the set of trajectories and, for large $N$ and $L$, $\left(\widetilde{\boldom}_0,\widetilde{\boldom}^{N-n}\right)$ must be in that ${\cal K}_l$ whose contribution to the above sum is maximum, so  $H\left(\widetilde{\boldom}_0,\widetilde{\boldom}^{N-n}\right)$ is relatively small and the entropy of ${\cal K}_l$ is relatively large.

\vspace{10pt}
\noindent
(ii) Deeper insight is provided by the fact that
$$e^{-\beta H\left(\boldom_0,\boldom^{N-n}\right)}\mu^{n\beta}_1\left(\d\boldom_0\right) \mu^\beta_{N-n}\left(\d\boldom^{N-n}\right)$$
is the restriction of $\mu^\beta_{N}\left(\d\boldom^N\right)$, the Gibbs measure on $\Omega^\beta_N$, to the subset of trajectories in which particle 1 is in a cycle of length $n$ starting and ending at the origin. Recall that $\boldom_0$ is a composite trajectory of $n$ particles,
\be\label{ntuple0}
\boldom_0 =(\boldom_{n,0},\boldom_{n,1},\ldots,\boldom_{n,n-1}),
\ee
where $\boldom_{n,k}\in\Omega^\beta$ and
\be\label{ntuple}
\boldom_{n,k}(t)=\boldom_0(k\beta+t)\quad(k=0,\ldots,n-1)
\ee
or
\be\label{ntuple2}
\boldom_{n,0}(0)=\boldom_{n,n-1}(\beta)=\0,\qquad \boldom_{n,k}(\beta)=\boldom_{n,k+1}(0)\quad(k=0,\ldots,n-2).
\ee
Although we treat $\boldom_0$ separately, there is nothing exceptional in it:
$n$-tuples of one-particle trajectories like (\ref{ntuple0})-(\ref{ntuple2}) occur also in $\boldom^{N-n}$ with a positive frequency. Their expected number is $\frac{N-n}{n} P_{N-n,L}(\xi_1=n)$ [Su1], and their self-energy and energy of interaction with the rest is similar to (\ref{self+int}).
It is the total energy $H\left(\boldom_0,\boldom^{N-n}\right)$ and not the separate terms ${\cal U}_{n}(\boldom_0)$ and $V_{n}\left(\boldom_0,\boldom^{N-n}\right)$ that determine the distribution of particles. Because $\mu^\beta_{N}$ is translation invariant, the interaction is superstable, and because $\left(\left\{\widetilde{\boldom}_{n,k}\right\}_{k=0}^{n-1},\widetilde{\boldom}^{N-n}\right)$ represents an average, large density fluctuations are absent in it. Moreover, if $u$ is not integrable, close encounters are also excluded. That is, in the case of a non-integrable $u$ one can choose
$\left(\left\{\widetilde{\boldom}_{n,k}\right\}_{k=0}^{n-1},\widetilde{\boldom}^{N-n}\right)$ to be a Delone (Delaunay) set at every instant: there exists some $d_2>\rho^{-1/d}>d_1>d_{\rm hc}$
such that at any fixed $t\in [0,\beta]$, every open ball of diameter $d_1$ contains at most one, and every ball of diameter $d_2$ contains at least one
point of $\left\{\widetilde{\boldom}_{n,k}(t)\right\}_{k=0}^{n-1} \bigcup\left\{\widetilde{\boldom}_i(t)\right\}_{i=1}^{N-n}$.
Here $d_{\rm hc}\geq 0$ is the hard-core diameter. The diameters characterize the whole set of trajectories, which contains cycles of very different lengths. Therefore $d_1$ and $d_2$ are not specific to $n$, neither do they depend on $\Lambda$; their only dependence is on $\rho$ and $\beta$.
If $u$ is integrable, trajectories may be close to each other for short times, they may even cross each other. This can be accounted for by setting $d_1=0$:
in this case $\left\{\widetilde{\boldom}_{n,k}(t)\right\}_{k=0}^{n-1} \bigcup\left\{\widetilde{\boldom}_i(t)\right\}_{i=1}^{N-n}$ is only relatively dense.

\vspace{10pt}\noindent
(iii) Inspired by what we know from experiments and hydrodynamic theory about the real-time dynamics of atoms in fluids we make an assumption about their imaginary-time dynamics. In a fluid at rest in thermal equilibrium particles evolve through diffusion, their typical displacement in time $t$ is proportional to $\sqrt{t}$. Moreover, the time evolution of distant particles is independent, resulting the fast decay of positional correlations and the absence of diagonal long-range order. {\em In analogy, we suppose that the increments of the representative one-particle trajectories, both of the individuals $\left\{\widetilde{\boldom}_{n,k}(t)\right\}_{k=0}^{n-1}$, $\left\{\widetilde{\boldom}_i(t)\right\}_{i=1}^{N-n}$ and of the composite ones, increase with $t$, subject to the condition that particles in 1-loops must return to their original position after time $\beta$. Furthermore, if two particles are in different cycles, the correlation between their increments decreases with time.} These properties form the basis of the ergodic hypothesis that we will add in the following section.

\vspace{10pt}\noindent
(iv) Consider the case when $0<\lambda_\beta\ll\rho^{-1/d}$. From this the classical limit $\lambda_\beta=0$ can be attained continuously. In that limit only $P_{N,L}(\xi_1=1)$ differs from zero. However, for any $\lambda_\beta>0$, both $P_{N,L}(\xi_1=n)$ and its thermodynamic limit $P_{\rho,\beta}(\xi_1=n)$ are positive for all $n$. To be consistent with the classical limit one must choose $\left\{\widetilde{\boldom}_{n,k}\right\}_{k=0}^{n-1}$ and $\left\{\widetilde{\boldom}_i\right\}_{i=1}^{N-n}$ in such a way that their increments in $[0,\beta]$ are $O(\lambda_\beta)$. Then, for the formation of $n$-cycles, the shorter trajectories must be combined with larger density fluctuations. This means an increased $d_2$ and a smaller $d_1=O(\lambda_\beta)$. The classical limit can be approached by raising the temperature. A more interesting case is when both $\beta$ and $\rho$ are relatively large, and $\lambda_\beta\ll\rho^{-1/d}$ holds because the mass of the particles is large. This is the setting for crystallization. Although translation invariance is broken, the extremal states have the same density, and $G_N(n)$ can be represented by a Delone set of trajectories with the mean position of the particles showing diagonal long-range order. Decreasing the temperature in the crystalline phase, ultimately $\lambda_\beta>\rho^{-1/d}$, the crystal enters the quantum regime. However, the increments of individual trajectories are hampered by high potential barriers, resulting cycles that are formed mostly by nearest neighbors. The consequence is, perhaps, many-body localization.


\vspace{10pt}\noindent
Because
\be\label{PNL}
P_{N,L}(\xi_1=n)=\frac{q_{n}}{N}\frac{Q_{N-n}}{Q_N}G_N(n)
\ee
as functions of $n$ are uniformly bounded, by the diagonal process one can select a subsequence $\{N_i,L_i\}$ on which they converge,
\be
P_{\rho,\beta}(\xi_1=n)=\lim_{N_i,L_i\to\infty, N_i/L_i^d=\rho}P_{N,L}(\xi_1=n)\qquad(n\geq 1).
\ee
If the equilibrium state in infinite volume is a mixture of pure states of different densities, the above limit may depend on the chosen subsequence. If, however, the homogeneous phase condition is met, there is a unique limit. To see this, note that due to the rapid decrease of the interaction, $\widetilde{\boldom}_0$ and all those $\widetilde{\boldom}_i$ giving a non-negligible contribution to $\sum_{k=0}^{n-1}V_{n,k}\left(\widetilde{\boldom}_0,\widetilde{\boldom}^{N-n}\right)$ can be chosen to be independent of $L$, once $L\gg n\beta$. More precisely, in the limit when $L$ goes to infinity the measure $\mu^{n\beta}_1$ goes to $P^{n\beta}_{\0\0}$ with normalization, c.f. Eq.~(\ref{P-W-continuum}), and $\widetilde{\boldom}_0$ can be chosen to converge as $L^{-\eta}$, c.f. Eq.~(\ref{tempered}), to a trajectory
\be\label{gamma-n}
~\gamma_n=\left(~\gamma_{n,0},~\gamma_{n,1},\ldots,~\gamma_{n,n-1}\right)
\ee
in the domain of $P^{n\beta}_{\0\0}$, where $~\gamma_{n,k}\in\Omega^\beta$,
\be
~\gamma_{n,k}(t)=~\gamma_{n}(k\beta+t)\quad(k=0,\ldots,n-1)
\ee
or
\be
~\gamma_{n,0}(0)=~\gamma_{n,n-1}(\beta)=\0,\qquad ~\gamma_{n,k}(\beta)=~\gamma_{n,k+1}(0)\quad(k=0,\ldots,n-2).
\ee
Furthermore, we can number $\widetilde{\boldom}_i\equiv\widetilde{\boldom}_i[N]$ in such a way that $i<j$ if
\be\label{numbering}
\int_0^\beta|\widetilde{\boldom}_i(t)|\d t<\int_0^\beta|\widetilde{\boldom}_j(t)|\d t.
\ee
This permits to choose $\widetilde{\boldom}_i[N+1]$ as a small modification of $\widetilde{\boldom}_i[N]$ and to conclude that for any fixed $i$, $\widetilde{\boldom}_i[N]$ converges with the same speed to a trajectory $~\nu_{n,i}$. Then,
\be\label{phi-n-beta}
\lim_{N,L\to\infty,N/L^d=\rho} {\cal U}_{n,k}(\widetilde{\boldom}_0)
=\frac{1}{\beta}\int_0^\beta\frac{1}{2}\sum_{l\neq k}u\left(~\gamma_{n,l}(t)-~\gamma_{n,k}(t)\right)\d t
\equiv u_{n,k}(\beta)
\ee
and
\be\label{vnk}
\lim_{N,L\to\infty,N/L^d=\rho}V_{n,k}\left(\widetilde{\boldom}_0,\widetilde{\boldom}^{N-n}\right)
=\frac{1}{\beta}\sum_{i=1}^\infty\int_0^\beta u\left(~\nu_{n,i}(t)-~\gamma_{n,k}(t)\right)\d t
\equiv v_{n,k}(\rho,\beta).
\ee
The correlation between $~\gamma_n$ and $\{~\nu_{n,i}\}_{i=1}^\infty$ is inherited from that of $\widetilde{\boldom}_0$ and $\{\widetilde{\boldom}_i\}$. So $$\left\{~\gamma_{n,k}(t)\right\}_{k=0}^{n-1}\bigcup\{~\nu_{n,i}(t)\}_{i=1}^\infty$$ is a Delone set of density $\rho$, but now in $\Rr^d$. Note also that with the numbering (\ref{numbering}) the distance of $\widetilde{\boldom}_i$ to the origin increases with $i$, therefore the correlation between $\widetilde{\boldom}_i$ and $\widetilde{\boldom}_0$ decreases, and this will hold for the correlation between $~\gamma_n$ and $~\nu_{n,i}$. By the assumption on the decay of $u$ and due to the Delone property the infinite sum in (\ref{vnk}) is convergent and the sequences $u_{n,k}$ and $v_{n,k}$ are bounded. Thus,
\be\label{Psi-n}
\Psi_{n}(\rho,\beta)=\lim_{N,L\to\infty,N/L^d=\rho}\Psi_{n,N-n}
=n\left[\phi_n(\beta)+\psi_n(\rho,\beta)\right]
\ee
with bounded sequences
\be\label{phi-psi}
\phi_{n}(\beta)=\frac{1}{n}\sum_{k=0}^{n-1}u_{n,k}(\beta),\quad
\psi_n(\rho,\beta)=\frac{1}{n}\sum_{k=0}^{n-1}v_{n,k}(\rho,\beta).
\ee
In an Appendix we present explicit upper bounds for $\phi_n(\beta)$, $\psi_n(\rho,\beta)$ and also for the free energy density in the case when $u$ is stable and integrable. Besides (\ref{Psi-n}) we have
\be\label{qn-lim}
\lim_{N,L\to\infty,N/L^d=\rho}\frac{q_{n}}{N}=\frac{1}{\rho\lambda_\beta^d n^{d/2}}
\ee
and
\be\label{QN-n/QN}
\lim_{N,L\to\infty,N/L^d=\rho}\frac{Q_{N-n}}{Q_N}=e^{n\beta\mu(\rho,\beta)}.
\ee
Here $\mu(\rho,\beta)$ is the chemical potential in infinite volume, computed in the canonical ensemble. For $N$ finite $\mu$ is by definition the cost of free energy for adding a particle to the system, and can be read off from the ratio $Q_{N-1}/Q_N$ [Van]. Equations~(\ref{Psi-n}), (\ref{qn-lim}) and (\ref{QN-n/QN}) together yield
\begin{proposition}
If the homogeneous phase condition is satisfied,
\be\label{Prho}
\lim_{N,L\to\infty, N/L^d=\rho}P_{N,L}(\xi_1=n)
=\frac{1}{\rho\lambda_\beta^d n^{d/2}}e^{n\beta\left[\mu(\rho,\beta)-\phi_n(\beta)-\psi_n(\rho,\beta)\right]}
=P_{\rho,\beta}(\xi_1=n).
\ee
\end{proposition}
We conclude this section with two simple results about the chemical potential.

\begin{proposition}
If $\beta\to 0$, the asymptotic formula
\be\label{mu-asymp}
\beta\mu(\rho,\beta)\asymp\ln\left(\rho\lambda_\beta^d\right)\quad \mbox{or}\quad z\asymp\rho\lambda_\beta^d
\ee
holds true, where $z$ is the fugacity. If $u$ cannot give rise to bound states, the same formula is valid also when $\rho\to 0$.
\end{proposition}

\noindent
{\bf Proof.}
If $\beta\to 0$ then $\beta\psi_n(\rho,\beta)\to 0$ for all $n$. Also, $\psi_n(\rho,\beta)\to 0$ for all $n$ if $\rho\to 0$ and no bound state can be formed. Because $\phi_1=0$, for $P_{\rho,\beta}(\xi_1=1)\leq 1$ to hold, $\beta\mu(\rho,\beta)$ must go to $-\infty$ at least as fast as $\ln\rho\lambda_\beta^d$. If $n\geq 2$, $\beta\phi_n(\beta)$ is still bounded below. Thus, the exponential in (\ref{Prho}) goes to zero at least as fast as $e^{n\beta\mu(\rho,\beta)}$, implying $\sum_{n=2}^\infty P_{\rho,\beta}(\xi_1=n)=o\left(P_{\rho,\beta}(\xi_1=1)\right)$, so  $P_{\rho,\beta}(\xi_1<\infty)=P_{\rho,\beta}(\xi_1=1)+o(1)$. Since there are no infinite cycles either $\beta$ or $\rho$ goes to zero, $P_{\rho,\beta}(\xi_1=1)$ converges to 1, which implies (\ref{mu-asymp}). $\quad\Box$

\vspace{10pt}
The asymptotic formula (\ref{mu-asymp}) can be compared with the upper bound of Adams et al. [Ad] for the free energy density in the case of an integrable $u$,
\be\label{Adams}
f(\rho,\beta)\leq \rho^2\int u(\x)\d\x+\frac{\rho}{\beta}\ln\left(\rho\lambda_\beta^d\right).
\ee
For $0<\rho\lambda_\beta^d\ll 1$ the bound takes the form
\be\label{Ad-bis}
f(\rho,\beta)\leq \rho^2\int u(\x)\d\x+\rho\ \mu(\rho,\beta).
\ee
The strength of (\ref{Adams}) is that it holds for all $\rho$ and $\beta$. Equation~(\ref{Ad-bis}) is, however, trivial. Recall that
\be
f(\rho,\beta)=\rho\mu^*-p(\mu^*,\beta),
\ee
where $p$ is the pressure and $\mu^*=\partial f(\rho,\beta)/\partial\rho=\mu(\rho,\beta)$. Because the pressure is positive, we always have
\be
f(\rho,\beta)\leq\rho\mu(\rho,\beta).
\ee

\begin{lemma}
The chemical potential satisfies the relation
\be\label{mu}
\mu(\rho,\beta)\leq\liminf_{n\to\infty}\,\left[\phi_n(\beta)+\psi_n(\rho,\beta)\right].
\ee
\end{lemma}

\noindent
\vspace{10pt}
{\bf Proof.}
$P_{\rho,\beta}(\xi_1<\infty)\leq 1$, therefore $P_{\rho,\beta}(\xi_1=n)$ tends to 0 as $n\to\infty$. Thus, for any $\epsilon>0$ there can only be a finite number of $n$ such that
\be
\mu(\rho,\beta)-\phi_n(\beta)-\psi_n(\rho,\beta)>\epsilon,
\ee
meaning (\ref{mu}).

\subsection{Fluids}\label{Bose-fluids}

We show that macroscopic cycle percolation is a robust phenomenon occurring in Bose systems that remain fluid down to zero temperature. The examples of liquid helium and a repulsively interacting Bose gas will be presented. Consider
\be\label{sum-prob-He}
P_{\rho,\beta}(\xi_1<\infty) =\frac{1}{\rho\lambda_\beta^d}\sum_{n=1}^\infty \frac{1}{n^{d/2}}\,e^{n\beta\left[\mu(\rho,\beta)-\chi_n(\rho,\beta)\right]},
\ee
where
\be
\chi_n(\rho,\beta)=n^{-1}\Psi_n(\rho,\beta)=\phi_n(\beta)+\psi_n(\rho,\beta),
\ee
the mean potential energy of a particle in a cycle of length $n$ in the field of all the other particles. Its detailed form is
\bea\label{chi-n-1}
\chi_n(\rho,\beta)
&=&\frac{1}{n}\sum_{k=0}^{n-1}\frac{1}{\beta}\int_0^\beta\d t \left[\frac{1}{2}\sum_{l\neq k}u(~\gamma_{n,l}(t)-~\gamma_{n,k}(t))+\sum_{i=1}^\infty u(~\nu_{n,i}(t)-~\gamma_{n,k}(t))\right], \nonumber\\
&\equiv& \frac{1}{n}\sum_{k=0}^{n-1}\frac{1}{\beta}\int_0^\beta\d t\  \chi_{n,k}(\rho,\beta;t),
\eea
so the average is taken over the particles in the cycle and over time. Equation~(\ref{chi-n-1}) can be rewritten as a unique time average,
\be\label{ch-n-2}
\chi_n(\rho,\beta)=\frac{1}{n\beta}\int_0^{n\beta}\d t \left[\frac{1}{2}\sum_{k=1}^{n-1}u\left(~\gamma_{n}(t)-~\gamma_{n}((t+k\beta)\,\mbox{mod}\,n\beta)\right) +\sum_{i=1}^\infty u\left(~\gamma_{n}(t)-~\nu_{n,i}(t\,\mbox{mod}\,\beta)\right)\right].
\ee
This suggests that the limit of $\chi_n(\rho,\beta)$ as $n$ goes to infinity actually exists. Moreover, we do not expect any trend in the convergence: only the division of the mean energy per particle between the two sums changes systematically with $n$, the $n$-dependence of the whole is a random fluctuation about the limit. Therefore, the convergence to the limit should not be slower than $(n\beta)^{-1}$.

\vspace{10pt}
\noindent
{\bf Ergodic hypothesis.} {\em For any $\beta>0$ the limit $\chi(\rho,\beta)=\lim_{n\to\infty}\chi_n(\rho,\beta)$ exists. For any $\beta_0>0$
\be\label{ergod2}
\sup_{\beta\geq\beta_0,n\geq 1}\left\{n\beta\left|\chi(\rho,\beta)-\chi_n(\rho,\beta)\right|\right\} \equiv C_0(\rho,\beta_0)<\infty.
\ee}
The hypothesis implies also that
\be
\lim_{\beta\to\infty}\left[\chi(\rho,\beta)-\chi_n(\rho,\beta)\right]=0.
\ee
The fluctuations of inter-particle distances and the potential energy with them decrease as the temperature decreases ($d_1$ increases, $d_2$ decreases), therefore $\chi(\rho,\beta)$ is a monotone decreasing function of $\beta$. If
\be
\chi_\rho=\lim_{\beta\to\infty}\chi(\rho,\beta),
\ee
then a consequence of the ergodic hypothesis is that $\lim_{\beta\to\infty}\chi_n(\rho,\beta)=\chi_\rho$. The average within the cycles seems also to be superfluous, so
\be\label{ergod1}
\lim_{\beta\to\infty}\frac{1}{\beta}\int_0^\beta\d t\  \chi_{n,k}(\rho,\beta;t)=
\lim_{\beta\to\infty}\chi_n(\rho,\beta)=\chi_\rho\qquad\mbox{all}\ n,\,k.
\ee
From (\ref{ergod2}) it follows that
\be
\chi(\rho,\beta)-\chi_n(\rho,\beta)=\frac{C_0(\rho,\beta_0)-C_n(\rho,\beta_0)}{n\beta},
\ee
where $C_n(\rho,\beta_0)\geq 0$ and is larger than $C_0(\rho,\beta_0)$ when the difference on the left is negative.

Before $P_{\rho,\beta}(\xi_1<\infty)$ starts to decrease, it stays at 1 when $\beta$ increases from small to rather high values while $\rho$ is fixed. There are two sources to compensate the initial decrease of $(\rho\lambda_\beta^d)^{-1}$: the increase of $\mu(\rho,\beta)$ from $-\infty$, c.f. Eq.~(\ref{mu-asymp}), and the decrease of $\chi(\rho,\beta)$ from $+\infty$ or some large positive value (because $d_1= 0$) at $\beta=0$. $P_{\rho,0}(\xi_1=n)=\delta_{n,1}$, and the overall phenomenon is a flow of probability towards larger $n$ as $\beta$ increases. Meanwhile, the rate of decrease of $(\rho\lambda_\beta^d)^{-1}$ and together with it the variation of $\mu(\rho,\beta)$ and $\chi(\rho,\beta)$ slows down. The transition is approached when these two are close to each other and the bound
\be\label{mu-bis}
\mu(\rho,\beta)\leq\chi(\rho,\beta),
\ee
c.f. (\ref{mu}), becomes effective. Now
\be
\mu(\rho,\beta)-\chi_n(\rho,\beta) =\mu(\rho,\beta)-\chi(\rho,\beta)+\frac{C_0(\rho,\beta_0)-C_n(\rho,\beta_0)}{n\beta},
\ee
so
\be
P_{\rho,\beta}(\xi_1<\infty)= \frac{e^{C_0(\rho,\beta_0)}}{\rho\lambda_\beta^d}\sum_{n=1}^\infty \frac{1}{n^{d/2}}\,e^{n\beta\left[\mu(\rho,\beta)-\chi(\rho,\beta)\right]-C_n(\rho,\beta_0)} \leq e^{C_0(\rho,\beta_0)}\frac{\zeta(d/2)}{\rho\lambda_\beta^d}.
\ee
The conclusion is that at a sufficiently large value of $\beta$, $P_{\rho,\beta}(\xi_1<\infty)$ starts to decrease and tends to zero as $\beta$ goes to infinity. By definition, $C_0(\rho,\beta)$ is a monotone decreasing function of $\beta$. So $\beta_c$ is the largest value for $\beta_0$ to cover the whole interval $\beta\geq\beta_c$ with the smallest $C_0(\rho,\beta_0)$ and $C_n(\rho,\beta_0)$. Depending on whether
$$e^{C_0(\rho,\beta_c)}\sum_{n=1}^\infty \frac{1}{n^{d/2}}\,e^{n\beta\left[\mu(\rho,\beta_c)-\chi(\rho,\beta_c)\right]-C_n(\rho,\beta_c)}$$
is smaller or larger than $\zeta(d/2)$, $\beta_c$ can be smaller or larger than $\beta^0_c$. Let
\be
\mu_\rho=\lim_{\beta\to\infty}\mu(\rho,\beta),
\ee
the chemical potential in the ground state. An important question is whether $\mu_\rho<\chi_\rho$ or $\mu_\rho=\chi_\rho$. The alternative is related to the sign of $\Delta T_c$. Although one cannot conclude without a more precise knowledge of $C_0(\rho,\beta_c)$, there is better chance to $\Delta T_c>0$ if $\mu_\rho<\chi_\rho$ and to $\Delta T_c<0$ if $\mu_\rho=\chi_\rho$.

We still have to find the rate of increase in finite systems of those trajectories that eventually contribute to cycle percolation. Let $\{K_N\}$ be any diverging sequence such that $K_N/N\to 0$. From Proposition~\ref{PKN} we know that the limit of $P_{N,L}(\xi_1\leq K_N)$ cannot be smaller than $P_{\rho,\beta}(\xi_1<\infty)$. That it cannot be larger depends on the existence of a uniform upper bound on
$
\frac{Q_{N-n}}{Q_N}G_N(n)
$
for $n\leq K_N$, c.f. Eq.~(\ref{QN-2}). If $n$ is fixed, this quantity tends to $\exp\{n\beta[\mu(\rho,\beta)-\chi_n(\rho,\beta)]\}\leq \exp C_0(\rho,\beta_0)$ for $\beta\geq\beta_0$ in the thermodynamic limit. Now
\be
\lim_{N,L\to\infty,N/L^d=\rho}Q_{N-n}/Q_{N-n+1}=e^{\beta\mu(\rho,\beta)}\quad\mbox{if}\ n=o(N),
\ee
and the convergence to the limit is from below. The reason is that $\mu(\rho,\beta)$ is an increasing function of $\rho$, and $(N-n)/L^d$ tends to $\rho$ from below. Thus,
\be
\lim_{N,L\to\infty,N/L^d=\rho}n^{-1}\ln Q_{N-n}/Q_N
=\beta\mu(\rho,\beta)\quad\mbox{if}\ n=o(N)
\ee
with a convergence to the limit from below. The average
\be
n^{-1}\ln Q_{N-n}/Q_N=n^{-1}\sum_{k=1}^n\ln Q_{N-k}/Q_{N-k+1}
\ee
is dominated by the terms with $k$ of order $n$. In lowest order the increase of the chemical potential with the density is linear, so for some $C'>0$
\be\label{mu-finite}
\frac{1}{n\beta}\ln Q_{N-n}/Q_N\leq\mu(\rho,\beta)-C'\frac{n}{L^d}.
\ee
On the other hand, due to temperedness (\ref{tempered}),
\be\label{chi-finite}
\frac{1}{n}\sum_{k=0}^{n-1}\frac{1}{\beta}\int_0^\beta\d t \left[\frac{1}{2}\sum_{l\neq k}u_L(\widetilde{\boldom}_{n,l}(t)-\widetilde{\boldom}_{n,k}(t))+\sum_{i=1}^{N-n} u(\widetilde{\boldom}_{i}(t)-\widetilde{\boldom}_{n,k}(t))\right] =\chi_n(\rho,\beta)+\frac{\delta_{n,N,L}(\beta)}{L^\eta},
\ee
where $\delta_{n,N,L}$ is bounded and its sign is that of the tail of the interaction. Combining Eqs.~(\ref{mu-finite}) and (\ref{chi-finite}) and recalling that for interactions with a negative tail we asked $\eta\geq d$, it is seen that
\be\label{finite}
\frac{Q_{N-n}}{Q_N}G_N(n)\leq e^{C_0(\rho,\beta_0)}
\ee
holds for any $n=o(N)$.

In Eq.~(\ref{P0NL(KN)}) we replace $\frac{Q^0_{N-n}}{Q^0_N}$ by $\frac{Q_{N-n}}{Q_N}G_N(n)$, bound it above by $e^{C_0(\rho,\beta_0)}$ in the second sum, and find that this one goes to zero for any $K_N=o(N)$.
Hence,
\be\label{epsilonL^2}
\lim_{N,L\to\infty, N/L^d=\rho}P_{N,L}\left(\xi_1\leq K_N\right)= P_{\rho,\beta}(\xi_1<\infty)\quad\mbox{for any $K_N\to\infty$, $K_N/N\to 0$}.
\ee
The equality (\ref{epsilonL^2}) means that the actual length of long cycles is of order $N$ as in the ideal gas. We summarize the result.

\begin{theorem}\label{thm-cycle-percol}
In the fluid phase of a system of interacting bosons macroscopic cycle percolation takes place at low enough temperatures, provided that the system does not crystallize. The probability $P_{\rho,\beta}(\xi_1~<\infty)$ of the occurrence of finite cycles converges to zero as the temperature decreases.
\end{theorem}

Note that the theorem gives no information about
\be\label{epsilon-bar}
\varepsilon_s=\inf\left\{\varepsilon: \lim_{N,L\to\infty,N/L^d=\rho}P_{N,L}(\xi_1>\varepsilon N)=0\right\}.
\ee
In the ideal gas the infimum is attained, $\varepsilon^0_s=P^0_{\rho,\beta}(\xi_1=\infty)=\rho_0/\rho$.

\vspace{5pt}
\noindent
We think that the condition for macroscopic cycle percolation, ergodicity in the form (\ref{ergod2}), is not satisfied in solids.

\subsubsection{Liquid helium}\label{liq-helium}

The interaction between noble gas atoms and small molecules was extensively studied in the past. An important contribution to its quantum-mechanical theory is due to London [Lon3]. The interaction has a strongly repulsive core caused by the overlap of atomic electron clouds and an attractive tail whose form in leading order is $\sim -|\x|^{-6}$, corresponding to the interaction of two mutually induced dipole momenta; thus, the interaction is tempered and $\eta=d$.  For precise computations the short-range repulsion is fitted by an exponential [Sl, Buc, Az]. The Lennard-Jones potential represents it by a $\sim |\x|^{-12}$ divergence at the origin [M, Lenn, Boe]; between 9 and 14 any power would do, 12 is chosen for convenience [Boe].

Under the effect of such an interaction, in real Bose systems vapor-liquid transition takes place at low temperatures. Even this classical phase transition is poorly understood beyond the mean-field theory; rigorous proofs could be done only for some wisely devised model interactions; see [Leb] and references therein. The van der Waals theory of fluids predicts that at a fixed density the chemical potential increases under cooling till the boiling point, below which it decreases [J], down to -7.2K in the case of liquid helium [Toe]. Below we list some relevant data for the helium liquid, including the parameters of the Lennard-Jones potential
\be\label{L-J}
u(\x)=4\epsilon\left[\left(\frac{d_0}{|\x|}\right)^{12}-\left(\frac{d_0}{|\x|}\right)^6\right].
\ee

\begin{itemize}
\item
$u(\x)=0$ at $|\x|=d_0=2.6${\rm \AA}.
\item
$u(\x)$ is minimal at $|\x|=d_m=2^{1/6}d_0=2.9${\rm \AA}, the depth of the minimum is $\epsilon\approx 11$K.
\item
The critical point is (5.2 K, 2.3 bar).
\item
The mass density is approximately constant, 0.147 g cm$^{-3}$, below 4.2K at saturated vapor pressure [Sn], [So]. It corresponds to $\rho=2.2\times10^{22}{\rm cm}^{-3}$, $\rho^{-1/3}=3.57${\rm \AA}.
\item
The system is in a liquid state for pressures below 25 bar or mass densities below 0.190 g cm$^{-3}$. The vapor pressure at the lambda-point is less than 0.1 bar.
\item
$T_{\rm boiling}=4.22$K, $T_c=2.17$K (superfluid transition), $T^0_c=3.12$K.
\item
$\lambda_\beta\equiv\lambda_{T}=\frac{8.67}{\sqrt{T[{\rm K}]}}${\rm \AA} which gives $\lambda_{4.22{\rm K}}=4.22{\rm \AA}$, $\lambda_{T_c}=5.89{\rm \AA}$, $\lambda_{1.5{\rm K}}=7.05{\rm \AA}$.
\item
$\rho\lambda_{T_c}^3=4.49$, $\rho\lambda_{T^0_c}^3=\zeta(3/2)=2.612$.
\end{itemize}

The helium gas can be cooled down to zero temperature so that the system passes above the critical point, and the only phase boundary it crosses is between He I and He II; thus, the homogeneous phase condition is satisfied at all temperatures. If cooling is done under ambient pressure, the system enters the He I phase through ordinary condensation, and our description is valid below the boiling point. The fact that the transition temperature is lower than that of the ideal gas suggests that $\chi_\rho$ agrees with the chemical potential in the ground state.

Liquid helium is the only real system which does not crystallize at ambient pressure. For comparison we cite the relevant data of liquid H$_2$ and liquid Ne. The boiling point of liquid H$_2$ is 20.27K, $\lambda_{20{\rm K}}=2.74{\rm \AA}$; the freezing point is 14K, $\lambda_{14{\rm K}}=3.28{\rm \AA}$. This is to be compared with $\rho^{-1/3}=3.61{\rm \AA}$ and $d_0=2.9{\rm \AA}$, $d_m=3.26{\rm \AA}$; for $u$ these are Lennard-Jones' data [Boe]. The depth of the potential well is more than four times the value for helium. These numbers show that in the respective liquids the attraction between H$_2$ molecules is much stronger than between He atoms. Meanwhile, the maximum of $\rho\lambda_\beta^3$ is 0.748, less than a third of $\zeta(3/2)$. As the temperature of the liquid decreases, the inter-molecular attraction becomes stronger, and before $\rho\lambda_\beta^3$ could be large enough, the system crystallizes. Neon is in liquid phase between 24.6K and 27.1K. Its mean inter-atomic distance is $\rho^{-1/3}=3.026{\rm \AA}$, while $\lambda_{26{\rm K}}=0.76{\rm \AA}$: $\rho\lambda_\beta^3$ is far too small to give any chance to cycle percolation.


\subsubsection{Bose gas with a nonnegative interaction}

If the interaction is nonnegative, $\mu<0$ prevents cycle percolation [Uel2]. An interesting consequence of it is as follows.

\begin{proposition}\label{mu-lemma}
For a nonnegative $u$ let $\rho_m(\beta)$ denote the smallest density at which $f(\rho,\beta)$ is minimum, that is, $\mu(\rho_m(\beta),\beta)=0$. Then $\rho_m(\beta)\leq\rho_c^0(\beta)$, or equivalently, $\mu(\rho,\beta)\geq 0$ for $\rho\geq\rho_c^0(\beta)$. If $u$ is also superstable, then $\mu(\rho,\beta)> 0$ for $\rho>\rho_c^0(\beta)$.
\end{proposition}

\noindent
\vspace{5pt}
{\bf Proof.}
Suppose that $\rho_m(\beta)>\rho_c^0(\beta)$. For $\rho_c^0(\beta)<\rho<\rho_m(\beta)$ and any $L$ large enough, there is some $N$ in a $o(L^d)$ neighborhood of $\rho L^d$ such that $Q_N\geq Q_{N'}$, all $N'<N$, and therefore
$$
Q_{N-n}G_N(n)/Q_N\leq 1,\quad 1\leq n\leq N.
$$
As a consequence,
$
P_{N,L}(\xi_1<\varepsilon N)\leq\sum_{n<\varepsilon N}q_n/N,
$
just as in the case of the ideal gas. Because $\rho\lambda_\beta^d>\zeta(d/2)$, Proposition~\ref{thm-ideal-2} applies, with the result that there is cycle percolation. However, $\mu(\rho,\beta)<0$ at $\rho$, and this contradicts the existence of infinite cycles. $\quad\Box$

\vspace{10pt}
As we supposed that $u$ is superstable, $\rho_m(\beta)$ is the unique solution for $\rho$ of the equation $\mu(\rho,\beta)=0$, and $\rho_m(\beta)\lambda_\beta^{d}\leq \zeta(d/2)$. From here we can obtain a strong indication that cycle percolation cannot occur even at $\mu=0$.

\begin{proposition}\label{mu=0-no-BEC}
If $u$ is nonnegative, then
\be
\lim_{\beta\to\infty} P_{\rho_m(\beta),\beta}(\xi_1<\infty)=1.
\ee
\end{proposition}

\vspace{10pt}
\noindent
{\bf Proof.} Consider
\be
\sum_{n=1}^\infty P_{\rho_m(\beta),\beta}(\xi_1=n) =\frac{1}{\rho_m(\beta)\lambda_\beta^d}\sum_{n=1}^\infty \frac{e^{-n\beta\left[\phi_n(\beta)+\psi_n(\rho_m(\beta),\beta)\right]}}{n^{d/2}}.
\ee
Because $n^{-d/2}$ is a summable upper bound, we can interchange the limit $\beta\to\infty$ and the summation. Suppose first that $u$ is integrable. We use the upper bounds (\ref{upper2}) and (\ref{psi}). Recalling from Eq.~(\ref{alpha-nk}) that
\be\label{alpha-nk'}
\alpha_{n,k} =\frac{1}{\lambda_\beta^2}\left(\frac{1}{k}+\frac{1}{n-k}\right),
\ee
they yield
\be
\phi_n(\beta)
\leq \frac{1}{2}\int u(\x)\sum_{k=1}^{n-1}\alpha_{n,k}^{d/2}e^{-\pi\alpha_{n,k}\x^2}\d\x
\leq\frac{\zeta(d/2)}{\lambda_\beta^d}\int u(\x)\d\x
\ee
and
\be\label{psi}
\psi_n(\rho,\beta)\leq\rho\int u(\x)\d\x.
\ee
So both
\be
\beta\phi_n(\beta)\leq \frac{\beta\zeta(d/2)}{\lambda_\beta^d}\int u(\x)\d\x
\ee
and
\be
\beta\psi_n(\rho_m(\beta),\beta)\leq\beta\rho_m(\beta)\int u(\x)\d\x
\ee
go to zero when $\beta$ goes to infinity. The second holds because
$$\beta\rho_m(\beta)\leq \frac{\beta}{\lambda_\beta^d}\zeta(d/2).$$
If $u$ is not integrable at the origin, based on the Delone property it is still possible to bound $\chi_n(\rho,\beta)$ from both sides by a positive multiple of $\rho$; therefore $\beta\chi_n(\rho_m(\beta),\beta)$ goes to zero as $\beta$ increases.
Thus,
\be
1\geq\lim_{\beta\to\infty}\sum_{n=1}^\infty P_{\rho_m(\beta),\beta}(\xi_1=n) =\frac{\zeta(d/2)}{\lim_{\beta\to\infty}\rho_m(\beta)\lambda_\beta^{d}}\geq 1.
\ee
This shows also that $\lim_{\beta\to\infty}\rho_m(\beta)\lambda_\beta^{d}=\zeta(d/2)$. $\quad\Box$

\vspace{10pt}
A repulsively interacting gas does not condense to a liquid but can crystallize at high density; this is known for pure hard-core interactions [Ald, Bow]. If $\rho$ is small enough, the system remains a gas down to zero temperature. For sufficiently low temperatures $\rho_m(\beta)<\rho$, $\mu(\rho,\beta)$ becomes positive, and macroscopic cycles can appear.


\subsection{Solids}

When cooling without applied pressure, the ultimate phase of rare gases, except for helium, is a close-packed crystal. Helium crystallizes at 25 bar in hcp structure; the fcc structure appears in its phase diagram only at about 1 kbar, and it is restricted to relatively high temperatures ($\geq$15K) [Gl]. All the other rare gases crystallize in fcc structure. Computing lattice sums for the hcp and fcc structures with the Lennard-Jones potential, the hcp structure appeared to be favored, and this remained true for more precise two-body forces. The conclusion was that in the case of the heavier rare gas solids three-body forces are necessary to reproduce the correct crystal structure [Ni]. Keeping this in mind, one still may ponder about the possibility of cycle percolation or something else in a Bose crystal, where only two-body forces are in action.

When cooling, the solids of Ne, Ar, Kr, and Xe are formed due to increasingly strong cohesion forces: the depth of the potential well of the pair potential between Ne atoms is about six times the value between He atoms, and this factor is roughly 100 for argon, 150 for krypton and 200 for xenon [Gl]. All this means that crystallization is basically a classical phase transition. Still, $P_{\rho,\beta}(\xi_1=n)$ is nonzero for each $n$, cycles of arbitrary length exist, but they are composed mostly of nearest neighbors. With the increase of $\lambda_\beta$ a transition to the realm of quantum mechanics takes place. We see two possible accompanying phenomena of this transition. The first is no cycle percolation but a cross-over to quantum-mechanical localization. The second possibility, more exotic, is that (non-macroscopic) cycle percolation appears [Uel1]. This would imply that the particles in infinite cycles are in extended states. Apparently, neither localization nor cycle percolation has thermodynamic consequences -- or one should look for them more carefully. The case of He4 is particular, the zero point energy almost precisely compensates the cohesion energy, and a crystal can be formed only under pressure. Cycle percolation with BEC seems possible, superflow in solid helium up to 66 bar was already reported [Kim]. Looking at the phase diagram, Fig.~1 in [Gl], maybe the easiest to detect a supersolid is in the loosest crystalline phase of solid He4, the tiny bcc phase bordered by three other phases, He I, He II and the hcp solid.


\newsec{Bose-Einstein condensation}\label{proof-BEC}

We prove the analogue of Theorems \ref{thm-ideal-1} and \ref{thm-ideal-3}.

\begin{theorem}\label{thm-interacting-1}
\noindent
(i) In an interacting system of bosons Bose-Einstein condensation implies cycle percolation.

\noindent
(ii) Finite cycles do not contribute to the condensate.

\noindent
(iii) Macroscopic cycle percolation implies Bose-Einstein condensation. With probability 1 every particle in the condensate is in a macroscopic cycle.
\end{theorem}

\vspace{10pt}
\noindent
{\bf Proof.} The partition function can be brought to a form in which each particle of the cycle of 1 is in the same one-particle state. As a result, if the cycle is long then a large number of particles occupy the same state. Because $\varphi_\0(\x)\equiv L^{-d/2}$ is the eigenvector belonging to the largest eigenvalue of the one-particle reduced density matrix, $\langle N_\0\rangle$ cannot be smaller than the occupation number of any other one-particle state [Su3], so we could conclude about BEC. However, complication arises from the fact that single-particle momenta are not conserved, as they were in the absence of interaction. When there is cycle percolation in the ideal gas, with probability 1 all the particles of zero momentum -- and only them -- are in one of the infinite cycles [Su2]. This is why the probability of cycle percolation equals the condensate fraction, c.f. Eq.~(\ref{ideal-concise}). No such equality holds for interacting bosons. Instead, what we will find is that the probability that a one-particle state is macroscopically occupied spreads over an infinity of states $\varphi$, and the probability of occupation of $\varphi_\0$ results from integrating over $\varphi$ the infinitesimal contributions weighted by $|\langle\varphi,\varphi_\0\rangle |^2$.

Application of the first mean-value theorem in Eq.~(\ref{recur-gen}) slightly differently as in Section~\ref{Gordian} yields
\be
Q_N =\frac{1}{N}\sum_{n=1}^N e^{-\beta{\cal U}_{n}(\widehat{\boldom}_0)} \int_\Lambda\d\x\int W^{n\beta}_{\x\x}(\d\boldom_0) Q_{N-n}(\boldom_0).
\ee
We divide and multiply with $Q_{N-n}$ to find
\bea\label{QN-new}
Q_N &=& \frac{1}{N}\sum_{n=1}^N e^{-\beta{\cal U}_{n}(\widehat{\boldom}_0)}Q_{N-n}\int_\Lambda\d\x\int W^{n\beta}_{\x\x}(\d\boldom_0)\left\langle e^{-\beta V_{n}\left(\boldom_0,\,\cdot\right)}\right\rangle_{\mu^\beta_{N-n}}
\nonumber
\\&=& \frac{1}{N}\sum_{n=1}^N e^{-\beta{\cal U}_{n}(\widehat{\boldom}_0)}Q_{N-n} \int_\Lambda\d\x\int W^{n\beta}_{\x\x}(\d\boldom_0)e^{-\beta V_{n}\left(\boldom_0,\widehat{\boldom}^{N-n}\right)},
\eea
see Eqs.~(\ref{QN-2}), (\ref{self+int}) and (\ref{Gibbs-omM}). In the second line we replaced the average over $\boldom^{N-n}$ by a representative term. $\widehat{\boldom}^{N-n}$ is first of all determined by the interaction among the $N-n$ particles, causing it to be a Delone set in itself at all $t\in[0,\beta]$. There is, moreover the interaction with the particles in the cycle of 1, making $\widehat{\boldom}^{N-n}$ dependent on $\boldom_0=\left\{\boldom_{n,k}\right\}_{k=0}^{n-1}$. However, this dependence is weak: $\boldom_0$ may be typical or atypical according to $W^{n\beta}_{\x\x}$, this has little influence on the average. What counts is that $\widehat{\boldom}^{N-n}(t)$ is neither too close to nor too far from $\boldom_0(t)$. Let us rewrite $\beta V_n$ as we did with $\chi_n(\rho,\beta)$ in Eq.~(\ref{ch-n-2}):
\be
\beta V_{n}\left(\boldom_0,\widehat{\boldom}^{N-n}\right) =\int_0^{n\beta}\d t \sum_{i=1}^{N-n} u_L(\boldom_0(t)-\widehat{\boldom}_i(t\,\mbox{mod}\,\beta)).
\ee
At every instant the important contribution to the sum comes from particles close to $\boldom_0(t)$.
We can obtain the same value for $\beta V_{n}\left(\boldom_0,\widehat{\boldom}^{N-n}\right)$ by a suitable choice of $\widehat{\boldom}^{N-n}$ from a family of continuously varying Delone sets of the form $\boldom^{N-n}[\boldom_0(t)]$, in which $\boldom^{N-n}$ as a $(N-n)$-point set depends on $t$ through the instantaneous position of $\boldom_0$. The set $\widehat{\boldom}^{N-n}$ thus chosen must have the properties that for every $\x\in\Lambda$ and every $i$, $\left|\widehat{\boldom}_j[\x]-\widehat{\boldom}_i[\x]\right|\geq d_1$ for all $j\neq i$ but there is some $j\neq i$ such that $\left|\widehat{\boldom}_j[\x]-\widehat{\boldom}_i[\x]\right|\leq d_2$, and that
$
d_1\leq \mbox{dist}(\x,\widehat{\boldom}^{N-n}[\x])\leq d_2.
$
Moreover, $\widehat{\boldom}^{N-n}[\x]$ is a continuous function of $\x$. With
\be
\beta V_{n}\left(\boldom_0,\widehat{\boldom}^{N-n}\right) =\int_0^{n\beta}\d t \sum_{i=1}^{N-n} u_L(\boldom_0(t)-\widehat{\boldom}_i[\boldom_0(t)])
\ee
Eq.~(\ref{QN-new}) becomes
\be
Q_N=\frac{1}{N}\sum_{n=1}^N e^{-\beta{\cal U}_{n}(\widehat{\boldom}_0)}Q_{N-n} \int_\Lambda\langle\x|e^{-n\beta h^\beta_{n,N,L}}|\x\rangle\d\x =\frac{1}{N}\sum_{n=1}^N e^{-\beta{\cal U}_{n}(\widehat{\boldom}_0)}Q_{N-n}\tr e^{-n\beta h^\beta_{n,N,L}},
\ee
where $h^\beta_{n,N,L}$ is a one-body Hamiltonian,
\be\label{h-nNL}
h^\beta_{n,N,L}=-\frac{\hbar^2}{2m}\Delta+V_{\widehat{\boldom}^{N-n}},\qquad V_{\widehat{\boldom}^{N-n}}(\x)=\sum_{i=1}^{N-n}u_L\left(\widehat{\boldom}_i[\x]-\x\right).
\ee
Let
\be
\epsilon_{0,\widehat{\boldom}^{N-n}}<\epsilon_{1,\widehat{\boldom}^{N-n}} \leq\epsilon_{2,\widehat{\boldom}^{N-n}}\leq\cdots
\ee
be the eigenvalues, $\varphi_{j,\widehat{\boldom}^{N-n}}$ the corresponding eigenvectors of $h^\beta_{n,N,L}$. The $n$ particles in the cycle of 1 occupy $\varphi_{i,\widehat{\boldom}^{N-n}}$ with the probability $e^{-n\beta\epsilon_{i,\widehat{\boldom}^{N-n}}}\left[\sum_{j=0}^\infty e^{-n\beta\epsilon_{j,\widehat{\boldom}^{N-n}}}\right]^{-1}$. In fluids $V_{\widehat{\boldom}^{N-n}}(\x)$ is an unordered field, so the eigenvalues are actually single apart from accidental degeneracy. The ultimate form of the partition function is
\be\label{QN-new-form}
Q_N =\frac{1}{N}\sum_{n=1}^N e^{-\beta {\cal U}_{n}(\widehat{\boldom}_0)} Q_{N-n}\sum_{j=0}^\infty e^{-n\beta\epsilon_{j,\widehat{\boldom}^{N-n}}}.
\ee
Furthermore,
\be
P_{N,L}(\xi_1=n)= \frac{1}{N}\frac{Q_{N-n}}{Q_N}\ e^{-\beta {\cal U}_{n}(\widehat{\boldom}_0)}\sum_{j=0}^\infty e^{-n\beta\epsilon_{j,\widehat{\boldom}^{N-n}}},
\ee
whose comparison with Eq.~(\ref{PNL}) yields
$$
e^{-\beta {\cal U}_{n}(\widehat{\boldom}_0)}\sum_{j=0}^\infty e^{-n\beta\epsilon_{j,\widehat{\boldom}^{N-n}}}=q_n G_N(n) = q_n e^{-\beta\left[{\cal U}_{n}(\widetilde{\boldom}_0)+ V_{n}\left(\widetilde{\boldom}_0,\widetilde{\boldom}^{N-n}\right)\right]}
$$
or
\be\label{trace-hat}
\sum_{j=0}^\infty e^{-n\beta\epsilon_{j,\widehat{\boldom}^{N-n}}}=q_n e^{-\beta\left[{\cal U}_n(\widetilde{\boldom}_0)-{\cal U}_n(\widehat{\boldom}_0) + V_{n}\left(\widetilde{\boldom}_0,\widetilde{\boldom}^{N-n}\right)\right]}.
\ee
This equation must be satisfied and sets thereby a condition on $\widehat{\boldom}^{N-n}$. Utilizing
\be
\tr |\k\rangle\langle\k|\,e^{-n\beta h^\beta_{n,N,L}} =\sum_{j=0}^\infty e^{-n\beta\epsilon_{j,\widehat{\boldom}^{N-n}}} \left|\langle\varphi_{j,\widehat{\boldom}^{N-n}},\varphi_{\k}\rangle\right|^2,
\ee
the probability of the joint event that particle 1 is in a cycle of length $n$ and has wave vector $\k$ reads
\bea P_{N,L}(\xi_1=n,\k_1=\k)&=& \frac{1}{N}\frac{Q_{N-n}}{Q_N}\ e^{-\beta{\cal U}_n(\widehat{\boldom}_0)} \sum_{j=0}^\infty e^{-n\beta\epsilon_{j,\widehat{\boldom}^{N-n}}} \left|\langle\varphi_{j,\widehat{\boldom}^{N-n}},\varphi_{\k}\rangle\right|^2\nonumber\\&=& \frac{\sum_{j=0}^\infty e^{-n\beta\epsilon_{j,\widehat{\boldom}^{N-n}}} \left|\langle\varphi_{j,\widehat{\boldom}^{N-n}},\varphi_{\k}\rangle\right|^2}{\sum_{j=0}^\infty e^{-n\beta\epsilon_{j,\widehat{\boldom}^{N-n}}}}P_{N,L}(\xi_1=n).\eea
From here
\be
\label{av-Nk}\langle N_\k\rangle=N P_{N,L}(\k_1=\k)=N\sum_{n=1}^N \frac{\sum_{j=0}^\infty e^{-n\beta\epsilon_{j,\widehat{\boldom}^{N-n}}} \left|\langle\varphi_{j,\widehat{\boldom}^{N-n}},\varphi_{\k}\rangle\right|^2}{\sum_{j=0}^\infty e^{-n\beta\epsilon_{j,\widehat{\boldom}^{N-n}}}}P_{N,L}(\xi_1=n).
\ee
In particular,
\bea\label{cond-fraction-new}
\frac{\langle N_\0\rangle}{N}&=& \sum_{n=1}^N\frac{\sum_{j=0}^\infty e^{-n\beta\epsilon_{j,\widehat{\boldom}^{N-n}}} \left|\langle\varphi_{j,\widehat{\boldom}^{N-n}},\varphi_{\0}\rangle\right|^2}{\sum_{j=0}^\infty e^{-n\beta\epsilon_{j,\widehat{\boldom}^{N-n}}}}P_{N,L}(\xi_1=n)\nonumber\\&=& \sum_{n=1}^N\frac{\int_\Lambda\d\x\int W^{n\beta}_{\0\x}(\d\boldom_0)e^{-\beta{\cal U}_n(\boldom_\0)}Q_{N-n}(\boldom_\0)}{L^d\int W^{n\beta}_{\0\0}(\d\boldom_0)e^{-\beta{\cal U}_n(\boldom_\0)}Q_{N-n}(\boldom_\0)}P_{N,L}(\xi_1=n).
\eea
The second line comes from Eq.~(\ref{N0-path}).

\noindent
(i)-(ii) We proceed with the first line by bounding the summand,
\be\label{eigenstate-av}
\frac{\sum_{j=0}^\infty e^{-n\beta\epsilon_{j,\widehat{\boldom}^{N-n}} \left|\langle\varphi_{j,\widehat{\boldom}^{N-n}},\varphi_{\0}\rangle\right|^2}}{\sum_{j=0}^\infty e^{-n\beta\epsilon_{j,\widehat{\boldom}^{N-n}}}} <\frac{e^{-n\beta\epsilon_{0,\widehat{\boldom}^{N-n}}}}{\sum_{j=0}^\infty e^{-n\beta\epsilon_{j,\widehat{\boldom}^{N-n}}}}.
\ee
$\epsilon_{0,\widehat{\boldom}^{N-n}}$ may be negative if the interaction has an attractive tail. However, it remains bounded below as $N,L\to\infty$, $N/L^d=\rho$. Therefore,  $e^{-n\beta\epsilon_{0,\widehat{\boldom}^{N-n}}}$ is bounded for $n$ fixed, while the denominator goes to infinity. In effect, for $n$ fixed,
\be
0<\lim_{N,L\to\infty, \frac{N}{L^d}=\rho}G_N(n)=e^{-n\beta\chi_n(\rho,\beta)}<\infty,
\ee
the limit of $e^{\beta{\cal U}_n(\widehat{\boldom}_0)}$ is also finite nonzero, while $q_n$ tends to infinity as $L^d$. So the average (\ref{eigenstate-av}) goes to zero, both for nonnegative and for Lennard-Jones type interactions. We fix an integer $M$, cut the sum (\ref{cond-fraction-new}) into two, from 1 to $M$ and from $M+1$ to $N$, take the thermodynamic limit, and find that the finite sum goes to zero. Sending $M$ to infinity results
\bea\label{BEC->CP}
\frac{\rho_0}{\rho}&=&\lim_{N,L\to\infty, \frac{N}{L^d}=\rho}\frac{\langle N_\0\rangle}{N} = \lim_{M\to\infty}\lim_{N,L\to\infty, \frac{N}{L^d}=\rho}\sum_{n=M+1}^N \frac{\sum_{j=0}^\infty e^{-n\beta\epsilon_{j,\widehat{\boldom}^{N-n}}} \left|\langle\varphi_{j,\widehat{\boldom}^{N-n}},\varphi_{\0}\rangle\right|^2}{\sum_{j=0}^\infty e^{-n\beta\epsilon_{j,\widehat{\boldom}^{N-n}}}}P_{N,L}(\xi_1=n)\nonumber\\
&\leq& P_{\rho,\beta}(\xi_1=\infty);
\eea
that is, BEC implies cycle percolation. We obtained also that finite cycles do not contribute to the condensate.

\noindent
(iii) Suppose that there is macroscopic cycle percolation. In this case
\be
\frac{\rho_0}{\rho}=\lim_{\varepsilon\to 0}\ \lim_{N,L\to\infty,N/L^d=\rho}\sum_{n>\varepsilon N} \frac{\sum_{j=0}^\infty e^{-n\beta\epsilon_{j,\widehat{\boldom}^{N-n}}} \left|\langle\varphi_{j,\widehat{\boldom}^{N-n}},\varphi_{\0}\rangle\right|^2}{\sum_{j=0}^\infty e^{-n\beta\epsilon_{j,\widehat{\boldom}^{N-n}}}}P_{N,L}(\xi_1=n),
\ee
so only particles in macroscopic cycles, if any, can contribute to the condensate. Macroscopic cycles extend to the entire available volume, and the same must be true for each $\varphi_{j,\widehat{\boldom}^{N-n}}$. A quantitative form of this property can be the following: there exist $\delta,\delta'>0$ independent of $N,L$ such that
the domain $\{\x\in\Lambda:|\varphi_{0,\widehat{\boldom}^{N-n}}(\x)|^2>\delta L^{-d}\}$ is connected and of volume
\be\label{extended}
\left|\{\x\in\Lambda: |\varphi_{0,\widehat{\boldom}^{N-n}}(\x)|^2>\delta L^{-d}\}\right|\geq \delta' L^d; \ee
or, in a semiclassical picture,
\be
|\{\x\in\Lambda:V_{\widehat{\boldom}^{N-n}}(\x)<\epsilon_{0,\widehat{\boldom}^{N-n}}\}|\geq \delta'' L^d
\ee
where $\{\x\in\Lambda:V_{\widehat{\boldom}^{N-n}}(\x)<\epsilon_{0,\widehat{\boldom}^{N-n}}\}$ is connected and $\delta''>0$. This can hold for a slowly varying bounded external field as $V_{\widehat{\boldom}^{N-n}}$, and suggests that the $\varphi_{j,\widehat{\boldom}^{N-n}}$ are weak perturbations of plane waves. Now
\be
\lim_{N,L\to\infty,N/L^d=\rho}\sum_{\k\in\frac{2\pi}{L}\Zz^d\setminus\{\0\}}e^{-\varepsilon N\beta\epsilon_\k}=0,
\ee
and in analogy we formulate the following

\vspace{10pt}
\noindent
{\bf Spectral conjecture.} {\em For any $\varepsilon>0$
\be\label{Spec-con}
\lim_{N,L\to\infty,N/L^d=\rho}\sum_{j=1}^\infty e^{-\varepsilon N\beta\left(\epsilon_{j,\widehat{\boldom}^{N(1-\varepsilon)}} -\epsilon_{0,\widehat{\boldom}^{N(1-\varepsilon)}}\right)}=0.
\ee}
The weaker assumption that $\sum_{j=J}^\infty e^{-\varepsilon N\beta\left(\epsilon_{j,\widehat{\boldom}^{N(1-\varepsilon)}} -\epsilon_{0,\widehat{\boldom}^{N(1-\varepsilon)}}\right)}$ has a finite limit for some finite $J$ would be enough: it would lead to a more complicated final formula but the conclusion about BEC would not change. From Eq.~(\ref{extended}) and $\varphi_{0,\widehat{\boldom}^{N-n}}~>0$ it follows that $\langle\varphi_{0,\widehat{\boldom}^{N-n}},\varphi_{\0}\rangle$ has a positive limit, and
\bea\label{MCP-implication}
\frac{\rho_0}{\rho} &=& \lim_{\varepsilon\to 0}\ \lim_{N,L\to\infty, \frac{N}{L^d}=\rho}\sum_{n>\varepsilon N} \left|\left\langle\varphi_{0,\widehat{\boldom}^{N-n}},\varphi_{\0}\right\rangle\right|^2 P_{N,L}(\xi_1=n)\nonumber\\&=& \int_0^{\varepsilon_s(\rho,\beta)}\lim_{N,L\to\infty,\frac{N}{L^d}=\rho} \left|\left\langle\varphi_{0,\widehat{\boldom}^{N(1-\varepsilon)}},\varphi_{\0}\right\rangle\right|^2 p_{\rho,\beta}(\d\varepsilon)>0.
\eea
In the second line $\varepsilon_s(\rho,\beta)$ is given by Eq~(\ref{epsilon-bar}) and $p_{\rho,\beta}$ is a measure of norm
\be
\int_0^{\varepsilon_s}p_{\rho,\beta}(\d\varepsilon)=P_{\rho,\beta}(\xi_1=\infty).
\ee
From the second line of Eq~(\ref{cond-fraction-new}) one can arrive -- less precisely, more intuitively -- at the same conclusion: if $n>\varepsilon N$, attaining from $\0$ any $\x\in\Lambda$ by a random walk is not much less probable than to return to $\0$; so the numerator and the denominator are of the same order of magnitude. $\quad\Box$

\vspace{10pt}
In the case of the ideal gas Eq.~(\ref{MCP-implication}) reproduces Eq.~(\ref{ideal-concise}). For liquid helium and the dilute repulsive Bose gas it is seen that the condensate fraction does not go to 1 when the temperature goes to 0, in contrast to cycle percolation whose probability tends to 1. This latter must have to do with the fact that superfluidity becomes complete in the limit of zero temperature.

\newsec{Reduced density matrix and macroscopic wave function}\label{MWF}

From Eqs.~(\ref{sigma1}) and (\ref{av-Nk}), the one-particle reduced density matrix can be obtained. Let $\tau_\y$ denote the shift by $\y$,
$
(\tau_\y f)(\x)=f(\x-\y).
$
Then
\be\label{<x|sigma|y>}
\langle \x|\sigma_1|\y\rangle =\rho\sum_{n=1}^N P_{N,L}(\xi_1=n)\frac{\sum_{j=0}^\infty e^{-n\beta\epsilon_{j,\widehat{\boldom}^{N-n}}} \left\langle\tau_\x\,\varphi_{j,\widehat{\boldom}^{N-n}},\tau_\y\,\varphi_{j,\widehat{\boldom}^{N-n}} \right\rangle}{\sum_{j=0}^\infty e^{-n\beta\epsilon_{j,\widehat{\boldom}^{N-n}}}}
\ee
where we used
\be
\sum_{\k\in\frac{2\pi}{L}\Zz^d}\left|\langle\varphi_{j,\widehat{\boldom}^{N-n}},\varphi_{\k}\rangle\right|^2 e^{i\k\cdot(\x-\y)}
=\left\langle\tau_\x\,\varphi_{j,\widehat{\boldom}^{N-n}},\tau_\y\,\varphi_{j,\widehat{\boldom}^{N-n}} \right\rangle.
\ee
The numerator in (\ref{<x|sigma|y>}) can be rewritten as
\bea
\sum_{j=0}^\infty e^{-n\beta\epsilon_{j,\widehat{\boldom}^{N-n}}} \left\langle\tau_\x\,\varphi_{j,\widehat{\boldom}^{N-n}},\tau_\y\,\varphi_{j,\widehat{\boldom}^{N-n}} \right\rangle
= \Tr \tau_{\x-\y}\,e^{-n\beta h^\beta_{n,N,L}}
\nonumber\\
= \int_\Lambda\d\z\int W^{n\beta}_{\z+\x,\z+\y}(\d\boldom) e^{-\int_0^{n\beta}V_{\widehat{\boldom}^{N-n}}(\boldom(t))\d t}>0.
\eea
Therefore $\langle \x|\sigma_1|\y\rangle$ can be bounded below by dropping any number of terms of the sum over $n$ in (\ref{<x|sigma|y>}). In the case of macroscopic cycle percolation this gives
\bea
\Sigma(\x,\y)\equiv \lim_{N,L\to\infty, \frac{N}{L^d}=\rho}\langle \x|\sigma_1|\y\rangle
&\geq&  \rho \lim_{\varepsilon\to 0}\lim_{N,L\to\infty, \frac{N}{L^d}=\rho} \sum_{n>\varepsilon N} P_{N,L}(\xi_1=n) \left\langle\tau_\x\,\varphi_{0,\widehat{\boldom}^{N-n}},\tau_\y\,\varphi_{0,\widehat{\boldom}^{N-n}} \right\rangle
\nonumber\\
&=& \rho\int_0^{\varepsilon_s(\rho,\beta) }\lim_{N,L\to\infty,\frac{N}{L^d}=\rho} \left\langle \tau_\x\, \varphi_{0,\widehat{\boldom}^{N(1-\varepsilon)}},\tau_\y\,\varphi_{0,\widehat{\boldom}^{N(1-\varepsilon)}} \right\rangle p_{\rho,\beta}(\d\varepsilon).
\nonumber\\
\phantom{a}
\eea
For any $\varepsilon\in(0,\varepsilon_s)$ we can define a macroscopic wave function. Let
\be
~\nu_{\rho(1-\varepsilon)}=\left\{~\nu_{\rho(1-\varepsilon),i}\right\}_{i=1}^\infty\qquad \left(~\nu_{\rho(1-\varepsilon),i}\in\Omega^\beta\right)
\ee
be the infinite volume limit of $\widehat{\boldom}^{N(1-\varepsilon)}=\left\{\widehat{\boldom}_i\right\}_{i=1}^{N(1-\varepsilon)}$, c.f. the obtention of $\left\{~\nu_{n,i}\right\}_{i=1}^\infty$ from $\left\{\widetilde{\boldom}_i\right\}_{i=1}^{N-n}$. There is an external field
\be
V_{~\nu_{\rho(1-\varepsilon)}}(\x) =\lim_{N,L\to\infty,N/L^d=\rho}V_{\widehat{\boldom}^{N(1-\varepsilon)}}(\x) =\sum_{i=1}^\infty u\left(~\nu_{\rho(1-\varepsilon),i}[\x]-\x\right)
\ee
associated with $~\nu_{\rho(1-\varepsilon)}$. Because of cycle percolation,
\be
h^\beta_{~\nu_{\rho(1-\varepsilon)}}=-\frac{\hbar^2}{2m}\Delta+V_{~\nu_{\rho(1-\varepsilon)}}
\ee
cannot have bound states. Let
\bea
E_{0,~\nu_{\rho(1-\varepsilon)}}&=&\inf{\rm spec}\left\{h^\beta_{~\nu_{\rho(1-\varepsilon)}}\right\} = \lim_{N,L\to\infty,N/L^d=\rho}\epsilon_{0,\widehat{\boldom}^{N(1-\varepsilon)}}\nonumber\\
&=&\lim_{N,L\to\infty,N/L^d=\rho} \frac{1}{\varepsilon N}\left[{\cal U}_{\varepsilon N}(\widetilde{\boldom}_0)-{\cal U}_{\varepsilon N}(\widehat{\boldom}_0) + V_{\varepsilon N}\left(\widetilde{\boldom}_0,\widetilde{\boldom}^{N(1-\varepsilon)}\right)\right];
\eea
the last equality follows from Eqs.~(\ref{trace-hat}) and (\ref{Spec-con}). Let $\Phi_{0,~\nu_{\rho(1-\varepsilon)}}$ be the lowest lying generalized eigenvector,
\be
h^\beta_{~\nu_{\rho(1-\varepsilon)}}\Phi_{0,~\nu_{\rho(1-\varepsilon)}} =E_{0,~\nu_{\rho(1-\varepsilon)}}\Phi_{0,~\nu_{\rho(1-\varepsilon)}},
\ee
with normalization
\be
\left\|\Phi_{0,~\nu_{\rho(1-\varepsilon)}}\right\|^2_2 =\lim_{L\to\infty}\frac{1}{L^d}\int_{[-L/2,L/2]^d}\left|\Phi_{0,~\nu_{\rho(1-\varepsilon)}}(\x)\right|^2\d\x =1.
\ee
$\Phi_{0,~\nu_{\rho(1-\varepsilon)}}$ is real nonnegative, and we expect it to be obtainable also as the pointwise limit
\be\label{pointwise}
\Phi_{0,~\nu_{\rho(1-\varepsilon)}}(\x) =\lim_{N,L\to\infty, \frac{N}{L^d}=\rho}L^{d/2} \varphi_{0,\widehat{\boldom}^{N(1-\varepsilon)}}(\x).
\ee
Define
\be
\left\|\Phi_{0,~\nu_{\rho(1-\varepsilon)}}\right\|_1 =\lim_{L\to\infty}\frac{1}{L^d}\int_{[-L/2,L/2]^d}\left|\Phi_{0,~\nu_{\rho(1-\varepsilon)}}(\x)\right|\d\x. \ee
Equation~(\ref{MCP-implication}) can be rewritten as
\be\label{cond-fraction-macroscopic}
\frac{\rho_0}{\rho}=\int_0^{\varepsilon_s(\rho,\beta)} \left\|\Phi_{0,~\nu_{\rho(1-\varepsilon)}}\right\|_1^2\ p_{\rho,\beta}(\d\varepsilon)
= \left\|\Phi_{0,~\nu_{\widehat{\rho}}}\right\|_1^2 P_{\rho,\beta}(\xi_1=\infty),
\ee
where $\widehat{\rho}$ is some intermediate density, $\rho[1-\varepsilon_s(\rho,\beta)]<\widehat{\rho}<\rho$. Moreover,
\be\label{Sigma(x,y)}
\Sigma(\x,\y)
\geq \rho\int_0^{\varepsilon_s(\rho,\beta) } \left\langle\tau_\x\,\Phi_{0,~\nu_{\rho(1-\varepsilon)}},\tau_\y\,\Phi_{0,~\nu_{\rho(1-\varepsilon)}} \right\rangle p_{\rho,\beta}(\d\varepsilon)
\ee
with
\be
\left\langle\Phi,\Psi\right\rangle=\lim_{L\to\infty}\frac{1}{L^d}\int_{[-L/2,L/2]^d}\Phi^*(\x)\Psi(\x)\d\x.
\ee
If there is no (diagonal) long-range order, with $|\x-\y|$ going to infinity $\tau_\x\,\Phi_{0,~\nu_{\rho(1-\varepsilon)}}$ and $\tau_\y\,\Phi_{0,~\nu_{\rho(1-\varepsilon)}}$ become asymptotically independent with respect to the uniform measure in $\Rr^d$,
\be
\left\langle\tau_\x\,\Phi_{0,~\nu_{\rho(1-\varepsilon)}},\tau_\y\,\Phi_{0,~\nu_{\rho(1-\varepsilon)}} \right\rangle
\asymp \left\langle\tau_\x\,\Phi_{0,~\nu_{\rho(1-\varepsilon)}},1\right\rangle \left\langle 1,\tau_\y\,\Phi_{0,~\nu_{\rho(1-\varepsilon)}} \right\rangle
=\left\|\Phi_{0,~\nu_{\rho(1-\varepsilon)}}\right\|_1^2.
\ee
Thus, the lower bound in (\ref{Sigma(x,y)}) tends to the condensate density. However, $\Sigma(\x,\y)$ also converges to the condensate density [Su7]. So as $|\x-\y|\to\infty$,
\bea
\Sigma(\x,\y)
&=& \rho \int_0^{\varepsilon_s(\rho,\beta) } \left\langle\tau_\x\,\Phi_{0,~\nu_{\rho(1-\varepsilon)}},\tau_\y\,\Phi_{0,~\nu_{\rho(1-\varepsilon)}} \right\rangle p_{\rho,\beta}(\d\varepsilon)+o(1)\nonumber\\
&= & \rho \left\langle\tau_\x\,\Phi_{0,~\nu_{\widehat{\rho}}},\tau_\y\,\Phi_{0,~\nu_{\widehat{\rho}}} \right\rangle P_{\rho,\beta}(\xi_1=\infty)+o(1)\nonumber\\
&\xrightarrow[|\x-\y|\to\infty]{}& \rho \left\|\Phi_{0,~\nu_{\widehat{\rho}}}\right\|_1^2 P_{\rho,\beta}(\xi_1=\infty)= \rho_0.
\eea
To summarize, there is a unique macroscopic wave function by which we can characterize the Bose-condensed fluid. For the fluid at rest, this function is real and nonnegative: it is the ground state (the lowest lying generalized eigenvector) of the operator $-\frac{\hbar^2}{2m}\Delta+V_{~\nu_{\widehat{\rho}}}$. In the ideal gas $\left\|\Phi_{0,~\nu_{\widehat{\rho}}}\right\|_1^2=1$, and we rediscover Eq.~(\ref{ideal-concise}). Because $P_{\rho,\beta}(\xi_1=\infty)$ tends to 1 when approaching zero temperature, in liquid helium $\left\|\Phi_{0,~\nu_{\widehat{\rho}}}\right\|_1^2\approx 0.09$ in the ground state [Pen, Ce, Sn].


\newsec{Phase transition in isotropic and axially anisotropic spin-1/2 Heisenberg models}

Lattice spins of quantum number 1/2 are equivalent to hard-core bosons. This equivalence can be used to apply what we learned about BEC in the continuum for the proof of phase transition in spin models.

Let $L$ be an even integer and
\be
\Lambda=\left[-\frac{L}{2}+1,-\frac{L}{2}+2,\ldots,\frac{L}{2}\right]^d,\quad d\geq 3.
\ee
The spin models will be defined in $\Lambda$ with periodic boundary conditions. We shall write down the Hamiltonians in terms of the Pauli matrices
\be
\sigma^1=\begin{pmatrix}
0 & 1\\1 & 0
\end{pmatrix}, \quad
\sigma^2=-i\begin{pmatrix}
0 & 1\\-1 & 0
\end{pmatrix}, \quad
\sigma^3=\begin{pmatrix}
1 & 0\\0 & -1
\end{pmatrix}.
\ee
When passing to lattice gas representation,
\be
\sigma^+=\begin{pmatrix}
0 & 1\\0 & 0
\end{pmatrix}, \quad
\sigma^-=\begin{pmatrix}
0 & 0\\1 & 0
\end{pmatrix}, \quad
\widehat{n}=\begin{pmatrix}
1 & 0\\0 & 0
\end{pmatrix}
\ee
will be used. Expressed with them,
\be\label{sigmas}
\sigma^1=\sigma^++\sigma^-,\quad \sigma^2=-i(\sigma^+-\sigma^-),\quad \sigma^3=2\sigma^+\sigma^--1
= 2\widehat{n}-1.
\ee
There is a set of these matrices assigned to each lattice site $\x$: $~\sigma_\x=(\sigma^1_\x,\sigma^2_\x,\sigma^3_\x)$, $\sigma^\pm_\x$ and $\widehat{n}_\x$. The commutation relations among $\left\{\sigma^\pm_\x\right\}_{\x\in\Lambda}$ are those of boson creation and annihilation operators, except for $\sigma^+_\x\sigma^-_\x+\sigma^-_\x\sigma^+_\x=1$, which is fermionic. However, for hard-core bosons, characterized by $(\sigma^+_\x)^2=(\sigma^-_\x)^2=0$, this is the right one, not $a_\x a^\dagger_\x-a^\dagger_\x a_\x=1$.

The spin at site $\x$ is $\s_\x=\frac{1}{2}~\sigma_\x$; the total spin is $\S=\sum_\x\s_\x$. The total spin quantum number is $S\in\left\{\frac{L^d}{2},\frac{L^d}{2}-1,\ldots,-\frac{L^d}{2}\right\}$, and for given $S$, $S^3\in \left\{S,S-1,\ldots,-S\right\}$ and $\left(S^1\right)^2+\left(S^2\right)^2$ are conserved quantities in each model. The number of particles in the equivalent boson gas is
\be\label{N-S3}
N=\sum_{\x\in\Lambda}\widehat{n}_\x=S^3+\frac{L^d}{2}.
\ee
The particle-hole transformation in the Fock space is implemented by the unitary operator
\be
U=\prod_{\x\in\Lambda}U_\x,\qquad U_\x=e^{i(\pi/2)\sigma^1_\x}=i\sigma^1_\x,\qquad U^{-1}_\x\sigma^\pm_\x U_\x=\sigma^\mp_\x.
\ee
In terms of the spin model, $U$ performs a rotation through $\pi$ about the first spin axis at each site. We shall make use of the following lemma.

\begin{lemma}\label{f(1/2)}
Let the Hamiltonian $H$ for hard-core bosons be invariant under the particle-hole transformation,
\be\label{UHU}
U^{-1}HU=H.
\ee

\noindent
(i) If $H\Psi=E\Psi$, then $H(U\Psi)=E(U\Psi)$.

\noindent
(ii) For all $\beta$ the free energy density $f(\rho,\beta)$ of the lattice gas in the canonical ensemble is minimum at $\rho=N/L^d=1/2$, and the corresponding chemical potential $\mu(1/2,\beta)=0$ for all $\beta$.
\end{lemma}

\noindent
{\bf Proof.} (i) is trivial. (ii) $H$ and $\sum_{\x\in\Lambda}\widehat{n}_\x$ have a common complete set of eigenvectors, and to each eigenvector in the $N$-particle subspace there corresponds an eigenvector in the $L^d-N$-particle subspace with the same eigenvalue. Therefore $f(1-\rho,\beta)=f(\rho,\beta)$ and, because $f$ is convex in $\rho$, it has a minimum at $\rho=1/2$. Using $\partial f(\rho,\beta)/\partial\rho=\mu(\rho,\beta)$, we conclude that $\mu(1/2,\beta)\equiv 0$. $\quad\Box$

\vspace{10pt}
The convexity of the free energy density $f(\rho,\beta)$ may not be strict, therefore the minimum at $\rho=1/2$ may not be unique; this happens when equilibrium states of different densities coexist. However, 1/2 is the only density minimizing the free energy at all temperatures. This can be understood also through a physical argument. The $S^3=0$ subspace is the largest among the $S^3$-eigensubspaces, so the entropy is the largest here. Furthermore, this is the only one in which the whole spectrum of $H$ is represented, so here can the energy attain the lowest value. Together, the two tells us that in this subspace the free energy is at its smallest value.

In models whose Hamiltonian satisfies Lemma~\ref{f(1/2)} Bose-Einstein condensation is always partial. Let, as earlier, $\rho_0$ denote the condensate density.

\begin{proposition}
In a hard-core lattice gas defined by a Hamiltonian invariant under particle-hole transformation, $\rho_0\leq \rho(1-\rho)$; thus, $\rho_0\leq 1/4$.
\end{proposition}

\noindent
{\bf Proof.} For $\k\in\frac{2\pi}{L}\Lambda$ define
\be
\sigma_\k=\frac{1}{L^{d/2}}\sum_{\x\in\Lambda}e^{i\k\cdot\x}\sigma^-_\x, \quad \sigma^\dagger_\k=\frac{1}{L^{d/2}}\sum_{\x\in\Lambda}e^{-i\k\cdot\x}\sigma^+_\x.
\ee
The occupation number operator for the $\k=\0$ state is
\be\label{N0-lattice}
N_\0=\sigma^\dagger_\0\sigma_\0=\frac{1}{L^d}\sum_{\x\in\Lambda}\sigma^+_\x \sum_{\y\in\Lambda}\sigma^-_\y.
\ee
On the other hand, from Eq.~(\ref{sigmas}) by direct computation
\be
\left(S^1\right)^2+\left(S^2\right)^2=\frac{1}{4}\left[\left(\sum_\x\sigma^1_\x\right)^2 +\left(\sum_\x\sigma^2_\x\right)^2\right]=\sum_{\x\in\Lambda}\sigma^+_\x \sum_{\y\in\Lambda}\sigma^-_\y,
\ee
whose comparison with (\ref{N0-lattice}) gives
\be\label{S-planar}
\left(S^1\right)^2+\left(S^2\right)^2=N_\0 L^d.
\ee
Furthermore, from Eq.~(\ref{N-S3}),
\be
\left(S^3\right)^2=N^2+\frac{L^{2d}}{4}-NL^d,
\ee
so
\be
\S^2=N_0L^d+N^2+\frac{L^{2d}}{4}-NL^d= S(S+1)\leq \frac{L^{2d}}{4}+\frac{L^{d}}{2}.
\ee
Dividing by $L^{2d}$, taking the canonical expectation value for given $N$ and then the thermodynamic limit with $N/L^d$ going to $\rho$, for the limit $\rho_0$ of $\langle N_0\rangle/L^d$ we find
\be
\rho_0+\rho^2-\rho\leq 0, \quad \rho_0\leq \rho(1-\rho). \qquad\Box
\ee

\vspace{10pt}
The family of models that we consider is defined by
\be
H=-\sum_{\langle\x\y\rangle} \left[\sigma^1_\x\sigma^1_\y+\sigma^2_\x\sigma^2_\y+c(\sigma^3_\x\sigma^3_\y-1)\right],
\ee
where $c$ is any real number and the summation is over nearest neighbor pairs in $\Lambda$. If $c<0$, the unitary operator
$
\prod_{\x:\sum_{j=1}^d x_j\ {\rm odd}}\exp\{i(\pi/2)\sigma^3_\x\}
$
transforms $H$ to be antiferromagnetic. The lattice-gas form of the Hamiltonian is
\be\label{Hspin-2nd-quant}
H
=-\sum_\x\sum_{\y:|\y-\x|=1}(\sigma^+_\x\sigma^-_\y+\sigma^+_\y\sigma^-_\x-2\sigma^+_\x\sigma^-_\x) -4c\sum_{\langle\x\y\rangle}\widehat{n}_\x\widehat{n}_\y-4d(1-c)\sum_{\x\in\Lambda}\widehat{n}_\x.
\ee
Invariance under particle-hole transformation can be checked on either form of $H$. For $c>0$ the hard-core on-site repulsion together with the nearest neighbor attraction mimics the interaction between noble gas atoms in the continuum, and $H$ in its form (\ref{Hspin-2nd-quant}) is a model for a noble gas in the grand-canonical ensemble.

Conservation of $S^3$ and $\left(S^1\right)^2+\left(S^2\right)^2$ corresponds to conservation of $N$ and $N_0$. Lemma~\ref{f(1/2)} applies to $H$, so $N=L^d/2$ minimizes the canonical free energy. In the sequel we shall work at half filling or $S^3=0$. In this subspace spin ordering can follow two different routes. Spins may align in direction 3, and then for $c>0$ there is phase separation into a sea of spins pointing upwards with islands of downwards spins and another sea of oppositely oriented spins, so that globally $S^3=0$. In principle, this does not exclude simultaneous planar ordering, that is, a non-vanishing limit of $\left\langle (S^1)^2+(S^2)^2\right\rangle/NL^d$. However, formulated in terms of the lattice gas, a high-density phase is separated from a low-density one and, because of spin symmetry, $\left\langle (S^1)^2+(S^2)^2\right\rangle$ and therefore $\langle N_0\rangle$ is the same in the two phases. So simultaneous axial and planar ordering would be equivalent to having the same amount of Bose-condensate in two phases of different density. For $c<0$ axial spin ordering results in phase separation into two antiferromagnets with opposite sublattice magnetization. The other route is that there is no axial ordering, the spins order in the 1-2 plane; in the language of the lattice gas this means BEC. Now the equilibrium state is a uniform mixture of pure states with different planar spin orientation or, in the lattice gas, with different gauge, and the homogeneous phase condition is satisfied.

The first quantized form of $H$, restricted to a subspace with $N$ fixed, is
\be\label{H-1st-quant}
H_N=-\sum_{i=1}^N\Delta_i+\sum_{1\leq i<j\leq N}u(\x_i-\x_j)-4d(1-c))N,
\ee
where $\Delta_i$ is the second difference operator acting in the variable $\x_i$, and
\be\label{u-discrete}
u(\x)=\left\{\begin{array}{cl}
+\infty, & \x=\0\\
-4c, & |\x|=1\\
0, & |\x|>1.
\end{array}\right.
\ee
The spectrum of $-\Delta$ is nonnegative, its eigenvalues are
\be\label{eps-k-lattice}
\epsilon_\k=2\sum_{j=1}^d(1-\cos k_j).
\ee
The one-particle partition function at inverse temperature $n\beta$ reads
\be
q_n=\sum_{\k\in\frac{2\pi}{L}\Lambda}e^{-n\beta\epsilon_\k} \asymp L^d\left[\frac{1}{\pi}\int_0^\pi e^{-2n\beta(1-\cos k)}\d k\right]^d \quad (L\to\infty).
\ee
We will need
\be\label{h-upper}
q(\rho,n\beta)=\lim_{N,L\to\infty,N/L^d\to\rho}\frac{q_n}{N} =\frac{1}{\rho}\left[\frac{1}{\pi}\int_0^\pi e^{-2n\beta(1-\cos k)}\d k\right]^d\leq \frac{1}{\rho(n\beta)^{d/2}}.
\ee
The upper bound is obtained by expanding $\cos k$ up to fourth order. For the path integral representation of the partition function we use the formula applied by Aizenman and Lieb to the Hubbard model [Ai1], and best adapted also to our problem:
\be
Q_N=\frac{e^{4\beta d(1-c)N}}{N!}\sum_{\pi\in S_N}\prod_{i=1}^N\sum_{\x_i\in\Lambda}\int W^\beta_{\x_i\x_{\pi(i)}}(\d\boldom_i)e^{-\beta{\cal U}(\boldom^N)}=\frac{1}{N}\sum_{n=1}^N e^{4\beta d(1-c)n} q_n Q_{N-n}G_N(n).
\ee
Here all the entries are formally the same as in Eqs.~(\ref{QN-2})-(\ref{Gibbs-omM}), but now $\boldom_i$, the trajectory of particle $i$, is a piecewise constant function of $t\in [0,\beta]$. Governed by the Brownian bridge measure $W^\beta_{\x\y}$ on the discrete torus, the particles hop with equal probability to one of the neighboring sites, independently and at random times, following a Poisson process. $W^\beta_{\x\y}$ is determined by the kinetic energy operator via
\be\label{W-lattice}
\int W^\beta_{\x\y}(\d\boldom)=\langle\y|e^{\beta\Delta}|\x\rangle =\frac{1}{L^d}\sum_{\k\in\frac{2\pi}{L}\Lambda}e^{i\k\cdot(\y-\x)-\beta\epsilon_\k}.
\ee
After summation over $\{\x_i\}$ only cycles are present in the partition function, and the measure of $\{\boldom|\boldom(0)=\boldom(n\beta)=\x\}$ is obviously positive,
\be
\int W^{n\beta}_{\x\x}(\d\boldom)=\langle\0|e^{n\beta\Delta}|\0\rangle =\frac{1}{L^d}\sum_{\k\in\frac{2\pi}{L}\Lambda}e^{-n\beta\epsilon_\k}=\frac{q_n}{L^d}.
\ee
The positivity of $\int W^\beta_{\0\x}(\d\boldom)$, the measure of the set $\{\boldom|\boldom(0)=\0,\boldom(\beta)=\x\}$, for all $\x\in\Lambda$ is less obvious. It is easier to prove first that the Brownian bridge measure on $\Zz^d$,
\be
\int P^{\beta}_{\0\x}(\d\boldom)=\lim_{L\to\infty}\int W^{\beta}_{\0\x}(\d\boldom) =\frac{e^{-2d\beta}}{\pi^d}\prod_{j=1}^d \int_0^\pi e^{2\beta\cos k}\cos kx_j \d k
\ee
is positive, and then to use the identity
\be\label{Brownian-identity}
\int W^{\beta}_{\0\x}(\d\boldom) =\sum_{\z\in\Zz^d}\int P^{\beta}_{\0,\x+L\z}(\d\boldom)
\ee
for the positivity of the measure on the torus. The identity (\ref{Brownian-identity}) itself can be shown with the help of the Poisson summation formula for distributions.

\begin{lemma}
Let $f(k)$ be any bounded positive strictly decreasing function in $[0,\pi]$. Then for any integer $x\geq 1$,
\be
\int_0^\pi f(k)\cos kx \d k>0.
\ee
\end{lemma}

\noindent
{\bf Proof.}
\be
\int_0^\pi f(k)\cos kx \d k = \frac{1}{x}\int_0^{x\pi}f(k/x)\cos k \d k.
\ee
Now
\be
\int_0^{x\pi}f(k/x)\cos k \d k =\sum_{l=0}^{2x-1}\int_{l\pi/2}^{(l+1)\pi/2}f(k/x)\cos k \d k =\int_0^{\pi/2}\sum_{l=0}^{2x-1}f\left(\frac{k+l\pi/2}{x}\right)\cos(k+l\pi/2)\d k.
\ee
Using
\be
\cos(k+l\pi/2)=\left\{\begin{array}{rl}
\cos k,& l=4n\\
-\sin k,& l=4n+1\\
-\cos k, & l=4n+2\\
\sin k, & l=4n+3
\end{array}\right.
\ee
and $\sin k=\cos(\pi/2-k)$,
\bea
\int_0^{x\pi}f(k/x)\cos k \d k =\int_0^{\pi/2} \sum_{n=0}^{n_{\rm max}}\left[f\left(\frac{k+2n\pi}{x}\right)+f\left(\frac{-k+(2n+1)\pi}{x}\right)\right. \nonumber\\ \left.-f\left(\frac{k+(2n+1)\pi}{x}\right)-f\left(\frac{-k+(2n+2)\pi}{x}\right)\right]\cos k \d k
\eea
if $x$ is even and then $n_{\rm max}=x/2-1$; if $x$ is odd, then $n_{\rm max}=(x-1)/2$, and the two negative terms for $n=n_{\rm max}$ are missing. In either case the sum over $n$ is term by term positive. $\quad\Box$

\begin{proposition}
\be
\int P^{\beta}_{\0\x}(\d\boldom)>0\quad\mbox{and}\quad \int W^{\beta}_{\0\x}(\d\boldom)>0
\ee
for any integer vector $\x$.
\end{proposition}

\vspace{10pt}
\noindent
{\bf Proof.} The first inequality is obtained by applying the previous lemma with $f(k)=e^{2\beta\cos k}$, the second follows from Eq.~(\ref{Brownian-identity}). $\quad\Box$

\vspace{10pt}
\noindent
For the free Bose gas BEC is qualitatively the same as in the continuum. Macroscopic cycle percolation, the analogue of Proposition~\ref{suppl-ideal}, is the consequence of the following lemma.

\begin{lemma}\label{KN=o(N)-lattice}
For any sequence $K_N\to\infty$, $K_N/N\to 0$,
\be
\lim_{N,L\to\infty,N/L^d\to\rho}\sum_{n=1}^{K_N}|q_n/N-q(\rho,n\beta)|=0.
\ee
\end{lemma}

\vspace{10pt}
\noindent
{\bf Proof.} $q_n/N-q(\rho,n\beta)$ is the analogue of $\frac{1}{\rho\lambda_{n\beta}^d}\sum_{\z\neq\0}e^{-\pi L^2\z^2/\lambda_{n\beta}^2}$. First, by expanding $\cos k$ up to second order and applying a bound for the error function, one finds
\be\label{h-lower}
q(\rho,n\beta)^{1/d}\geq \frac{1}{2\sqrt{\pi}\rho^{1/d}\sqrt{n\beta}}\left(1-\frac{e^{-\pi^2 n\beta}}{\pi^{3/2}\sqrt{n\beta}}\right)\geq \frac{1}{4\rho^{1/d}\sqrt{n\beta}},
\ee
where the last inequality holds for $\beta\geq 1$. Second, from a comparison of
\be
q_n^{1/d}=1-e^{-4n\beta}+2e^{2n\beta}\sum_{l=1}^{L/2}e^{2n\beta\cos 2\pi l/L}
\ee
and $q(\rho,n\beta)^{1/d}$, a straightforward computation gives
\be
q(\rho,n\beta)\left[1-\left(\frac{1}{N q(\rho,n\beta)}\right)^{1/d}\right]^d\leq\frac{q_n}{N}
\leq q(\rho,n\beta)\left[1+\left(\frac{1}{N q(\rho,n\beta)}\right)^{1/d}\right]^d.
\ee
This, combined with the bounds (\ref{h-upper}) and (\ref{h-lower}), yields
\be
q(\rho,n\beta)\left[1-\left(\frac{4\sqrt{n\beta}}{L}\right)^{1/d}\right]^d\leq\frac{q_n}{N}
\leq q(\rho,n\beta)\left[1+\left(\frac{4\sqrt{n\beta}}{L}\right)^{1/d}\right]^d.
\ee
From here we obtain
\be
|q_n/N-q(\rho,n\beta)|\leq \frac{1}{\rho(n\beta)^{d/2}}\sum_{l=0}^{d-1}{d\choose l}\left(\frac{4\sqrt{n\beta}}{L}\right)^{d-l}
\ee
and
\be
\sum_{n=1}^{K_N}|q_n/N-q(\rho,n\beta)|\leq \frac{1}{\rho\beta^{d/2}}\sum_{l=0}^{d-1}{d\choose l}\left(\frac{4\sqrt{n\beta}}{L}\right)^{d-l}\sum_{n=1}^{K_N}\frac{1}{n^{l/2}}.
\ee
Each term of the sum over $l$ is $o(1)$ if $K_N=o(N)$. $\quad\Box$

Now we return to the problem of the Heisenberg model. If the first route is followed, our task is finished. Suppose that there is no phase separation. With $\rho=1/2$ and $\mu(\rho,\beta)=0$ we can go further up to
\be\label{Prhobeta-lattice}
P_{\rho,\beta}(\xi_1=n)=\lim_{N,L\to\infty,N/L^d=\rho}e^{4\beta d(1-c)n}\frac{q_{n}}{N}\frac{Q_{N-n}}{Q_N}G_N(n) = e^{4\beta d(1-c)n}q(\rho,n\beta)e^{-\beta\Psi_n(\rho,\beta)}.
\ee
A representative set of trajectories can be introduced only after having replaced $u(\0)=\infty$ by a finite $u_0$. This modifies the relation between the density and the chemical potential. The density $\rho=\lim N/L^d$ is chosen so that $\mu(\rho,\beta)=0$. For $u_0$ very large but finite $\rho$ may be somewhat larger than but close to $1/2$. Expressed in terms of the representative trajectories, $\Psi_n(\rho,\beta)$ is given by Eqs.~(\ref{phi-n-beta})-(\ref{phi-psi}). The analysis we made in Section~\ref{Gordian} is relevant with the exception that $d_1=0$ must be chosen. However, we can safely bound the number of particles at a given site by some $m_0\geq 2$. Let
\be
u_0\gg 4d\max\{2m_0c,1-c\}.
\ee
Then, Eq.~(\ref{sum-prob-He}) reads
\be
P_{\rho,\beta}(\xi_1<\infty) =\sum_{n=1}^\infty q(\rho,n\beta)\,e^{n\beta\left[4d(1-c)-\chi_n(\rho,\beta)\right]}.
\ee
At high temperatures the exponent is negative, and it can increase only through the decrease of $\chi_n(\rho,\beta)$ from a large positive value at small $\beta$, to compensate for a while the decrease of $q(\rho,n\beta)$ as $\beta$ increases. Instead of Eq.~(\ref{mu}) or (\ref{mu-bis}), the convergence of the sum imposes
\be
\liminf_{n\to\infty}\chi_n(\rho,\beta)=\chi(\rho,\beta)\geq 4d(1-c).
\ee
Although Theorem~\ref{thm-cycle-percol} applies, because $u(\x)$ takes on only three different values, $u_0$, $-4c$ and 0, we can directly analyze $\chi_n(\rho,\beta)$. Let us redefine $u_{n,k}(\beta)$ as
\be
u_{n,k}(\beta)=\frac{1}{\beta}\int_0^\beta\sum_{l>k}u\left(~\gamma_{n,l}(t)-~\gamma_{n,k}(t)\right)\d t,
\ee
which does not alter the definition of $\phi_n(\beta)$ and $\chi_n(\rho,\beta)$. Now
\bea\label{u-v-lattice}
\beta u_{n,k}(\beta) &=& \int_0^\beta \left(\,u_0|\{l>k:~\gamma_{n,l}(t)=~\gamma_{n,k}(t)\}|
-4c |\{l>k:|~\gamma_{n,l}(t)-~\gamma_{n,k}(t)|=1\}|\,\right)\d t\nonumber\\
\beta v_{n,k}(\rho,\beta) &=& \int_0^\beta \left(\,u_0|\{i:~\nu_{n,i}(t)=~\gamma_{n,k}(t)\}|
-4c |\{i:|~\nu_{n,i}(t)-~\gamma_{n,k}(t)|=1\}|\,\right)\d t,
\eea
and one has
\bea
|\{l>k:~\gamma_{n,l}(t)=~\gamma_{n,k}(t)\}| &+& |\{i:~\nu_{n,i}(t)=~\gamma_{n,k}(t)\}|\leq m_0-1\nonumber\\
|\{l>k:|~\gamma_{n,l}(t)-~\gamma_{n,k}(t)|=1\}|&+& |\{i:|~\nu_{n,i}(t)-~\gamma_{n,k}(t)|=1\}|\leq 2m_0d. \eea
We divide $[0,\beta]$ into intervals where the sum of the two integrands in Eqs.~(\ref{u-v-lattice}) is constant. Let $\Delta^{l_1l_2}_{n,k}$ denote the total time spent by the $k$th particle of $~\gamma_n$ on the energy level $l_1u_0-4l_2c$. Then
\be\label{vnk-psink-lattice}
\beta \left[u_{n,k}(\beta)+v_{n,k}(\rho,\beta)\right]
=\sum_{l_1=1}^{m_0-1}\sum_{l_2=0}^{2m_0d} (l_1u_0-4l_2c)\Delta^{l_1l_2}_{n,k}-4c\sum_{l_2=1}^{2m_0d}l_2\Delta^{0l_2}_{n,k}
\ee
which after summation over $k$ gives
\be\label{psi-constraint-bis}
n\beta\,\chi_n(\rho,\beta)=\sum_{k=0}^{n-1}\left(\sum_{l_1=1}^{m_0-1}\sum_{l_2=0}^{2m_0d} (l_1u_0-4l_2c)\Delta^{l_1l_2}_{n,k} -4c\sum_{l_2=1}^{2m_0d}l_2\Delta^{0l_2}_{n,k}\right).
\ee
Besides,
\be\label{times}
\Delta^{00}_{n,k}+\sum_{l_1=1}^{m_0-1}\sum_{l_2=0}^{2m_0d}\Delta^{l_1l_2}_{n,k} +\sum_{l_2=1}^{2m_0d}\Delta^{0l_2}_{n,k}=\beta, \quad k=0,\ldots,n-1.
\ee
Each $\Delta^{l_1l_2}_{n,k}$ is the sum of local times of order 1 that particle $k$ spends on the energy level $l_1u_0-4l_2c$ at some site $\x$, with specified other particles on $\x$ and on its nearest neighbors. The local times of a given particle are independent except for the weak constraint that their sum is $\beta$. The choice of the representative set must reflect this fact. Therefore, $\Delta^{l_1l_2}_{n,k}$ are of order $\beta$ and only weakly correlated through (\ref{times}). Moreover, the local times of particles at a distance greater than 2 are independent. This is valid to the particles of $~\gamma_n$ if $\beta$ is large enough. Recall now that $P_{\rho,\beta}(\xi_1=n)\leq 1$ implies
\be\label{psi-constraint}
n\beta\left[4d(1-c)-\chi_n(\rho,\beta)\right] \ \leq\ \ln\left(4^d\rho(n\beta)^{d/2}\right),
\ee
and from high to quite low temperatures the inequality is satisfied with a negative left-hand side. Even if the left of (\ref{psi-constraint}) becomes positive for a finite number of $n$ at lower temperatures, it cannot remain positive for arbitrarily large $\beta$, because then it should be $O(\ln\beta)$. This, for a sum of terms of order $\beta$ could be possible only if the terms were strongly correlated. So for $\beta$ large enough it becomes negative for each $n$, and for further increase of $\beta$, $P_{\rho,\beta}(\xi_1<\infty)$ detaches from 1. As Proposition~\ref{suppl-ideal} for Bose fluids, now Lemma~\ref{KN=o(N)-lattice} can serve to conclude that cycle percolation is macroscopic; then, Theorem~\ref{thm-interacting-1} can be used for the proof of BEC. The only change in Theorem~\ref{thm-interacting-1} is that now the problem is finite dimensional, so $h^\beta_{n,N,L}$, c.f. Eq.~(\ref{h-nNL}), has a finite number of eigenstates. Similarly to (\ref{vnk-psink-lattice}), the external field becomes
\be
\beta V_{\boldom^{N-n}}(\x)=\sum_{l_1=1}^{m_0-1}\sum_{l_2=0}^{2m_0d} (l_1u_0-4l_2c)\Gamma^{l_1l_2}_{\boldom^{N-n}}(\x)-4c\sum_{l_2=1}^{2m_0d}l_2\Gamma^{0l_2}_{\boldom^{N-n}}(\x)
\ee
with
\be
\sum_{l_1=0}^{m_0-1}\sum_{l_2=0}^{2m_0d}\Gamma^{l_1l_2}_{\boldom^{N-n}}(\x)=\beta.
\ee
$\Gamma^{l_1l_2}_{\boldom^{N-n}}(\x)$ is the total time during which the external field at $\x$ is $l_1u_0-4l_2c$. In the passage to the hard-core repulsion the $v_{n,k}$ do not necessarily increase. The system can accommodate to a larger $u_0$ in two ways: for $l_1>0$, $\Delta^{l_1l_2}_{n,k}$ decreases and eventually tends to zero, and the density decreases to 1/2. Meanwhile, $u_0\sum_{l_1=1}^{m_0-1}\sum_{l_2=0}^{2m_0d}l_1\,\Delta^{l_1l_2}_{n,k}$ must have a positive limit, otherwise the effect of the on-site exclusion would disappear. In the $u_0\to\infty$ limit the partition function does not vanish: only $\boldom^{N-n}$ compatible with the hard-core condition survive. For $l_1>0$, $\Gamma^{l_1l_2}_{\boldom^{N-n}}(\x)$ tends to zero but $u_0\sum_{l_1=1}^{m_0-1}\sum_{l_2=0}^{2m_0d} l_1\Gamma^{l_1l_2}_{\boldom^{N-n}}(\x)$ has a nonzero limit. The critical $\beta$ tends to a finite value, and the conclusion about macroscopic cycle percolation and BEC remains valid. We obtained the following result.

\begin{theorem}
In $d\geq 3$ dimensional hypercubic lattices the spin-1/2 isotropic and axially anisotropic Heisenberg models, including the ferromagnetic, the antiferromagetic and the XY model, undergo spin ordering at low enough temperatures. With the restriction to the $S^3=0$ subspace, the phase transition can follow two different routes. There can be axial ordering in direction 3, and in this case there is two-phase coexistence. If there is no axial ordering, the spins order in the 1-2 plane.
\end{theorem}

\noindent
For $c\leq 0$ the theorem reproduces a part of the results of Dyson, Lieb and Simon [Dy]. We expect planar ordering for $|c|<1$ and axial ordering for $|c|> 1$. For the isotropic ferromagnet the breakdown of the full rotation symmetry can be obtained both through axial or through planar ordering by choosing axis 3 in arbitrary direction and constraining to $S^3=0$. As noted above, the two are mutually excluding, therefore majority (larger entropy) wins, and there will be planar ordering, i.e., BEC. Equations~(\ref{H-1st-quant}) and (\ref{u-discrete}) show that $c$ is proportional to the particle mass. Thereby a good analogy with Bose liquids is established for the models with $c>0$: $0<c\leq 1$ corresponds to light bosons that Bose-condense, while liquids of heavier bosons ($c>1$) crystallize under cooling. Finally, we recall that in the continuum BEC cannot take place at zero chemical potential if the interaction is nonnegative, c.f. Proposition~\ref{mu=0-no-BEC}. This is because the corresponding density goes to zero as $\beta$ goes to infinity. On the lattice the density corresponding to $\mu=0$ is 1/2 for all $\beta$, this is why BEC is possible also for $c\leq 0$.

\newsec{Summary}

Using the rigorous path integral formalism of Feynman and Kac we proved London's eighty years old conjecture that during the superfluid transition in liquid helium Bose-Einstein condensation (BEC) takes place. The result was obtained in two steps. First, based on an ergodic hypothesis we proved Feynman's conjecture, that at low enough temperatures infinite permutation cycles appear in the system. Second, we showed that BEC implies infinite cycles, and the presence of cycles that contain a positive fraction of the total number of particles implies BEC. For this latter we needed a conjecture concerning the spectrum of a Schr\"odinger operator on the torus. We found that in the limit of zero temperature the infinite cycles contain with probability 1 all the particles, while BEC remains partial. The proof provides a physical interpretation of the permutation cycles: being in the same cycle means being in the same one-particle state, even though this state varies according to a probability distribution. In effect, the lack of saturation of BEC was obtained by showing that particles in the same macroscopic cycle occupy with some probability $p_\varphi$ the same one-particle state $\varphi$ taken from an infinite set different from the plane waves, and the condensate fraction is given by $\sum_\varphi |\langle\varphi,\varphi_\0\rangle|^2 p_\varphi$, where $\varphi_\0$ is the zero-momentum plane wave state. This sum remains smaller than 1, while $\sum_\varphi p_\varphi$, the probability that a marked particle is in an infinite cycle, tends to 1 as the temperature goes to zero. The difference in the limit of the two sums is related to the curious fact that superfluidity becomes 100\% and BEC rests below 10\% in the ground state. It appears that superfluidity depends on the macroscopic occupation of one-particle states definitely different from $\varphi_\0$, for we know since Landau [Lan] that the Bose-condensed ideal gas is not a superfluid. It is possible to select one of these states and define with it a macroscopic wave function. The squared $L^1$-norm of this function multiplied with $\sum_\varphi p_\varphi$ gives the condensate fraction. In a final section we discussed the isotropic and the axially anisotropic spin-1/2 Heisenberg models. We proved a phase transition at low enough temperatures, during which either axial or planar spin ordering takes place.

\vspace{10pt}
\noindent
{\bf Acknowledgements.} I thank Aernout van Enter, Robert Seiringer and Daniel Ueltschi for critical remarks on the manuscript, and L\'aszl\'o Gr\'an\'asy for a  helpful discussion.
\setcounter{section}{0}
\renewcommand{\thesection}{\Alph{section}}
\setcounter{equation}{0}
\renewcommand{\theequation}{\Alph{section}.\arabic{equation}}
\section{Appendix. Some upper bounds}

Here we suppose that $u$ is stable and integrable, and derive upper bounds for $\phi_n$, $\psi_n$ and the free energy density. Jensen's inequality implies that
\bea\label{1st}
\left\langle e^{-\beta{\cal U}_n}\ e^{-\beta\sum_{k=0}^{n-1}V_{n,k}\left(\boldom_0,\boldom^{N-n}\right)}\right\rangle_{\mu^{n\beta}_1\times\mu^\beta_{N-n}}
\geq e^{-\beta\left\langle{\cal U}_n\right\rangle_{\mu^{n\beta}_1}}\ e^{-\beta\sum_{k=0}^{n-1} \left\langle V_{n,k}\left(\boldom_0,\boldom^{N-n}\right)\right\rangle_{\mu^{n\beta}_1\times\mu^\beta_{N-n}}} \nonumber\\
=\exp\left\{-\beta\left\langle{\cal U}_n\right\rangle_{\mu^{n\beta}_1}\right\}\ \exp\left\{-\beta \frac{n(N-n)}{L^d}\int u(\x)\d\x\right\}.
\eea
Here ${\cal U}_n=\sum_{k=1}^{n-1}{\cal U}_{n,k}$. The equality holds because the integral with $\mu^\beta_{N-n}$ includes integration over $(\x_1,\ldots,\x_{N-n})$, which is invariant under global shifts. Using the notation $(\boldom_i+\x)(t)=\boldom_i(t)+\x$, we can write
\bea
\int\mu^\beta_{N-n}(\d\boldom^{N-n})V_{n,k}\left(\boldom_0,\boldom^{N-n}\right) &=& \int\mu^\beta_{N-n}(\d\boldom^{N-n})\int_\Lambda\frac{\d\x}{L^d}\ V_{n,k}\left(\boldom_0,\{\boldom_i+\x\}_{i=1}^{N-n}\right)\nonumber\\
&=& \frac{N-n}{L^d}\int u(\x)\d\x
\eea
independently of $\boldom_0$, so that we have also
\be
\left\langle e^{-\beta\sum_{k=0}^{n-1}V_{n,k}\left(\boldom_0,\boldom^{N-n}\right)}\right\rangle_{\mu^\beta_{N-n}}\geq \exp\left\{-\beta \frac{n(N-n)}{L^d}\int u(\x)\d\x\right\}.
\ee

\begin{lemma} For $n$ fixed,
\be\label{2nd}
\left\langle e^{-\beta{\cal U}_n}\right\rangle_{\mu^{n\beta}_1}\geq \exp\left\{-\beta\left\langle{\cal U}_n\right\rangle_{\mu^{n\beta}_1}\right\}
\asymp\exp\left\{-\frac{n\beta}{2}\int u(\x)\sum_{k=1}^{n-1}\alpha_{n,k}^{d/2}e^{-\pi\alpha_{n,k}\x^2}\right\}\quad (L/\lambda_\beta\gg 1),
\ee
where
\be\label{alpha-nk}
\alpha_{n,k}=\frac{1}{\lambda_{k\beta}^2}+\frac{1}{\lambda_{(n-k)\beta}^2} =\frac{1}{\lambda_\beta^2}\left(\frac{1}{k}+\frac{1}{n-k}\right).
\ee
When $n\propto L^d$, the bound modifies to
\be\label{3rd}
\left\langle e^{-\beta{\cal U}_n}\right\rangle_{\mu^{n\beta}_1}\geq \exp\left\{-n\beta\left[\frac{1}{\lambda_\beta^d}\int u(\x)h_\beta(\x)\d\x+\frac{n-1}{2L^d}\right]\right\}
\ee
with
\be
h_\beta(\x)=\sum_{k=1}^\infty\frac{1}{k^{d/2}}\exp\left\{-\frac{\pi\x^2}{k\lambda_\beta^2}\right\}.
\ee
\end{lemma}

\vspace{5pt}
\noindent
{\bf Proof.}
\bea
\int W^{n\beta}_{\0\0}(\d\boldom){\cal U}_n(\boldom)&=&\sum_{\z\in\Zz^d}\int P^{n\beta}_{\0,L\z}(\d\boldom)\ {\cal U}_n(\boldom)\nonumber\\
&=& \frac{1}{\beta}\int_0^\beta \d t\sum_{0\leq k<l\leq n-1}\int P^{n\beta}_{\0,L\z}(\d\boldom)u_L(\boldom(l\beta+t)-\boldom(k\beta+t)).
\eea
This expression can be made simpler by noting that equal-time increments have the same distribution: Let  $0<t_1<t_2<\beta$, $\boldom(0)=\0$, $\boldom(\beta)=\y$, and consider any $f$ depending only on $\boldom(t_2)-\boldom(t_1)$. Then
\bea
\int P^\beta_{\0\y}(\d\boldom)f(\boldom(t_2)-\boldom(t_1)) &=&\int\d\x_1\d\x_2\ \psi_{t_1}(\x_1)\psi_{t_2-t_1}(\x_2-\x_1)\psi_{\beta-t_2}(\y-\x_2)f(\x_2-\x_1)\nonumber\\
&=&\int\d\x\ \psi_{t_2-t_1}(\x)f(\x)\int\d\x_1\ \psi_{t_1}(\x_1)\psi_{\beta-t_2}(\y-\x-\x_1)\\
&=&\int\d\x\ \psi_{t_2-t_1}(\x)f(\x)\psi_{\beta-(t_2-t_1)}(\y-\x)=\int P^\beta_{\0\y}(\d\boldom)f(\boldom(t_2-t_1)).\nonumber
\eea
Thus, shifting the time interval by $k\beta+t$ and using $\boldom(0)=\0$,
\be
\int P^{n\beta}_{\0,L\z}(\d\boldom) u_L(\boldom(l\beta+t)-\boldom(k\beta+t)) =\int P^{n\beta}_{\0,L\z}(\d\boldom) u_L(\boldom((l-k)\beta)),
\ee
the $t$-dependence has dropped. Furthermore,
\be\label{k/n-k}
\int P^{n\beta}_{\0,L\z}(\d\boldom) u_L(\boldom((n-k)\beta)) =\int P^{n\beta}_{\0,L\z}(\d\boldom) u_L(\boldom(k\beta)).
\ee
To see it, apply Eq.~(\ref{one-point}) with the suitable entries and $u_L(-\x)=u_L(\x)$, $u_L(\x+L\z)=u_L(\x)$. Thus,
\bea
\int P^{n\beta}_{\0,L\z}(\d\boldom){\cal U}_n(\boldom)&=&\int P^{n\beta}_{\0,L\z}(\d\boldom)\sum_{k=1}^{n-1}(n-k) u_L(\boldom(k\beta))=\int P^{n\beta}_{\0,L\z}(\d\boldom)\sum_{k=1}^{n-1}k u_L(\boldom(k\beta))\nonumber\\
&=&\frac{n}{2} \sum_{k=1}^{n-1}\int P^{n\beta}_{\0,L\z}(\d\boldom)u_L(\boldom(k\beta)).
\eea
The integral can be computed by using Eq.~(\ref{one-point}) with $\int P^{n\beta}_{\0,L\z}(\d\boldom)=e^{-\pi (L/\lambda_{n\beta})^2\z^2}/\lambda_{n\beta}^{d}$ and
\be
\psi_{k\beta}(\x)\psi_{(n-k)\beta}(L\z-\x)=\left(\int \d P^{n\beta}_{\0,L\z}(\boldom)\right)\alpha_{n,k}^{d/2}e^{-\pi\alpha_{n,k}(\x-Lk\z/n)^2}.
\ee
Summing over $k$ we obtain
\be
\int P^{n\beta}_{\0,L\z}(\d\boldom){\cal U}_n(\boldom)
=\left(\int P^{n\beta}_{\0,L\z}(\d\boldom)\right)\frac{n}{2}\int u_L(\x)\sum_{k=1}^{n-1}\alpha_{n,k}^{d/2}e^{-\pi\alpha_{n,k}(\x-L\{k\z/n\})^2}\d\x,
\ee
where
\be
\{k\z/n\}_i=\{kz_i/n\},\quad 1\leq i\leq d,
\ee
and $\{kz_i/n\}$ denotes the signed distance of $kz_i/n$ to the integers, so that $|\{kz_i/n\}|\leq 1/2$. We could replace $k\z/n$ with $\{k\z/n\}$ because of the periodicity of $u_L$. Expanding $u_L$ according to Eq.~(\ref{uL}),
\be\label{int-uL-expanded}
\int P^{n\beta}_{\0,L\z}(\d\boldom){\cal U}_n(\boldom) =\left(\int P^{n\beta}_{\0,L\z}(\d\boldom)\right)\frac{n}{2}\int u(\x)g_{L,n,\z}(\x),
\ee
where
\be\label{kernel}
g_{L,n,\z}(\x) =\sum_{k=1}^{n-1}\alpha_{n,k}^{d/2}\sum_{\v\in\Zz^d}\exp\left\{-\pi\alpha_{n,k} \left(\x+L\v-L\{k\z/n\}\right)^2\right\}.
\ee
The periodic heat kernel is positive definite,
\be
\alpha^{d/2}\sum_{\v\in\Zz^d}\exp\left\{-\pi\alpha(\y+L\v)^2\right\} =L^{-d}\sum_{\v\in\Zz^d}\exp\left\{-\frac{\pi\v^2}{\alpha L^2}\right\}\exp\left\{i \frac{2\pi}{L}\y\cdot\v\right\},
\ee
so its maximum is at $\y=\0$, therefore
\bea\label{g-upper}
g_{L,n,\z}(\x)\leq\sum_{k=1}^{n-1}\alpha_{n,k}^{d/2} \sum_{\v\in\Zz^d}\exp\left\{-\pi\alpha_{n,k}L^2\v^2\right\}\leq \sum_{k=1}^{n-1}\alpha_{n,k}^{d/2}\left(1+\frac{1}{L\sqrt{\alpha_{n,k}}}\right)^d\nonumber\\
\asymp \frac{2^{d/2+1}\zeta(d/2)}{\lambda_\beta^d}+\frac{n-1}{L^d}\quad(L/\lambda_\beta\gg 1).
\eea
If $\v$ or $\z$ are nonzero, $\left|L\left(\v-\{k\z/n\}\right)\right|\geq L/2$, and the exponential in (\ref{kernel}) is of order 1 only if $|\x|\sim L$; but then $u$ is vanishingly small. Therefore, for $n$ fixed and $L/\lambda_\beta\gg 1$ it suffices to consider $g_{L,n,\0}$ with the single term $\v=\0$. Substituting this into Eq.~(\ref{int-uL-expanded}), summing over $\z$ and dividing by $\int\d W^{n\beta}_{\0\0}(\boldom)$ provides (\ref{2nd}). On the other hand, if $n$ is of order $L^d$, we must keep the term $(n-1)/L^d$, the sum can be extended to infinity, and this gives rise to the inequality (\ref{3rd}), obviously valid also for $n$ finite. $\quad\Box$

\vspace{10pt}
We summarize what we have found above.
\begin{proposition} Suppose that $u$ is stable [see Eq.~(\ref{super})] and integrable. Then
\be\label{upper2}
-A\leq\phi_n(\beta) \equiv \phi_{n,~\gamma_n}(\beta)
\leq \frac{1}{2}\int u(\x)\sum_{k=1}^{n-1}\alpha_{n,k}^{d/2}e^{-\pi\alpha_{n,k}\x^2}\d\x =\left\langle\phi_{n,\cdot}(\beta)\right\rangle_{P^{n\beta}_{\0\0}}
\ee
and
\be\label{psi}
\psi_n(\rho,\beta)=\frac{1}{n}\sum_{k=0}^{n-1}v_{k,n}(\rho,\beta)\leq\rho\int u(\x)\d\x.
\ee
\end{proposition}

As a by-product, the inequalities just derived provide an upper bound for the free energy density $f(\rho,\beta)$.

\begin{proposition}\label{f-upperbound}
Suppose that $u$ is stable and integrable. Then
\be\label{f-bound}
f(\rho,\beta)\leq \frac{\rho^2}{2}\int u(\x)\d\x+
\frac{\rho}{\lambda_\beta^d}\int u(\x)h_\beta(\x)\d\x+f^0(\rho,\beta).
\ee
\end{proposition}

\noindent
{\bf Proof.} Applying the bounds (\ref{1st}) and (\ref{3rd}) in  Eq.~(\ref{QNint}) we obtain
\be\label{lower-for-QN}
Q_N \geq\frac{1}{N}\sum_{n=1}^N q_nQ_{N-n}\exp\left\{-C\left[n(N-n)+\frac{1}{2}n(n-1)\right]\right\} \exp\left\{-Dn\right\}
\ee
where
\be
C=\beta L^{-d}\int u(\x)\d\x,\quad D=\beta\lambda_\beta^{-d}\int u(\x)h_\beta(\x)\d\x.
\ee
We define an auxiliary function $Q_N^-$ recursively by $Q_0^-=1$ and
\be
Q_N^-=\frac{1}{N}\sum_{n=1}^N q_nQ_{N-n}^-\exp\left\{-C\left[n(N-n)+\frac{1}{2}n(n-1)\right]\right\} \exp\left\{-Dn\right\} \equiv\frac{1}{N}\sum_{n=1}^N q_nQ_{N-n}^-e^{-\beta\Psi^+_{n,N-n}}.
\ee
$Q_N^-$ has two properties relevant for our purpose.

\vspace{3pt}
(i) $Q_N^-\leq Q_N$. To see it we rewrite $Q_N$ as
\be\label{QN-bis}
Q_N=\frac{1}{N}\sum_{n=1}^N q_nQ_{N-n}e^{-\beta\Psi_{n,N-n}},
\ee
c.f. Eqs.~(\ref{QNint}), (\ref{QN-2}). We have just proved that $\Psi_{n,N-n}\leq\Psi^+_{n,N-n}$, therefore
\bea
Q_N-Q_N^-&=&\frac{1}{N}\sum_{n=1}^N q_n\left[Q_{N-n}e^{-\beta\Psi_{n,N-n}}-Q_{N-n}^-e^{-\beta\Psi^+_{n,N-n}}\right]\nonumber\\
&=&\frac{1}{N}\sum_{n=1}^N q_n e^{-\beta\Psi_{n,N-n}}\left[Q_{N-n}-Q_{N-n}^-e^{-\beta(\Psi^+_{n,N-n}-\Psi_{n,N-n})}\right]\nonumber\\ &\geq&\frac{1}{N}\sum_{n=1}^N q_n e^{-\beta\Psi_{n,N-n}}\left[Q_{N-n}-Q_{N-n}^-\right].
\eea
For $N=1$ this reads
\be
Q_1-Q_1^-\geq q_1 e^{-\beta\Psi_{1,N-1}}[Q_0-Q_0^-]=0,
\ee
and the result follows by induction.

\vspace{3pt}
(ii)
\be\label{QN-minus}
Q_N^-=\exp\left\{-\frac{1}{2}CN(N-1)-DN\right\}Q_N^0.
\ee
From the identities
\be
n(N-n)+\frac{1}{2}n(n-1)=\frac{1}{2}N(N-1)-\frac{1}{2}(N-n)(N-n-1),\quad n=N-(N-n)
\ee
one can see that $A_N=e^{\frac{1}{2}CN(N-1)+DN}Q_N^-$ satisfies the recurrence relation
\be
A_N=\frac{1}{N}\sum_{n=1}^N q_n A_{N-n}
\ee
with the initial condition $A_0=1$. Comparison with Eq.~(\ref{recur-free}) yields $A_N=Q_N^0$. According to Eq.~(\ref{QN-minus}), $Q_N^-$ is the partition function of a mean-field model whose Hamiltonian depends on the temperature. The proof is completed by combining this equation with $Q_N^-\leq Q_N$. $\Box$

\vspace{20pt}
\noindent{\Large\bf References}
\begin{enumerate}
\item[{[Ad]}] Adams S., Collevecchio A., and K\"onig W.: {\em A variational formula for the free energy of an interacting many-particle system.} Ann. Prob. {\bf 39}, 683-728 (2011).
\item[{[Ai1]}] Aizenman M. and Lieb E. H.: {\em Magnetic properties of some itinerant-electron systems at $T>0$.} Phys. Rev. Lett. {\bf 65}, 1470-1473 (1990).
\item[{[Ai2]}] Aizenman M. and Nachtergaele B.: {\em Geometric aspects of quantum spin states.} Commun. Math. Phys. {\bf 164}, 17-63 (1994).
\item[{[Ald]}] Alder B. J.and Wainwright T. E.: {\em Phase transition in elastic discs.} Phys. Rev. {\bf 127}, 359-361 (1962).
\item[{[All]}] Allen J. F. and Misener D.: {\em Flow of liquid helium II.} Nature {\bf 141},  75 (1938), and {\bf 142}, 643-644 (1938).
\item[{[Az]}] Aziz R. A., Nain V. P. S., Carley J. S., Taylor W. L. and McConville G. T.: {\em An accurate intermolecular potential for helium.} J. Chem Phys. {\bf 70}, 4330-4342 (1979).

\item[{[Ben]}] Benfatto G., Cassandro M., Merola I. and Presutti E.: {\em Limit theorems for statistics of combinatorial partitions with applications to mean field Bose gas.} J. Math. Phys. {\bf 46}, 033303 (2005).
\item[{[Ber1]}] van den Berg M., Lewis J. T and de Smedt Ph: {\em Condensation in the imperfect boson gas.} J. Stat. Phys. {\bf 37}, 697-707 (1984).
\item[{[Ber2]}] van den Berg M., Dorlas T. C., Lewis J. T. and Pul\'e J. V.: {\em A perturbed meanfield model of an interacting boson gas and the large deviation principle.} Commun. Math. Phys. {\bf 127}, 41-69 (1990).
\item[{[Bet1]}] Betz V. and Ueltschi D.: {\em Spatial random permutations and infinite cycles.} Commun. Math. Phys. {\bf 285}, 469-501 (2009).
\item[{[Bet2]}] Betz V. and Ueltschi D.: {\em Spatial random permutations and Poisson-Dirichlet law of cycle lengths.} Electr. J. Probab. {\bf 16}, 1173�1192 (2011).
\item[{[Bet3]}] Betz V., Ueltschi D. and Velenik Y.: {\em Random permutations with cycle weights.} Ann. Appl. Probab. {\bf 21}, 312�331 (2011).
\item[{[Boe]}] de Boer J. and Michels A.: {\em Contribution to the quantum-mechanical theory of the equation of state and the law of corresponding states. Determination of the law of force of helium.} Physica {\bf 5}, 945-957 (1938).
\item[{[Bog1]}] Bogoliubov N. N.: {\em On the theory of superfluidity.} J. Phys. USSR {\bf 11}, 23-32 (1947).
\item[{[Bog2]}] Bogoliubov N. N.: {\em Quasi-averages in problems of statistical mechanics.} Dubna Report No. D-781, (1961), Ch. II (in Russian); Phys. Abhandl. Sowijetunion, 1962, {\bf 6}, 1-110; ibid., 1962, {\bf 6}, 113-229 (in German); {\em Selected Works, vol.II: Quantum Statistical Mechanics.} Gordon and Breach (N.Y. 1991); Collection of Scientific Papers in 12 vols.: Statistical Mechanics, vol.6, Part II. Nauka (Moscow 2006).
\item[{[Bog3]}] Bogolyubov N. N. and Khatset B. I.: {\em On some mathematical problems of the theory of statistical equilibrium.} Dokl. Akad. Nauk SSSR {\bf 66}, 321 (1949).
\item[{[Bog4]}] Bogolyubov N. N., Petrina D. Ya. and Khatset B. I.: {\em Mathematical description of the equilibrium state of classical systems on the basis of the canonical ensemble formalism.} Teor. Mat. Fiz. {\bf 1}, 251-274 (1969) and Ukr. J. Phys. {\bf 53}, 168-184 (2008).
\item[{[Bos]}] Bose S. N.: {\em Plancks Gesetz und Lichtquantenhypothese.} Z. Phys. \textbf{26}, 178-181 (1924).
\item[{[Bou]}] Bouziane M. and Martin Ph. A.: {\em Bogoliubov inequality for unbounded operators and the Bose gas.} J. Math. Phys. {\bf 17}, 1848-1851 (1976).
\item[{[Bow]}] Bowen L., Lyons R., Radin Ch. and Winkler P.: {\em Fluid-solid transition in a hard-core system.} Phys. Rev. Lett. {\bf 96}, 025701 (2006).
\item[{[Buc]}] Buckingham R. A.: {\em The classical equation of state of gaseous helium, neon and argon.} Proc. R. Soc. {\bf A168}, 264-283 (1938).
\item[{[Buf]}] Buffet E. and Pul\'e J. V.: {\em Fluctuation properties of the imperfect Bose gas.} J. Math. Phys. {\bf 24}, 1608-1616 (1983).
\item[{[Bun]}] Bund S. and Schakel M. J.: {\em String picture of Bose-Einstein condensation.} Mod. Phys. Lett. B {\bf 13}, 349 (1999).

\item[{[Ce]}] Ceperley D. M.: {\em Path integrals in the theory of condensed helium.} Rev. Mod. Phys. {\bf 67}, 279-355 (1995).
\item[{[Cr]}] Creamer D. B., Thacker H. B. and Wilkinson D.: {\em A study of correlation functions for the delta-function Bose gas.} Physica {\bf 20D}, 155-186 (1986).

\item[{[Dav]}] Davies E. B.: {\em The thermodynamic limit for an imperfect boson gas.} Commun. Math. Phys. {\bf 28}, 69-86 (1972).
\item[{[Do1]}] Dorlas T. C., Lewis J. T. and Pul\'e J. V.:{\em Condensation in some perturbed meanfield models of a Bose gas.} Helv. Phys. Acta {\bf 64}, 1200-1224 (1991).
\item[{[Do2]}] Dorlas T. C., Lewis J. T. and Pul\'e J. V.:{\em The full diagonal model of a Bose gas.} Commun. Math. Phys. {\bf 156}, 37-65 (1993).
\item[{[Do3]}] Dorlas T. C., Martin Ph. A. and Pul\'e J. V.: {\em Long cycles in a perturbed mean field model of a boson gas.} J. Stat. Phys. {\bf 121}, 433-461 (2005).
\item[{[Dy]}] Dyson F. J., Lieb E. H. and Simon B.: {\em Phase transitions in quantum spin systems with isotropic and nonisotropic interactions.} J. Stat. Phys. {\bf 18}, 335-383 (1978).

\item[{[E1]}] Einstein A.: {\em Quantentheorie des einatomigen idealen Gases.} Sitz.ber. Preuss. Akad. Wiss. {\bf 1924}, 261-267.
\item[{[E2]}] Einstein A.: {\em Quantentheorie des einatomigen idealen Gases. II.} Sitz.ber. Preuss. Akad. Wiss. {\bf 1925}, 3-14.
\item[{[E3]}] Einstein A.: {\em Quantentheorie des idealen Gases.} Sitz.ber. Preuss. Akad. Wiss. {\bf 1925}, 18-25.
\item[{[E4]}] Einstein A.: {\em Answer to Schr\"odinger on 28 February 1925.} The collected papers of Albert Einstein, Vol. 14, Document 446, p. 438.

\item[{[Fa1]}] Fannes M. and Verbeure A.: {\em The condensed phase of the imperfect Bose gas.} J. Math. Phys. {\bf 21}, 1809-1818 (1980).
\item[{[Fa2]}] Fannes M., Pul\'e J. V. and Verbeure A.: {\em On Bose condensation.} Helv. Phys. Acta {\bf 55}, 391-399 (1982).
\item[{[Fe1]}] Feynman R. P.: {\em Space-time approach to non-relativistic quantum mechanics.} Rev. Mod. Phys. {\bf 20}, 367-387 (1948).
\item[{[Fe2]}] Feynman R. P.: {\em Atomic theory of the $\lambda$ transition in helium.} Phys. Rev. {\bf 91}, 1291-1301 (1953).
\item[{[Fe3]}] Feynman R. P.: {\em Atomic theory of liquid helium near absolute zero.} Phys. Rev. {\bf 91}, 1301-1308 (1953).
\item[{[Fe4]}] Feynman R. P.: {\em Atomic theory of the two-fluid model of liquid helium.} Phys. Rev. {\bf 94}, 262-277 (1954).
\item[{[Fi]}] Fisher M. E. and Essam J. W.: {\em Some cluster size and percolation problems.} J. Math. Phys. {\bf 2}, 609-619 (1961).
\item[{[Fr]}] Fr\"ohlich J., Simon B. and Spencer T.: {\em Infrared bounds, phase transitions and continuous symmetry breaking.} Commun. Math. Phys. {\bf 50}, 79-85 (1976).
\item[{[G1]}] Ginibre J.: {\em Some applications of functional integration in Statistical Mechanics.} In: {\em Statistical Mechanics and Quantum Field Theory}, eds. C. De Witt and R. Stora, Gordon and Breach (New York 1971).
\item[{[G2]}] Ginibre J.: {\em Reduced density matrices of quantum gases. I. Limit of infinite volume.} J. Math. Phys. {\bf 6}, 238-251 (1965).
\item[{[G3]}] Ginibre J.: {\em Reduced density matrices of quantum gases. II. Cluster property.} J. Math. Phys. {\bf 6}, 252-262 (1965).
\item[{[G4]}] Ginibre J.: {\em Reduced density matrices of quantum gases. III. Hard-core potentials.} J. Math. Phys. {\bf 6}, 1432-1446 (1965).
\item[{[G5]}] Ginibre J.: {\em On the asymptotic exactness of the Bogolyubov approximation for many boson systems.} Commun. Math. Phys. {\bf 8}, 26-51 (1968).
\item[{[Gi]}] Girardeau M.: {\em Relationship between systems of impenetrable bosons and fermions in one dimension.} J. Math. Phys. {\bf 1}, 516-523 (1960).
\item[{[Gl]}] Glyde H. R.: {\em Solid Helium. In: Rare gas solids, vol. I}, eds. M. L. Klein and J. A. Venables, Academic Press (London-New York-San Francisco, 1976), fig. 1.
\item[{[Gr]}] Griffin A.: {\em Excitations in a Bose-condensed liquid} (Cambridge University Press, 1993).

\item[{[Hal]}] Haldane F. D. M.: {\em Effective harmonic-fluid approach to low-energy properties of one-dimensional quantum fluids.} Phys. Rev. Lett. {\bf 47}, 1840-1843 (1981).
\item[{[Ho]}] Hohenberg P. C.: {\em Existence of long-range order in one and two dimensions.} Phys. Rev. {\bf 158}, 383-386 (1967).
\item[{[Hu1]}] Huang K.: {\em Imperfect Bose gas.} In: {\em Studies in Statistical Mechanics, Vol II.} eds. J. de Boer and G. E. Uhlenbeck, North-Holland (Amsterdam 1964), pp. 1-106.
\item[{[Hu2]}] Huang K.: {\em Statistical Mechanics.} 2nd ed. Wiley (New York 1987), p. 303, Prob. 12.7.
\item[{[Hu3]}] Huang K. and Yang C. N.: {\em Quantum-mechanical many-body problem with hard-sphere interaction.} Phys. Rev. {\bf 105}, 767-775 (1957).
\item[{[Hu4]}] Huang K., Yang C. N. and Luttinger J. M.: {\em Imperfect Bose with hard-sphere interactions.} Phys. Rev. {\bf 105}, 776-784 (1957).

\item[{[J]}] Johnston D. C.: {\em Thermodynamic properties of the van der Waals fluid.} arXiv:1402.1205 (2014) Fig. 7.

\item[{[Kac1]}] Kac M.: {\em On distributions of certain Wiener functionals.} Trans. Amer. Math. Soc. {\bf 65}, 1-13 (1949).
\item[{[Kac2]}] Kac M.: {\em On some connections between probability theory and differential and integral equations.} In: Proceedings of the Second Berkeley Symposium on Probability and Statistics, J. Neyman ed., Berkeley, University of California Press (1951).
\item[{[Kac3]}] Kac M. and Luttinger J. M.: {\em Bose-Einstein condensation in the presence of imputities.} J. Math. Phys. {\bf 14}, 1626-1628 (1973).
\item[{[Kad]}] Kadanoff L. P.: {\em Slippery wave functions.} J. Stat. Phys. {\bf 152}, 805-823 (2013).
\item[{[Kag]}] Kagan M. Yu.. {\em Modern trends in superconductivity and superfluidity} (Springer-Verlag, 2013).
\item[{[Kah]}] Kahn B. and Uhlenbeck G. E.: {\em On the theory of condensation.} Physica {\bf 5}, 399-416 (1938).
\item[{[Kap]}] Kapitza P.: {\em Viscosity of liquid helium below the $\lambda$-point.} Nature {\bf 141}, 74 (1938).
\item[{[Ken]}] Kennedy T., Lieb E. H. and Shastry B. S.: {\em The XY model has long-range order for all spins and all dimensions greater than one.} Phys. Rev. Lett. {\bf 61}, 2582-2584 (1988).
\item[{[Kim]}] Kim E. and Chan M. H. W.: {\em Observation of superflow in solid helium.} Science {\bf 305}, 1941-1944 (2004).
\item[{[Ko]}] Korepin V. E., Bogoliubov N. M. and Izergin A. G.: {\em Quantum inverse scattering method and correlation functions.} Cambridge University Press (1993) Ch. XVIII.2.
\item[{[Kub]}] Kubo K. and Kishi T.: {\em Existence of long-range order in the XXZ model.} Phys. Rev. Lett. {\bf 61}, 2585-2587 (1988).
\item[{[Kun]}] Kunz H. and Souillard B.: {\em Essential singularity in percolation problems and asymptotic behavior of cluster size distribution.} J. Stat. Phys. {\bf 19}, 77-106 (1978).

\item[{[Lan]}] Landau L.: {\em The theory of superfluidity of helium II.} J. Phys. USSR {\bf 5}, 71-90 (1941).
\item[{[Leb]}] Lebowitz J. L., Mazel A. and Presutti E.: {\em Liquid-vapor phase transitions for systems with finite-range interactions.} J. Stat. Phys. {\bf 94}, 955-1025 (1999).
\item[{[Leg]}] Leggett A. J.: {\em Quantum liquids} (Oxford University Press, 2006).
\item[{[Len]}] Lenard A.: {\em Momentum distribution in the ground state of the one-dimensional system of impenetrable bosons.} J. Math. Phys. {\bf 5}, 930-943 (1964).
\item[{[Lenn]}] Lennard-Jones J. E.: {\em On the determination of molecular fields.} Proc. Roy. Soc. Lond. A{\bf 106}, 463-477 (1924).
\item[{[Lew1]}] Lewis J. T., Zagrebnov V. A. and Pul\'e J. V.: {\em The large deviation principle for the Kac distribution.} Helv. Phys. Acta {\bf 61} 1063-1078 (1988).
\item[{[Li1]}] Lieb E. H. and Liniger W.: {\em Exact analysis of an interacting Bose Gas. I. The general solution and the ground state.} Phys. Rev. {\bf 130}, 1605-1616 (1963).
\item[{[Li2]}] Lieb E. H.: {\em Exact analysis of an interacting Bose gas. II. The excitation spectrum.} Phys. Rev. {\bf 130}, 1616-1624 (1963).
\item[{[Li3]}] Lieb E. H. and Yngvason J.: {\em Ground state energy of the low density Bose gas.} Phys. Rev. Lett. {\bf 80}, 2504-2507 (1998).
\item[{[Li4]}] Lieb E. H. and Yngvason J.: {\em The ground state energy of a dilute two-dimensional Bose gas.} J. Stat. Phys. {\bf 103}, 509-526 (2001).
\item[{[Li5]}] Lieb E. H., Seiringer R., Solovey J. P. and Yngvason J.: {\em The mathematics of the Bose gas and its condensation.}  Birkh\"auser Verlag (Basel-Boston-Berlin 2005), and arXiv:cond-mat/0610117.
\item[{[Li6]}] Lieb E. H., Seiringer R. and Yngvason J.: {\em Justification of {\rm c}-number substitution in bosonic Hamiltonians.} Phys. Rev. Lett. {\bf 94}, 080401 (2005).
\item[{[Lon1]}] London F.: {\em The $\lambda$-phenomenon of liquid helium and the Bose-Einstein degeneracy.} Nature {\bf 141}, 643-644 (1938).
\item[{[Lon2]}] London F.: {\em On the Bose-Einstein condensation.} Phys. Rev. {\bf 54}, 947-954 (1938).
\item[{[Lon3]}] London F.: {\em Zur Theorie und Systematik der Molekularkr\"afte.} Z. Phys. {\bf 63}, 245-279 (1930).
\item[{[Lon4]}] London F.: {\em Superfluids, Vol. 2: Macroscopic theory of superfluid helium.} (Wiley, New York, 1954).

\item[{[M]}] Mie G.: {\em Zur kinetischen Theorie der einatomigen K\"orper.} Annalen der Physik \textbf{316}, 657-697 (1903).

\item[{[Ni]}] Niebel K. F. and Venables J. A.: {\em The crystal structure problem. In: Rare gas solids, vol. I}, eds. M. L. Klein and J. A. Venables, Academic Press (London-New York-San Francisco, 1976).
\item[{[No]}] Nozi\`eres P. and Pines D.: {\em Theory of quantum liquids II: Superfluid Bose liquids.} Addison-Wesley (Redwood City, 1990).

\item[{[Pen]}] Penrose O. and Onsager L.: {\em Bose-Einstein condensation and liquid He.} Phys. Rev. {\bf 104}, 576-584 (1956).
\item[{[Pet]}] Pethick C. J. and Smith H.: {\em Bose-Einstein condensation in dilute gases.} Cambridge University Press (2002).
\item[{[Pi1]}] Pitaevskii L. and Stringari S.: {\em Bose-Einstein condensation.} Clarendon Press (Oxford 2003).
\item[{[Pi2]}] Pitaevskii L. and Stringari S.: {\em Uncertainty principle, quantum fluctuations, and broken symmetries.} J. Low Temp. Phys. {\bf 85}, 377-388 (1991).

\item[{[R1]}] Ruelle D.: {\em Classical statistical mechanics of a system of particles.} Helv. Phys. Acta \textbf{36}, 183-197 (1963).
\item[{[R2]}] Ruelle D.: \emph{Statistical Mechanics.} W. A. Benjamin (New York-Amsterdam 1969).
\item[{[R3]}] Ruelle D.: {\em Superstable interactions in classical statistical mechanics.} Commun. Math. Phys. {\bf 18}, 127-159 (1970).
\item[{[R4]}] Ruelle D.: {\em Correlation functions of classical gases.} Ann. Phys. {\bf 25}, 109-120 (1963).
\item[{[R5]}] Ruelle D.: {\em Cluster property of the correlation functions of classical gases.} Rev. Mod. Phys. {\bf 36}, 580-584 (1964).

\item[{[Scha]}] Schakel A. M. J.: {\em Percolation, Bose-Einstein condensation, and string proliferation.} Phys. Rev. E {\bf 63}, 026115 (2001).
\item[{[Schr]}] Schr\"odinger E.: {\em Letter to Einstein on 5 February 1925.} The collected papers of Albert Einstein, Vol. 14, Document 433, p. 429.
\item[{[Schu]}] Schultz T. D.: {\em Note on the one-dimensional gas of impenetrable point-particle bosons.} J. Math. Phys. {\bf 4}, 666-671 (1963).
\item[{[Se1]}] Seiringer R.: {\em Free energy of a dilute Bose gas: Lower bound.} Commun. Math. Phys. {\bf 279}, 595-636 (2008).
\item[{[Se2]}] Seiringer R. and Ueltschi D.: {\em Rigorous upper bound on the critical temperature of dilute Bose gases.} Phys. Rev. B {\bf 80}, 014502 (2009).
\item[{[Sew]}] Sewell G. L.: {\em Quantum mechanics and its emergent macrophysics.} Princeton University Press (2002).
\item[{[Sl]}] Slater J. C. and Kirkwood J. G.: {\em The van der Waals forces in gases.} Phys. Rev. {\bf 37}, 682-697 (1931).
\item[{[Sn]}] Snow W. M. and Sokol P. E.: {\em Density and temperature dependence of the momentum distribution in liquid Helium 4.} J. Low. Temp. Phys. {\bf 101}, 881-928 (1995).
\item[{[So]}] Sosnick T. R., Snow W. M. and Sokol P. E.: {\em Deep-inelastic neutron scattering from liquid He4.} Phys. Rev. B{\bf 41}, 11185-11202 (1990).
\item[{[Su1]}] S\"ut\H o A.: {\em Percolation transition in the Bose gas.} J. Phys. A: Math. Gen. {\bf 26}, 4689-4710 (1993).
\item[{[Su2]}] S\"ut\H o A.: {\em Percolation transition in the Bose gas: II.} J. Phys. A: Math. Gen. {\bf 35}, 6995-7002 (2002).
\item[{[Su3]}] S\"ut\H o A.: {\em Thermodynamic limit and proof of condensation for trapped bosons.} J. Stat. Phys. {\bf 112}, 375-396 (2003).
\item[{[Su4]}] S\"ut\H o A.: {\em Correlation inequalities for noninteracting Bose gases.} J. Phys. A: Math. Gen. {\bf 37}, 615�621 (2004).
\item[{[Su5]}] S\"ut\H o A.: {\em Ground state at high density.} Commun. Math. Phys. {\bf 305}, 657-710 (2011).
\item[{[Su6]}] S\"ut\H o A.: {\em The total momentum of quantum fluids.} J. Math. Phys. \textbf{56}, 081901 (2015), Section IV.
\item[{[Su7]}] S\"ut\H o A.: {\em A possible mechanism of concurring diagonal and off-diagonal long-range order for soft interactions.} J. Math. Phys. \textbf{50}, 032107 (2009).
\item[{[Su8]}] S\"ut\H o A.: {\em Equivalence of Bose-Einstein condensation and symmetry breaking.} Phys. Rev. Lett. {\bf 94}, 080402 (2005).
\item[{[Su9]}] S\"ut\H o A.: {\em Bose-Einstein condensation and symmetry breaking.} Phys. Rev. A {\bf 71}, 023602 (2005).
\item[{[Su10]}] S\"ut\H o A. and Sz\'epfalusy P.: {\em Variational wave functions for homogenous Bose systems.} Phys. Rev. {\bf A77}, 023606 (2008).

\item[{[T1]}] Tisza L.: {\em Transport phenomena in helium II.} Nature {\bf 141}, 913 (1938).
\item[{[T2]}] Tisza L.: {\em La viscosit\'e de h\'elium liquide et la statistique de Bose-Einstein.} C. R. Paris {\bf 207}, 1035-1186 (1938).
\item[{[Toe]}] Toennies J. P. and Vilesov A. F.: {\em Superfluid helium droplets: A uniquely cold nanomatrix for molecules and molecular complexes.} Angew. Chem. Int. Ed. {\bf 43}, 2622-2648 (2004) Table 2.
\item[{[Ton]}] Tonks L.: {\em The complete equation of state of one, two and three-dimensional gases of hard elastic spheres.} Phys. Rev. {\bf 50}, 955-963 (1936).
\item[{[Tot]}] T\'oth B.: {\em Improved lower bound on the thermodynamic pressure of the spin 1/2 Heisenberg ferromagnet.} Lett. Math. Phys. {\bf 28}, 75-84 (1993).

\item[{[Ued]}] Ueda M.: {\em Fundamentals and new frontiers of Bose-Einstein condensation.} World Scientific (2010).
\item[{[Uel1]}] Ueltschi D.: {\em Relation between Feynman cycles and off-diagonal long-range order.} Phys. Rev.Lett. {\bf 97}, 170601 (2006).
\item[{[Uel2]}] Ueltschi D.: {\em Feynman cycles in the Bose gas.} J. Math. Phys. {\bf 47}, 123303 (2006).
\item[{[Uh]}] Uhlenbeck G. E.: {\em Dissertation Leiden} 1927, p. 69.

\item[{[Van]}] Van der Linden J.: {\em On the asymptotic problem of statistical thermodynamics for a real system. III. Questions concerning chemical potential and pressure.} Physica {\bf 38}, 173-188 (1968).
\item[{[Ver]}] Verbeure A.: {\em Many body boson systems: half a century later.} Springer (2011).

\item[{[W]}] Wagner H.: {\em Long-wavelength excitations and the Goldstone theorem in many-particle systems with "broken symmetries".} Z. Phys. {\bf 195}, 273-299 (1966).

\item[{[Ya]}] Yang C. N.: {\em Concept of off-diagonal long-range order and the quantum phases of liquid He and of superconductors.} Rev. Mod. Phys. {\bf 34}, 694-704 (1962).
\item[{[Yi]}] Yin J.: {\em Free energies of dilute Bose gases: Upper bound.} J. Stat. Phys. {\bf 141}, 683-726 (2010).

\item[{[Z]}] Zagrebnov V. A. and Bru J.-B.: {\em The Bogoliubov model of weakly imperfect Bose gas.} Phys. Rep. {\bf 350}, 291-434 (2001).
\end{enumerate}

\end{document}